\documentclass[a4paper, twocolumn,usenatbib,useAMS]{mnras}

\PassOptionsToPackage{dvipsnames,svgnames,x11names}{xcolor}
\usepackage{xparse,soul}

\makeatletter
\NewDocumentCommand{\sotwo}{O{red}O{black}+m}
    {%
        \begingroup
        \color{#1}%
        \setul{-.5ex}{.4pt}%
        \def\SOUL@uleverysyllable{%
            \rlap{%
                \color{#2}\the\SOUL@syllable
                \SOUL@setkern\SOUL@charkern}%
            \SOUL@ulunderline{%
                \phantom{\the\SOUL@syllable}}%
        }%
        \ul{#3}%
        \endgroup
    }
\makeatother

\usepackage{array}
\usepackage{graphicx}
\usepackage{amsmath}
\usepackage{amssymb}
\usepackage{multicol}        
\usepackage{multirow}
\usepackage{bm}		
\usepackage{pdflscape}
\usepackage{caption}
\usepackage{subfigure}
\usepackage{empheq}
\usepackage[x11names, rgb]{xcolor}
\usepackage{tikz}
\usepackage[utf8]{inputenc}

\usepackage[normalem]{ulem}

\usepackage[T1]{fontenc}
\usepackage{ae,aecompl}

\usepackage{newtxtext,newtxmath}

\usepackage{ulem}

\title[Gravitational Potential from Gaia DR2]{Gravitational Potential from small-scale clustering in action space: Application to Gaia DR2}

\author[Yang et al.]{
Tianyi Yang$^{1,2,3}$\thanks{E-mail: t65yang@edu.uwaterloo.ca}
Supranta S. Boruah,$^{1,2,3,4}$\thanks{E-mail: ssarmabo@uwaterloo.ca}
Niayesh Afshordi,$^{1,2,3}$\thanks{E-mail: nafshordi@pitp.ca}
\\
$^{1}$Waterloo Centre for Astrophysics, University of Waterloo, Waterloo, ON, N2L 3G1, Canada \\
$^{2}$Department of Physics and Astronomy, University of Waterloo, Waterloo, ON, N2L 3G1, Canada\\
$^{3}$Perimeter Institute of Theoretical Physics, 31 Caroline St. N., Waterloo, ON, N2L 2Y5, Canada\\
$^{4}$Department of Applied Mathematics, University of Waterloo, Waterloo, ON, N2L 3G1, Canada
}

\date{\today}

\pubyear{2019}

\begin{document}
\label{firstpage}
\pagerange{\pageref{firstpage}--\pageref{lastpage}}
\maketitle

\begin{abstract}
Most measurements of mass in Astronomy that use kinematics of stars or gas rely on assumptions of equilibrium that are often hard to verify. Instead, we develop a novel idea that uses the clustering in action space, as a probe of underlying gravitational potential: the correct potential should maximize small-scale clustering in the action space. We provide a first-principle derivation of likelihood using the two-point correlation function in action space, and test it against simulations of stellar streams. We then apply this method to the 2nd data release of Gaia, and use it to measure the radial force fraction $f_h$ and logarithmic slope $\alpha$ of dark matter halo profile.  We investigate stars within  9-11 kpc and 11.5-15 kpc from Galactic centre, and find $(f_h,\alpha)= (0.391\pm 0.009,  1.835\pm 0.092) $ and $(0.351\pm 0.012,1.687\pm 0.079)$, respectively. We also confirm that the set of parameters that maximize the likelihood function do correspond to the most clustering in the action space. The best-fit circular velocity curve for Milky Way potential is consistent with past measurements (although it is  $\sim$ 5-10\% lower than previous methods that use masers or globular clusters). 
Our work provides a clear demonstration of the full statistical power that lies in the full phase space information, relieving the need for {\it ad hoc} assumptions such as virial equilibrium, circular motion, or steam-finding algorithms. 
\end{abstract}


\section{Introduction}\label{sec::intro}

Understanding the nature of dark matter is one of the most significant challenges in the 21st century for both physicists and astrophysicists. While the cold dark matter (CDM) paradigm the most popular model of dark matter, observational tensions on the galactic scale \citep{small_scale_CDM_tension_review, small_scale_CDM_tension_review_second} and non-detection of dark matter particles in the ground-based experiments have led scientists to examine alternative possibilities. Part of the reason why it is difficult to probe the nature of dark matter is that (as far as we can tell) it only interacts with ordinary matter via gravity; only affecting astrophysical observations on very large scales ($\gtrsim$ few kpc). On these scales, extracting the full 6-dimensional phase space information (that is necessary to infer dark matter mass unambiguously), has been a difficult task. 

Nonetheless, European Space Agency (ESA)'s Gaia mission has recently started probing the kinematics of the Milky Way stars with unprecedented precision \citep{Gaia_mission}. Gaia is a space-based observatory launched by European Space Agency (ESA) in December 2013, which aims at constructing the largest catalogue of 3d positions and velocities of the Milky Way stars, using Astrometric techniques. The first data set (Gaia DR1), released in September 2016, did not have the measurement of radial velocities. However, the Data Release 2 (Gaia DR2), which was released in 2018, with more complete magnitude measurement and longer span compared to Gaia DR1, contains the proper motion, parallax as well as the radial velocity information of more than 7 million stars \citep{Gaia_2_release}. The proliferation of data on the kinematics of Milky Way's stars has therefore opened a new avenue to probe the structure of the Milky Way potential and the nature of dark matter. 

In this work, we propose a new method to constrain the potential of the Milky Way and apply it to two six dimensional sub-samples of stars in Gaia DR2. Our method is based on maximizing the statistical clustering of the stars in the space of actions. The current theories and observational evidence suggest that the growth of structure in our universe is hierarchical, where smaller structures merge to form bigger ones. During the formation of galaxies, however, the smaller structures are tidally disrupted and due to various relaxation mechanisms at play, the memory of their common origin in configuration space is erased. This makes identifying stars with common origin nearly impossible. Nevertheless, the information regarding their common origin may still be present in the phase space of action variables. The action variables will remain conserved as long as the host potential evolves adiabatically. When the smaller structures are tidally disrupted in the Milky Way potential, their spread in the action space is much smaller than the rest of the stars in the Milky Way. Therefore, we expect the small scale structure of the action space to contain the hierarchical tidal disruption/assembly history of the Milky Way \citep{2009PhRvD..79h3526A}. 

Since action variables are conserved, due to their common origin, the action variables of the various structures would be clustered on small scales (in action space). This principle has previously been proposed to infer the potential by maximizing Kullback-Liebler divergence (KLD) or "relative entropy" in the action space \citep{KLD_original, action_space_clustering, Sanderson_action_clustering}. In these two studies, the viability of this idea was tested in simulated stellar distribution, where they successfully recover a spherical isochrone potential \citep{action_space_clustering} and a spherical NFW profile \citep{Sanderson_action_clustering} using this method. The same method has also been applied to constrain the parameters of the globular cluster. By minimizing the KL entropy in the phase-space, \citet{Applying_Gaia_louville_theorem} successfully constrain the mass and the King radius of the simulated M4 globular cluster, which provides another proof that the original parameters of a system can be recovered if the true phase-space information are found.  Maximizing the statistical clustering in the action space does not require identification of the membership of any star, which is one of the major merits for this method. Related methods have been proposed by \citet{energy_clustering} and \citet{Magorrian_paper}. However, neither have applied 2-point correlation function as a measure of clustering in the action space. Therefore, the significance of our work is to demonstrate the viability of using 2-point correlation function to measure the mass of a system.

This principle can be used to infer the potential of the Milky way. If the action variables are estimated using the incorrect potential, the resulting quantity will not be conserved with the dynamical evolution. Therefore, the clustering of the stars, in action space, on small scales will be destroyed if we use the wrong potential. Conversely, using the correct potential will maximize the small scale clustering in this space. We provide a first-principle derivation that the likelihood for the potential can be expressed as an integral (or KLD) over the 2-point correlation function in the action space, and test it using simulations of mock streams. Then, as an example, we fit a power-law dark matter profile (assuming a fixed form for bulge and disk component) to Gaia DR2, and compare our results to those that use other methods.

The paper is structured as follows: In Section \ref{sec::theory}, we introduce the required theoretical background for our methodology, including our parametrized models of the Milky Way potential and the computation of action variables. In Section \ref{sec::data}, we briefly discuss the data sets we used and the selection cuts imposed on the raw data. Details related to the two-point correlation function and the likelihood test we used in the action space are presented in Section \ref{sec::method}. To check the viability of our method, we first apply it to simulations with only stream stars, where the streams are simulated using the Python package \texttt{galpy} \citep[Version 1.3.0. See][for more details]{Galpy_Paper}. Then we proceed to apply our method to the real observations taken from Gaia DR2 with the selection cuts listed in Section \ref{sec::data} from two different radial bins. The results of our analysis is presented in Section \ref{sec::results}. In Section \ref{sec::discussion}, we discuss the shortcomings and future possibilities of our method before concluding in Section \ref{sec::conclusion}.

\section{Theory}\label{sec::theory}

\subsection{Modelling the Milky Way Potential}\label{ssec::potential}

We use a parametrized model for the Milky Way potential which can be regarded as a combination of the bulge, the disk and the dark matter halo. The specific model we are using is a slight modification to the \texttt{MWPotential2014} potential in \texttt{galpy}, which is assumed to be a good approximation to the Milky Way potential.

The bulge is modelled with a power-law density with an exponential cut-off: 

\begin{equation}\label{eqn::Bulge_pot}
	\rho_{b}(r) \propto r^{-1.8} \exp\bigg[-\bigg(\frac{r}{1.9~{\rm kpc}}\bigg)^2\bigg].
\end{equation}
The contribution from the central bulge is negligible at radius greater than 9 kpc, but we still include this component in the model for completeness.

The disk is modelled as a Miyamoto-Nagai Potential profile, but with fixed parameters \citep{Galpy_Paper}:
\begin{equation}\label{eqn::Disk_pot}
	\Phi_{d}(r,z) \propto -\frac{1}{\sqrt{R^2+\left[ 3~{\rm kpc}+\sqrt{z^2+(0.28 ~{\rm kpc})^2} \right]^2}},
\end{equation}
where $R$ and $z$ are the radial and vertical galactocentric cylindrical coordinates, respectively ($z = 0$ is the plane of the galaxy). 

Finally, we model the dark-matter halo profile as a spherical power-law profile
\begin{equation}\label{eqn::DM_profile}
	\rho_{dm}(r) \propto r^{-\alpha},
\end{equation}
which is also a built-in potential expression categorized as \texttt{PowerSphericalPotential} in \texttt{galpy} package. Note in \texttt{MWPotential2014}, the halo potential is characterized as NFW profile \citep[NFW: ][]{NFW_profile}. Instead, in this work, we use a power law potential for simplicity. Also, as we shall see later, the data set for constraining the potential does not span over a large range, so a localized power law potential should be a good approximation for NFW profile.

Finally, we set the normalizations of bulge, disk, and dark matter components, $\rho_b, \Phi_d, \rho_{dm}$, so that the fraction of radial force due to dark matter is $f_h$ at the position of the Sun, while the ratio of force due to stellar bulge to disk is fixed to be 1:12. This gives the normalization of halo component as 0.65$f_h$/(1-$f_h$), and the normalization of the bulge and disk component are fixed to be 0.05 and 0.60 respectively. Therefore, in the end we are left with two free parameters, $f_h$, and the power-law index of the density profile, $\alpha$, which we aim to constrain using our method.

\subsection{Action Variables}\label{ssec::actions}

Regular (i.e. non-chaotic) orbits in the galactic potential should admit 3 integrals of motion \citep[e.g.,][]{mo_white_vdBosch}. However, finding these integrals of motions in terms of the phase space coordinates could be a difficult task. Nevertheless, it may be possible to find canonical transformations so that, finding integrals of motions in these coordinates are easy. One particularly convenient system of canonical variables is the so-called \textit{action-angle variables} [denoted by (${\bm{\theta}}$, ${\bm{J} }$)], where the canonical momenta $\bm{J}$, or {\it actions}, are also the integrals of motion. The angle variables $\bm{\theta}$ are periodic in orbital torus, where an increase of $2\pi$ in the angle would be associated with the same point in phase space. The action conjugate to this angle is then defined as:
\begin{equation}\label{eqn::action_def}
	J_i = \frac{1}{2\pi}\oint_{\gamma_i} \bm{p} \cdot d\bm{q},
\end{equation}
where $\gamma_i$ is the orbit section where the $i$-th angle, $\theta_i$ increases from $0$ to $2\pi$.

Finding a canonical transformation to transform to the action-angle variables provide a convenient convention to define integrals of motion for an integrable potential.

\subsubsection{Calculating the action variables}\label{sssec::calc_actions}

The action-angle variables provide a convenient way to find the integrals of motion. However, finding closed analytic forms for the actions is only possible for a few special potentials. Therefore, in galactic dynamics, we often have to rely on approximate methods which involve integration of the orbits, e.g., adiabatic approximation \citep{adiabatic_method_citation} or the torus construction method \citep{torus_mapper_citation}. Here, we use the `St\"ackel approximation' \citep{Staeckel_approx}, which is implemented in \texttt{galpy}. 

St\"ackel potentials are a special class of potentials where the Hamiltonian can be written a separable form, using a canonical transformation. The St\"ackel potentials are expressed in the spheroidal coordinates, $(u,v)$ which are related to cylindrical coordinates through 
\begin{align}\label{eqn::spheroidal_coords}
	R = \Delta \sinh u \sin v; && z = \Delta \cosh u \cos v.
\end{align}
In these coordinates, the St\"ackel potential takes the form
\begin{equation}\label{eqn::Stackel_def}
	\Phi(u,v) = \frac{U(u)-V(v)}{\sinh^2u+\sin^2v}.
\end{equation}
The radial and the azimuthal actions can be expressed in closed analytic forms 
\begin{align}\label{eqn::Stackel_actions}
	J_r &= \frac{1}{\pi}\int_{u_{\textrm{min}}}^{u_{\textrm{max}}} p_u(u)du,  \\
	J_z &= \frac{2}{\pi}\int_{v_{\textrm{min}}}^{\pi/2} p_v(v) dv ,
\end{align}
where, 
\begin{align}\label{eqn::Stackel_momenta}
	\frac{p_u^2}{2\Delta^2} &= E\sinh^2u-I_3-U(u)-\frac{L^2_z}{2\Delta^2\sinh^2u}\\ 
	\frac{p_v^2}{2\Delta^2} &= E\sin^2v+I_3+V(v)-\frac{L^2_z}{2\Delta^2\sin^2v}.
\end{align}

In the above relations, $E$ is the energy of the orbit and $I_3$ is a third integral of motion (apart from the energy and the azimuthal angular momentum) which can be expressed analytically in terms of the conjugate variables and the momenta.

The St\"ackel approximation involves approximating any nominal potential of the Milky Way as a St\"ackel potential, and use this potential to find approximate action variables. To do this, we need to find the effective confocal length, $\Delta$ in Equation \ref{eqn::spheroidal_coords}.

A prescription for this was given by \citet{Sander_stackel}. To obtain the value of $\Delta$, we make use of the fact that $\Phi(\sinh^2u+\sin^2v)$ is a separable function of $u$ and $v$. Therefore, any mixed derivative of this quantity must vanish. Using the model potential, $\Phi_{\textrm{model}}$, we obtain
\begin{equation}
	\frac{\partial^2}{\partial u \partial v}\bigg[(\sinh^2u+\sin^2 v)\Phi_{\textrm{model}}\bigg] \approx 0.
\end{equation}
This equation can then be solved for $\Delta$ in terms of the derivatives of the potential, $\partial \Phi/\partial R$, $\partial \Phi/\partial z$, $\partial^2 \Phi/\partial R^2$, $\partial^2 \Phi/\partial z^2$ and $\partial^2 \Phi/\partial R \partial z$. The relation between $\Delta$ and the derivatives were obtained in \cite{Sander_stackel}: 
\begin{equation}\label{eqn::Delta_relation}
	\Delta^2 = z^2-R^2+\bigg(3R \frac{\partial \Phi}{\partial z}-3z \frac{\partial \Phi}{\partial R}+Rz\bigg(\frac{\partial^2\Phi}{\partial R^2}-\frac{\partial^2\Phi}{\partial z^2}\bigg)\bigg)/\frac{\partial^2\Phi}{\partial  R \partial z}.
\end{equation}
This algorithm is implemented using the python package \texttt{galpy}.

\section{Data Set}\label{sec::data}

The kinematics of stars in this paper are obtained from the Data Release 2 (DR2) of the Gaia mission, which was released in April 2018. All data can be accessed though Gaia Archive\footnote{Gaia Archive website: http://gea.esac.esa.int/archive/}. With the aid of the Gaia radial velocity spectrometer \citep{gaia_rvs}, we can directly obtain the full six-dimensional phase space information of approximately 7 million stars. For the sample we used for our analysis, we make a few simple quality cuts in order to avoid the stars with large errors on the parallax or proper motion measurement. We shall impose the cut that the relative error on the parallax, and the proper motion are less than $20\%$:
 
 \begin{equation}\label{eqn::selection_cut}
     \frac{\Delta p}{p},\frac{\Delta \mu_{\textrm{RA}}}{|\mu_{\textrm{RA}}|},\frac{\Delta{\mu}_{\textrm{dec}}}{|{\mu}_{\textrm{dec}}|}, \frac{\Delta V_{\textrm{radial}}}{|V_{\textrm{radial}}|}<0.2,
 \end{equation}
 where, $p, {\mu}_{RA}, {\mu}_{\textrm{dec}}$, $V_{\textrm{radial}}$ are the parallax, proper motion in the right ascension direction, the proper motion in the declination direction, and  radial velocity respectively. Here, $\Delta$' denotes the measurement error in each of these quantities. After applying the selection cuts mentioned above to the raw data, we are left with around 5.6 million stars in the data sample for further analysis\footnote{Note that for the actual data analysis, there are additional cuts for halo vs. all stars, and radial distribution, which will be discussed in Section \ref{ssec::real_data}.}. Even though the $20\%$ cut (while common as in \citet{action_space_clustering, Sanderson_action_clustering}, also in \citet{20_percent_error_cut, 20_percent_paper_2, 20_percent_paper_3}) is {\it ad hoc}, we further verify that this choice has little effect on our results, as stars with large uncertainties in their phase space  coordinates are unlikely to form close pairs in action space\footnote{We did change the choice of relative error cut to another value, for instance, to 10\%. However, this does not affect our final estimations.}. 
 
This data allow us to constrain the two-parameter power law potential in Equation \ref{eqn::DM_profile} by using the likelihood test, which we will discuss in the next section. As calculations are conducted by using the built-in functions in \texttt{galpy}, the inputs for most of the functions we used are in cylindrical coordinates. For the convenience of computation, all calculations are done in galactocentric coordinate system, and the coordinate transformation are as well handled by the built-in functions in \texttt{galpy} library \citep{Galpy_Paper}.

\section{Method}\label{sec::method}
We are interested in the small scale clustering of the stars in the action space. There are many different measures of clustering which are useful for different purposes. Here, we use two-point correlation function as our measure of clustering, which should be one of the most straightforward ones. However, to define the correlation function, we need to have a measure of the distance. While this choice is not unique, we shall use the following measure to find the distance of two stars in action space

\begin{equation}\label{eqn::distance_calculation_real_data}
	{D} = \sqrt{(\Delta {J_R}/\sigma_{J_R})^2 + (\Delta {J_{\phi}}/\sigma_{J_{\phi}})^2 + (\Delta {J_z}/\sigma_{J_z})^2},
\end{equation}
where $\Delta J_i$ denotes the difference in the action coordinates of the two stars, while $\sigma_{J_i}$'s are standard deviations of $J_i$'s over all stars. The reason why we normalize the difference in action by the standard deviation in the action space is that stream stars, due to their common origin, should be significantly more clustered than the background, i.e. $\Delta J \ll \sigma_J$. Since the background could be anisotropic in the action space, this normalization provides a more appropriate distance measure. We further discuss this choice in Section \ref{sec::discussion} below. 

The calculation of $\sigma_{J_i}$ can be affected by the outliers in the raw data or numerical artifacts in \texttt{galpy} action approximation. Therefore, another constraint is added to effectively exclude outliers out of the sample with $\frac{|J_{i}-\bar{J_{i}}|}{\sigma_{J_i}}>3$. This choice of 3$\sigma$ seems relatively arbitrary. However, we confirm that the constraints are not affected by this choice, as long as those outliers are safely removed. Also, there is only a small fraction of stars being cut off by the “3$\sigma$-cut” from the original catalogue. Therefore, we do not believe our results are significantly biased by this choice. We then use the remaining action variables that satisfy the above criterion to re-calculate the standard deviation.

For points distributed randomly with a uniform distribution in a three dimensional action space, the probability of finding pairs at a separation between $D$ and $D+dD$ is given by
\begin{align}\label{eqn::Correlation_eqn}
	\mathcal{P}(D)|_{\rm uniform}dD &\propto D^2 dD \nonumber  \\
	\implies \mathcal{P}(\ln D)|_{\rm uniform}d\ln D &\propto D^3 d \ln D. 
\end{align}
\begin{figure}
   \includegraphics[width=\columnwidth]{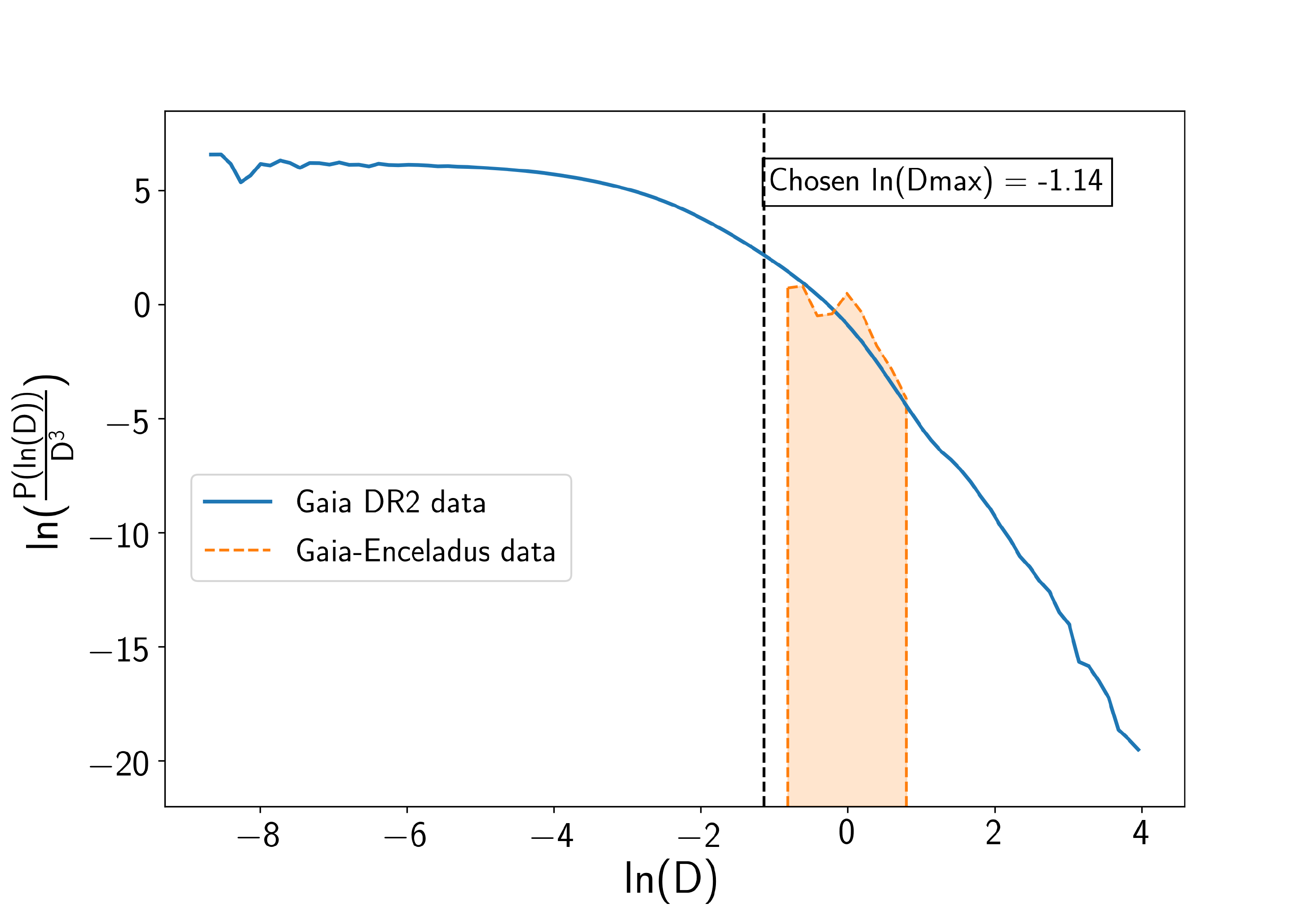}
   \caption{$\ln\left[\frac{P(\ln D)}{D^3}\right]$ versus $\ln (D)$ calculated by using the Gaia DR2 real data from galactocentric radius 11.5-15 kpc with {$f_h$ = 0.34, $\alpha$ = 1.66}. Here $D$ is the normalized distance of pairs of stars in the action space, while $P(\ln D)$ is its probability density over all pairs of Gaia DR 2 within our sample. The correlation function computed using $\emph{Gaia-Enceladus}$ data \citep{globular_cluster_data} is over-plotted on the same figure (orange dashed line with shaded area). For comparison, the optimum chosen value of $D_{\rm max}$ for likelihood estimate is also shown as black vertical dashed line. Please see Section \ref{sec::discussion} for more details.}
   \label{probability function}
\end{figure}

However, the actual probability distribution $\mathcal{P}(\ln D)$ will be different from $\mathcal{P}(\ln D)|_{\rm uniform} \propto D^3$ due to clustering in the action space. This clustering can be quantified using the 2-point correlation function $\xi(\ln D)$:
\begin{equation}\label{eqn::norm_factor}
1+\xi(\ln D) \equiv \frac{\mathcal{P}(\ln D)}{\mathcal{P}(\ln D)|_{\rm uniform}} = \frac{ D_{\textrm{max}}^3 \mathcal{P}(\ln D)}{3 D^3\int_{-\infty}^{\ln D_{\textrm{max}}}\mathcal{P}(\ln D')d \ln D'},    
\end{equation}
where we used the fact that both $\mathcal{P}$ and $\mathcal{P}|_{\rm uniform}$ should integrate to unity over the range $\ln D \in (-\infty,\ln D_{\textrm{max}})$. 

The blue solid line in Figure \ref{probability function} shows $\ln\left[\frac{P(\ln D)}{D^3}\right]$ as a function of $\ln (D)$ calculated by using the Gaia DR2 real data from galactocentric radius 11.5-15 kpc with [$f_h$ = 0.34, $\alpha$ = 1.66]\footnote{As shall be seen in Section \ref{ssec::real_data}, this corresponds to the best-fit halo potential recovered by our method}, where $D$ is the normalized distance of pairs of stars in the action space (as shown in Equation \ref{eqn::distance_calculation_real_data}), while $\mathcal{P}(\ln D)$ is its probability density over all pairs of Gaia DR 2 within our sample. According to Equation \ref{eqn::Correlation_eqn}, if stars are uniformly distributed, a plateau is expected at small values of $\ln D$, and deviations from this plateau would correspond to clustering. As can be seen from this figure, at small distance, $\mathcal{P}(\ln D)$ roughly obeys uniform distribution as stated in Equation \ref{eqn::Correlation_eqn}, and the probability drops significantly due to the lack of stellar pairs at large values of $\ln D$.

As it turns out, with certain assumptions, the statistical likelihood of any action-space distribution can be expressed in terms of $\xi(\ln D)$. The key idea here is to assume the star distribution in the action space is the Poisson sampling of a near-uniform background plus a random gaussian field. The correlation function of this random gaussian field encodes all the clustering information at small scale in the action space. This model is agnostic about the distribution function, $f({\bf J})$ and instead relates the likelihood to the correlation function in the action-space $\xi(\ln D)$, after marginalizing over all possible $f({\bf J})$'s. More explicitly, we find that the log-likelihood for a potential is given by 
\begin{eqnarray}
\label{eqn::likelihood}
    \ln \mathcal{L}({\rm data}|f_h, \alpha) \simeq \left\langle \sum_{\rm pairs} \ln\left[1+\xi(\ln D_{\rm pair})\right]\right\rangle_{\rm pairings} =\nonumber \\ N_{\rm pairs}\int_{-\infty}^{\ln D_{\textrm{max}}}\mathcal{P}(\ln D)\ln\left[1+\xi(\ln D)\right] d\ln D.
\end{eqnarray}
A detailed derivation of this expression is presented in Appendix \ref{sec::likelihood_modified}. In this equation, $N_{\rm pairs}$ is the total number of pairs, i.e. half of the number of stars in the sample. Also, as shown in Figure \ref{probability function}, $\mathcal{P}(\ln D)$ obeys the scaling of uniform distribution only at small values of $\ln D$. Therefore, when evaluating the value of likelihood function, integration is terminated at a chosen value of $D_{\textrm{max}}$. This $D_{\rm max}$ characterizes the scale of homogeneity in the action space background, and we shall discuss the choice of $D_{\rm max}$ in later section.

We further note that relative entropy \citep{KLD_original} of the distribution $\mathcal{P}(\ln D)$, with respect to the uniform distribution, is defined as 
\begin{equation}
S_{\rm relative} \equiv -\int \mathcal{P}(\ln D)\ln\left[\frac{\mathcal{P}}{\mathcal{P}|_{\rm uniform}}\right] d\ln D = - \frac{\ln \mathcal{L}({\rm data}|f_h,\alpha)}{N_{\rm pairs}}, 
\end{equation}
i.e.  the maximization of the likelihood function corresponds to minimizing the entropy relative to the uniform pair distribution. In other words, the best-fit values for the dark matter halo density produce the most non-uniform distribution of pairs in the action space. We should note that while this is similar to the criterion proposed by \cite{action_space_clustering}, their relative entropy is based on phase space density in the action space $f({\bf J})$, while our derivation in Appendix \ref{sec::likelihood_modified} shows that likelihood depends on the relative entropy of the pair distance probability distribution $\mathcal{P}(\ln D)$. 

\section{Results}\label{sec::results}
\subsection{Simulations}\label{ssec::simulation}
To validate our method, we simulate the orbits of a few stars in a known parametrized potential of the same form and then applied the above mentioned analysis to check if we can recover the true parameters of the potential. The simulation includes three groups of tidal stream stars with different initial conditions of progenitors. This is achieved by using the built-in modelling method in \texttt{galpy} package \citep{Stream_model}. One can specify the gravitational potential that stars evolve in, the method for action variables calculation, the initial conditions of progenitors, the velocity distribution of progenitors and the time when the disruption began. Initial conditions of the progenitors' orbit for three streams are tabulated in Table \ref{tab:init_table}. The header of the table is organized in the order of R, $\phi$, z, $v_R$, $v_T$, $v_z$, velocity dispersion ($\sigma_{v}$) and the disruption time ($t_{\rm disrupt}$). Each of the three streams consist of  3000 stream stars. With the simulated stellar trajectories, the action variables of stars can be calculated based on the St\"ackel approximation as explained in Section \ref{sec::theory}. We now wish to test whether our proposed likelihood function (Equation \ref{eqn::likelihood}) leads to constraints that are consistent with parameters that are used in our simulated host potential. 

\begin{table*}
\centering
\caption{Initial conditions of progenitor for the generation of stream stars (where galactocentric radius and velocity are normalized by solar radius value.} 
\begin{tabular}{c rrrrrrrr} 
\hline\hline 
Stream Number & \multicolumn{8}{c}{Initialization}
\\ [0.5ex]
\hline 
& R & $\phi$ & z & $V_R$ & $V_T$ & $V_Z$ & $\sigma_v(\rm km/s)$ & $t_{\rm disrupt}$ (Gyr) \\[0.5ex]
\hline

{Stream 1} & 1.56 & 0.12 & 0.89 & 0.35 & $-1.15$ & $-0.48$ & 0.3 & 2 \\[1.5ex]

{Stream 2} & 1.00 & $-0.05$ & 0.001 & $-0.60$ & 0.51 & 0.0086 & 0.3 & 2 \\[1.5ex]

{Stream 3} & 1.20 & $-0.05$ & $-1$ & $-0.30$ & 0.51 & 0.16 & 0.3 & 2 \\[1.5ex]
\hline 
\end{tabular}
\label{tab:init_table}
\end{table*}

\begin{figure*}
    \centering
    \begin{subfigure}
        \centering
        \includegraphics[width=.35\linewidth]{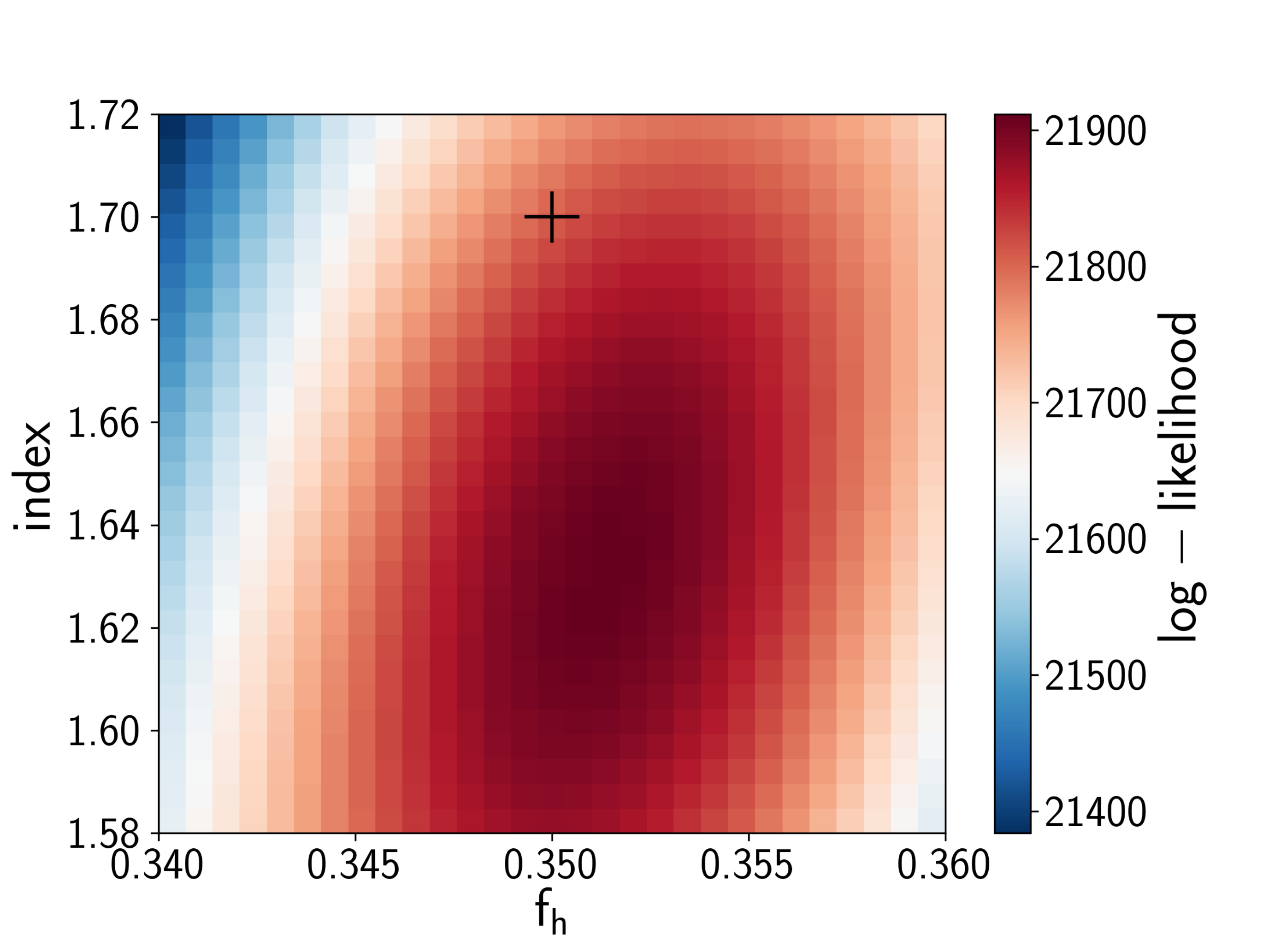}
        \label{sim_PL_3517}
    \end{subfigure}%
    \begin{subfigure}
        \centering
        \includegraphics[width=.6\linewidth]{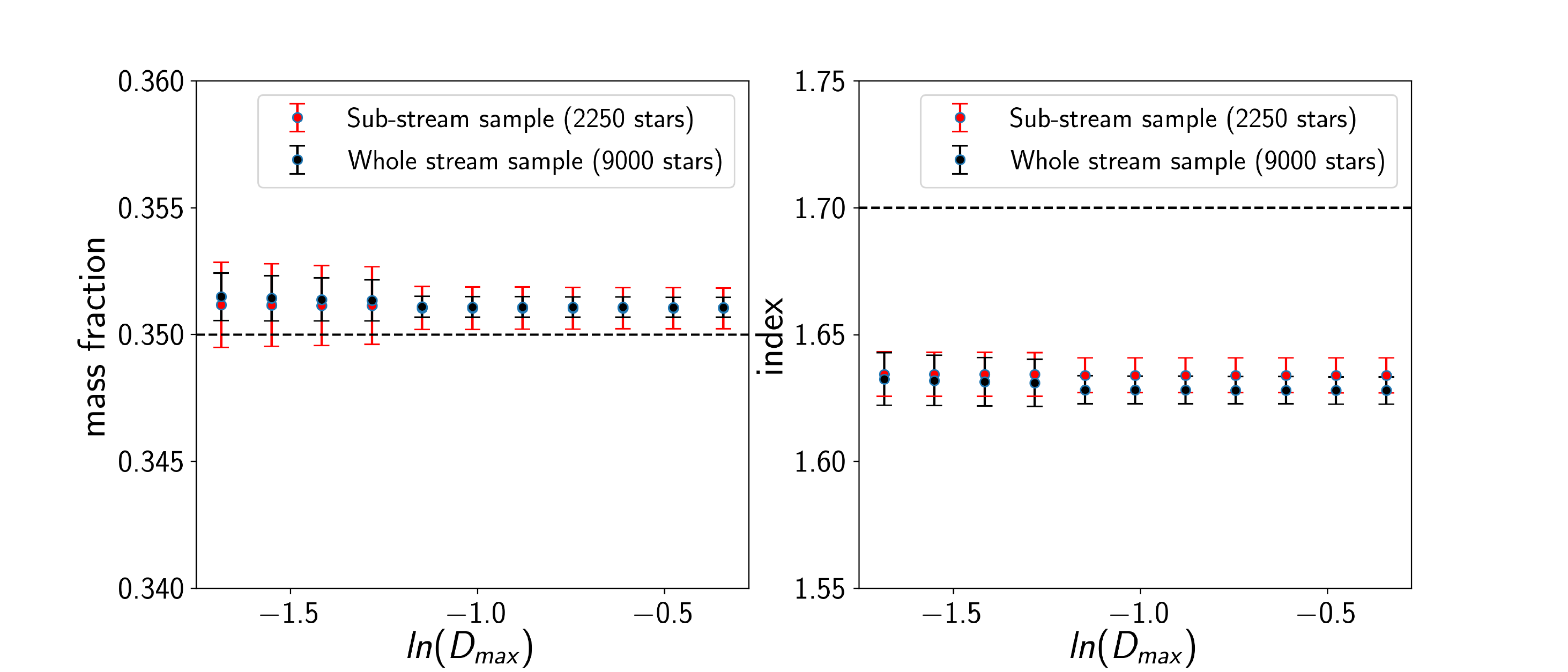}
        \label{error_bar_PL_3517}
    \end{subfigure}
    \begin{subfigure}
        \centering
        \includegraphics[width=.35\linewidth]{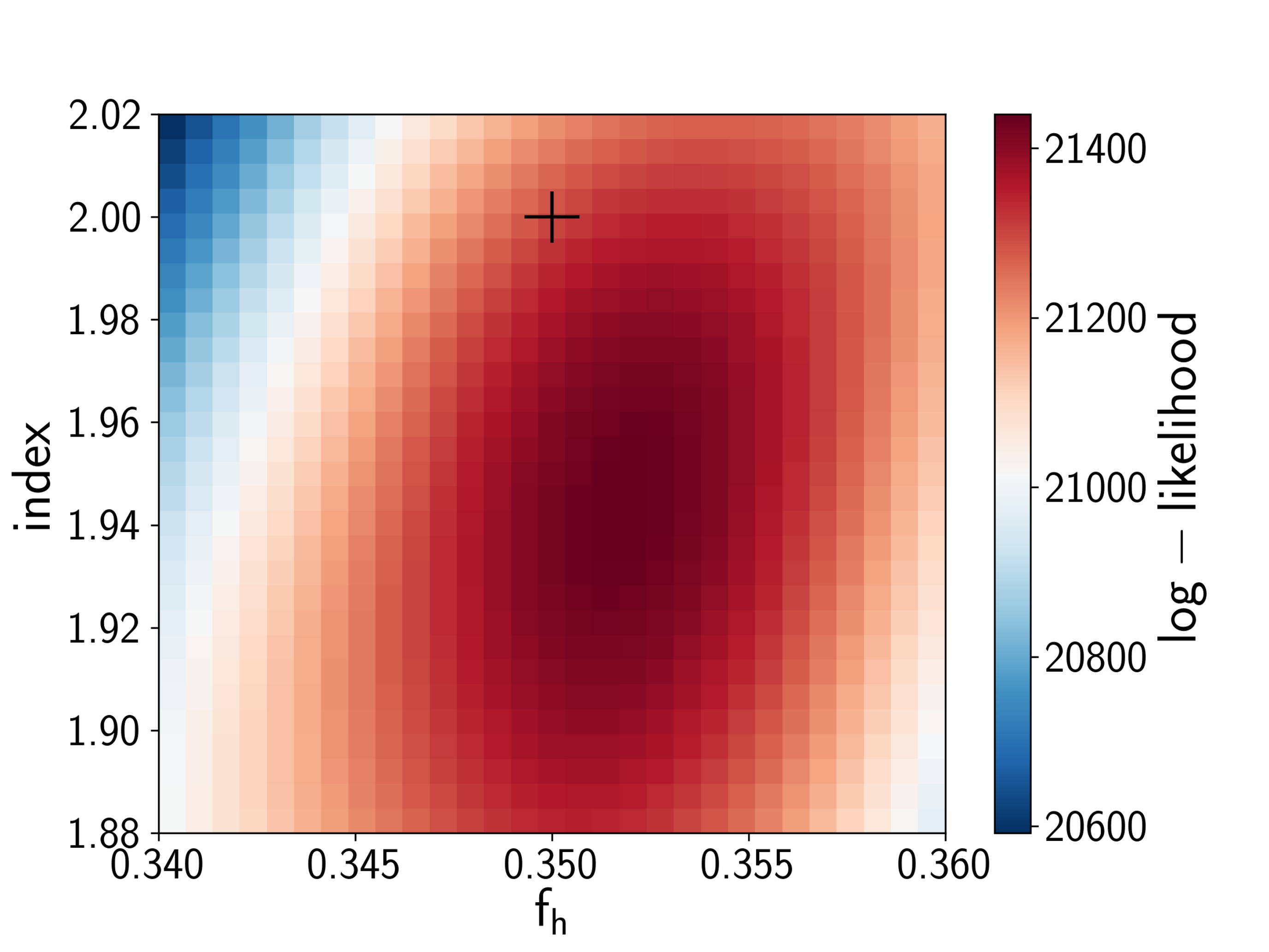}
        \label{sim_PL_3520}
    \end{subfigure}%
    \begin{subfigure}
        \centering
        \includegraphics[width=.6\linewidth]{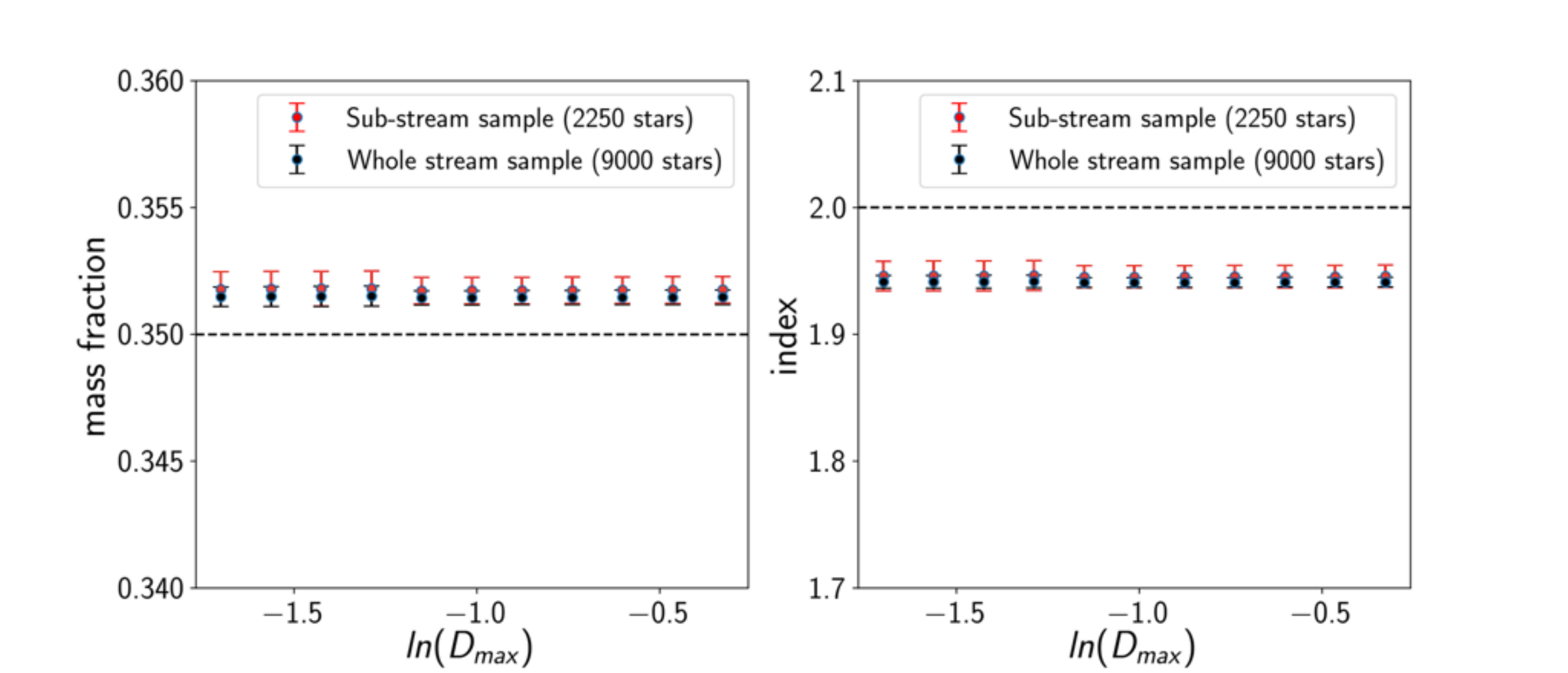}
        \label{error_bar_PL_3520}
    \end{subfigure}
\caption{Top panel: likelihood test and error bar plot for case [$f_h$ = 0.35, $\alpha$ = 1.70]. Bottom panel: likelihood test and error bar plot for case [$f_h$ = 0.35, $\alpha$ = 2.0]. The maximum likelihood gives constraints on the parameters $f_h$ = 0.35, $\alpha$ = 1.63 for the first case and $f_h$ = 0.35, $\alpha$ = 1.95 for the second case. The initial set of parameter is indicated as black plus sign on the likelihood plot for either case. Error bars are determined based on the paraboloid fitting, where the black points with wider black error bars are determined with whole 9000 stars simulated from three different streams (initial conditions are listed in Table \ref{tab:init_table}), while the red points with narrower red error bars are a randomly-chosen sub-sample of stars. The ultimate results for both parameters are not significantly changed with the variation of $D_{\textrm{max}}$ and the size of the sample. Based on this figure, an approximate $4\%$ systematic discrepancy might be expected between the likelihood evaluation and the actual value of $\alpha$, and a $1\%$ systematic discrepancy might exist in the $f_h$ evaluation. The $D_{\textrm{max}}$ values are taken as $\ln D_{\textrm{max}}$ = -1.0 for both cases.}
\label{sim_obs_PL}
\end{figure*}

There are two free parameters in the expression of the dark matter halo density profile (Equation \ref{eqn::DM_profile}), $f_h$ which fixes the normalization, and the logarithmic slope $\alpha$. We choose the mass fraction of the halo $f_h = 0.35$ based on Table 1 in \cite{Galpy_Paper}, and we produce two sets of stream simulations with different choices of $\alpha=$ 1.70 and 2.00 respectively. The progenitor stars are evolved in these two host gravitational potentials respectively. We then compute the action variables on a grid in the ($f_h$, $\alpha$) space, and compute the corresponding likelihood function using Equation \ref{eqn::likelihood}. The likelihood functions evaluated with 9000 simulated stream stars for both potentials are shown in Figure \ref{sim_obs_PL} (assuming $\ln D_{\textrm{max}} \sim -1$). For each case, we find clear constraints on both parameters as expected, in reasonable agreement with input parameters of the simulations, subject to caveats that we discuss next.  

To determine the location and the uncertainties of the measurements at each $D_{\textrm{max}}$, we fit the log-likelihood distribution with a quadratic function around its maximum. The assumption made by this procedure is that the likelihood only has a single peak that can be approximated by a gaussian distribution. In order to check the validity of this assumption, we plot the posterior of parameter at different values of $D_{\textrm{max}}$, which is shown in Appendix \ref{sec::sanity_check_simulation}. As can be seen from Figure \ref{posterior_sim_3517}, both of the posterior distributions for $f_h$ and $\alpha$ have a single peak that can be reasonably approximated by gaussian. Also, the probability distributions do not drastically vary with the choices of $D_{\textrm{max}}$. Therefore, we conclude that, at least for our simulated streams, our likelihood distribution is well approximated by gaussian statistics:
\begin{equation}\label{eqn::paraboloid}
    \chi^2 = \chi_{min}^2 + (X_{i} - \bar{X_i})F_{ij}(X_{i} - \bar{X_i})^T,
\end{equation}
where $\chi_{min}^2$ is given by the likelihood peak value by assuming $\mathcal{L} \propto e^{-\frac{\chi^2}{2}}$, and $F_{ij}$ represents the Fisher information matrix for $X_1=f_h$ and $X_2=\alpha$. The covariance of the parameters is then given by the inverse of the Fisher matrix,  $F_{ij}^{-1}$. There are six parameters to be determined in Equation \ref{eqn::paraboloid}. In practice, we fit for these parameters using a $3\times3$ grid around the peak of the likelihood. 

As can be seen from Figures \ref{sim_obs_PL},  the final measurements of mass fraction and index do not significantly vary with the choices of $D_{\textrm{max}}$, with small error bars that further shrink by increasing the number of stars or $D_{\textrm{max}}$.  Based on these plots, we conclude that the measurement error for the simulations are dominated by systematic error, which is at the level of $1\% $ for mass fraction $f_h$ and $4\%$ for the power law index $\alpha$. This systematic error, while small, might arise due to the use of the St\"ackel approximation to compute our action variables. We need to keep this in mind when we apply our method to real data, highlighting where further improvements may be needed when other sources of error are small.   

As a further sanity check, we can verify that the parameters of potential found by the maximum likelihood test do correspond to the most clustering in the action space. In other words, the two-point correlation function defined in Equation \ref{eqn::Correlation_eqn} should be maximized for the correct potential. To verify this, we show how a 2D projection of stellar distribution in the action space varies with different choices of potential for both simulations. Figures are shown in Appendix \ref{sec::sanity_check_simulation}. As expected, the most compact distributions occur when parameters approach the correct values for the simulation, which is also reflected in the behavior of the two-point correlation function.

To summarize, we have confirmed that our method to maximize likelihood (Equation \ref{eqn::likelihood}) based on clustering in action space can yield reasonable constraints on simulated potentials, subject to small systematic errors of $1\%$ ($4\%$) on normalization and logarithmic slope. There also does not seem to be any significant dependence on the maximum separation of included pairs in action space $D_{\rm max}$. A more exciting step is to apply our method to real Gaia DR2 data to see how well it can constrain the Milky Way potential, which we shall do next.

\subsection{Real data}\label{ssec::real_data}

After confirming the reliability of the method, we proceed with our analysis using real data from Gaia DR2. The criteria for data selection were already discussed in Section \ref{sec::data}. Let us now introduce some additional selection cuts. Recall that in the derivation of likelihood function (Appendix \ref{sec::likelihood_modified}), we assume the stellar distribution in the action space is a uniform background plus fluctuations. This assumption is more appropriate for halo stars in our galaxy, as disk stars have $J_z \simeq 0$. Additionally, as we are trying to constrain dark matter profile, which mostly occupies the Milky Way halo, halo stars should be better candidates compared to disk stars. Due to these considerations, we only select stars that have vertical distance to the galactic plane $>$ 1 kpc\footnote{We will discuss the effect of this cut, as well as the measurements error cuts on the final results later in Section \ref{sec::discussion}.}. 

\begin{figure*}
    \centering
    \begin{subfigure}
        \centering
        \includegraphics[width=.45\linewidth]{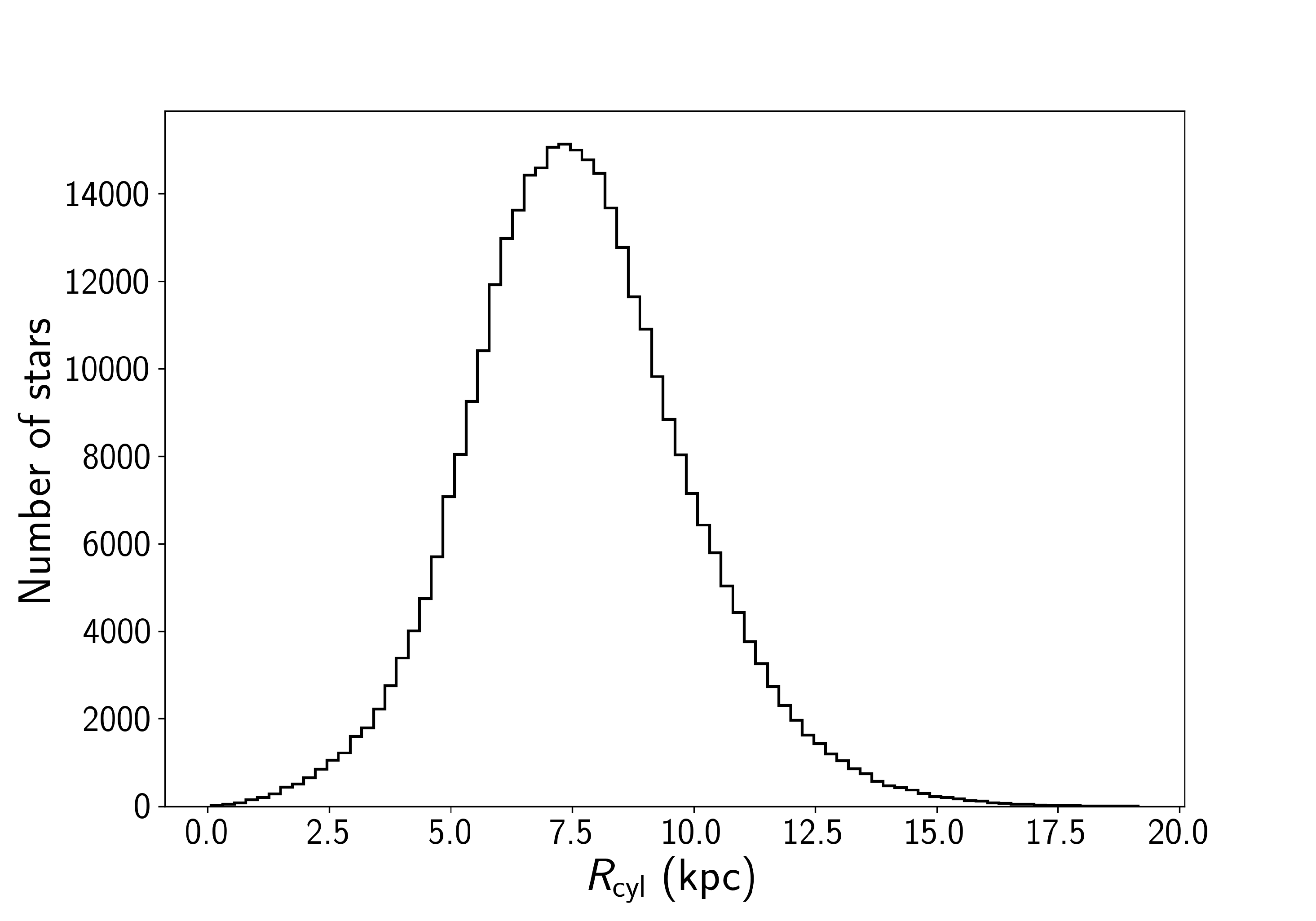}
        \label{R_distribution_real_data}
    \end{subfigure}%
    \begin{subfigure}
        \centering
        \includegraphics[width=.45\linewidth]{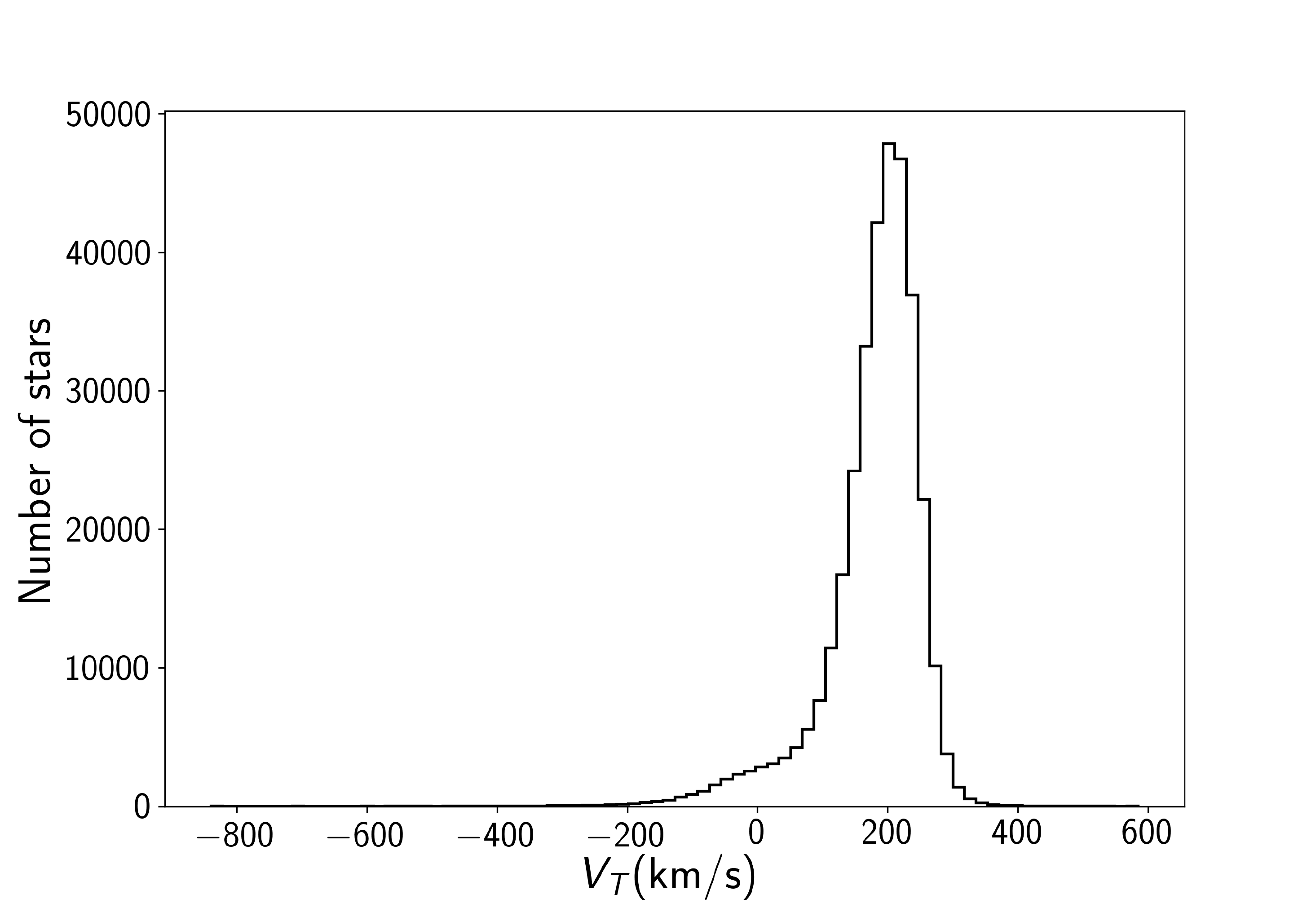}
        \label{VT_distribution_real_data}
    \end{subfigure}
\caption{The galactocentric radius and tangential velocity distribution in cylindrical coordinates for the selected Gaia DR2 catalogue. Calculations are all conducted in cylindrical coordinate.}
\label{real_data_distribution}
\end{figure*}

Figure \ref{real_data_distribution} shows the galactocentric distance and the tangential velocity distribution (in cylindrical coordinate) for all the data with relative measurement error smaller than 20$\%$ and $|z|$ > 1 kpc. There are around 337,022 stars in total. As expected, the peak of radial distribution is around solar radius and the peak of $V_{T}$ distribution is around the value of circular velocity at solar radius. Here, we assume $R_{\odot}$ = 8.122 kpc, the vertical distance to the galactic plane $z_{\odot}$ = 0.025 kpc, and the galactocentric velocity of the Sun $V_{x, \odot}$ = -11.1 km/s, $V_{y, \odot}$ = 245.8 km/s, $V_{z, \odot}$ = 7.8 km/s as taken in \citet{Eilers_paper}, but we shall discuss this choice further in Sec. \ref{sec::discussion}. Due to the limitation of the computational time, another galactocentric radius cut is also applied to the data: We choose two different radial ranges 9 kpc < R < 11 kpc (hereafter real-data-9-11) and 11.5 kpc < R < 15 kpc (hereafter real-data-115-15). After applying all of these cuts, there are approximately 61,000 and 16,000 stars in each sample, respectively. 

 Now, taking the NFW profile as reference, the expected value of the $\alpha$ in the power law density profile should be within 1 to 3: For $r \gg r_s$, the density is proportional to $r^{-3}$, while for $r \ll r_s$, it goes to $r^{-1}$. Furthermore, \cite{index_constraint} used the assumption of Jean's equilibrium for G-dwarfs from SEGUE survey to constrain $\alpha < 1.53$ (at 95\% confidence) between R=4 kpc and 9 kpc. Therefore, to allow for a conservative prior, we consider the range:
 \begin{eqnarray}
 \label{eqn::prior_range}
    0.5<&\alpha&<2.5,\nonumber \\
    0.25<&f_h&<0.55,
\end{eqnarray}
for our dataset within 9 kpc to 15 kpc.

\begin{figure*}
    \centering
    \begin{subfigure}
        \centering
        \includegraphics[width=.35\linewidth]{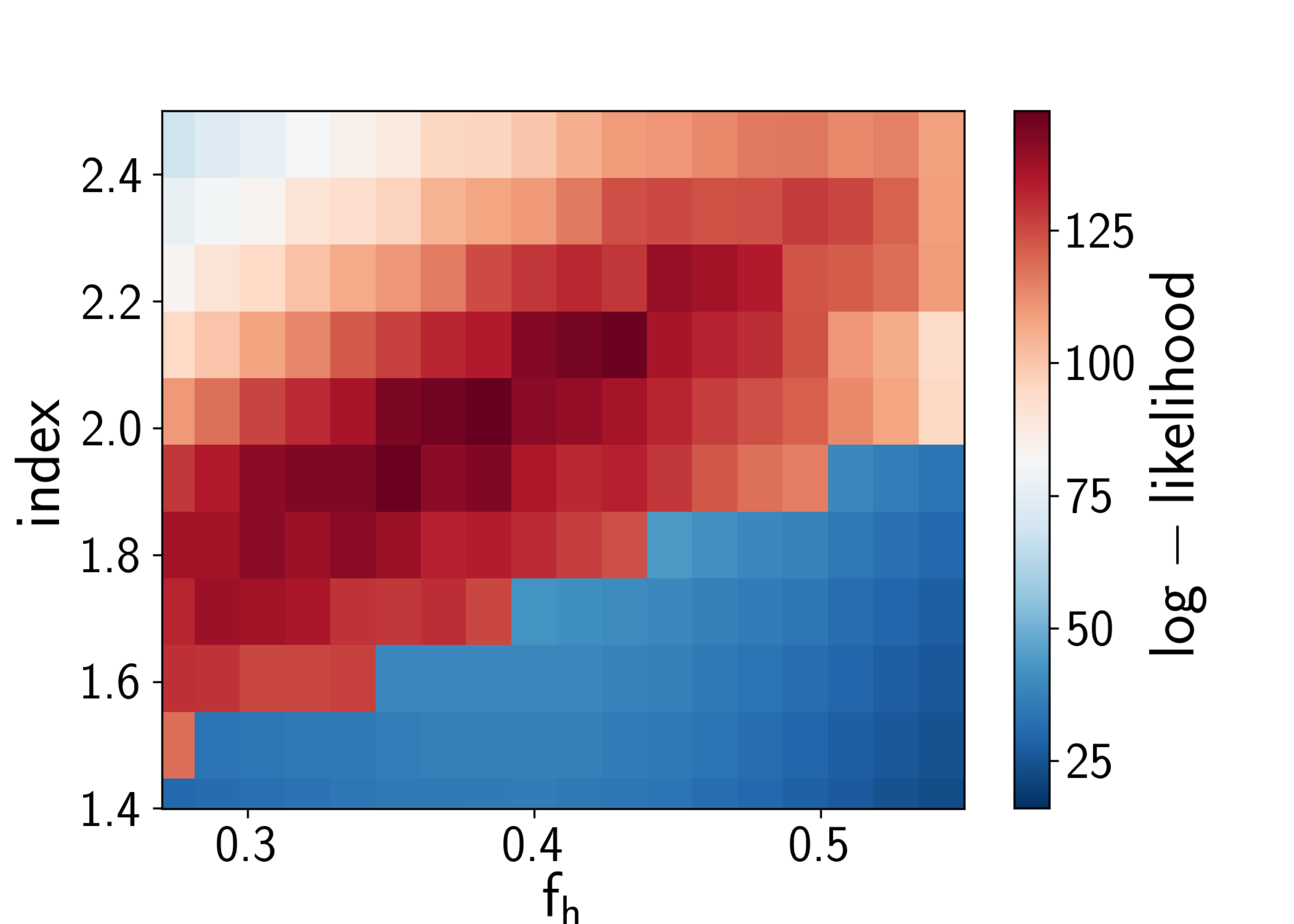}
        \label{real_data_likelihood_PL_9_11}
    \end{subfigure}%
    \begin{subfigure}
        \centering
        \includegraphics[width=.6\linewidth]{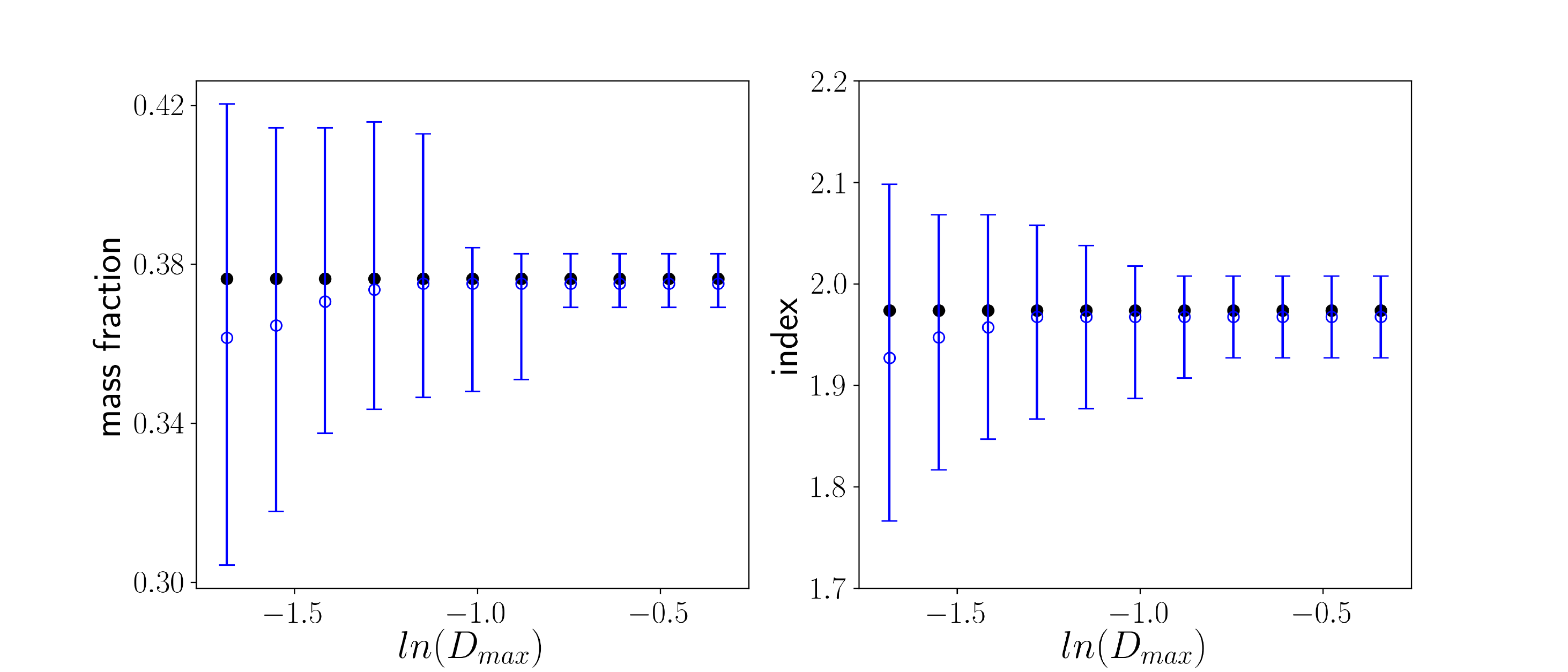}
        \label{error_bars_real_data_9_1}
    \end{subfigure}
    \begin{subfigure}
        \centering
         \includegraphics[width=.35\linewidth]{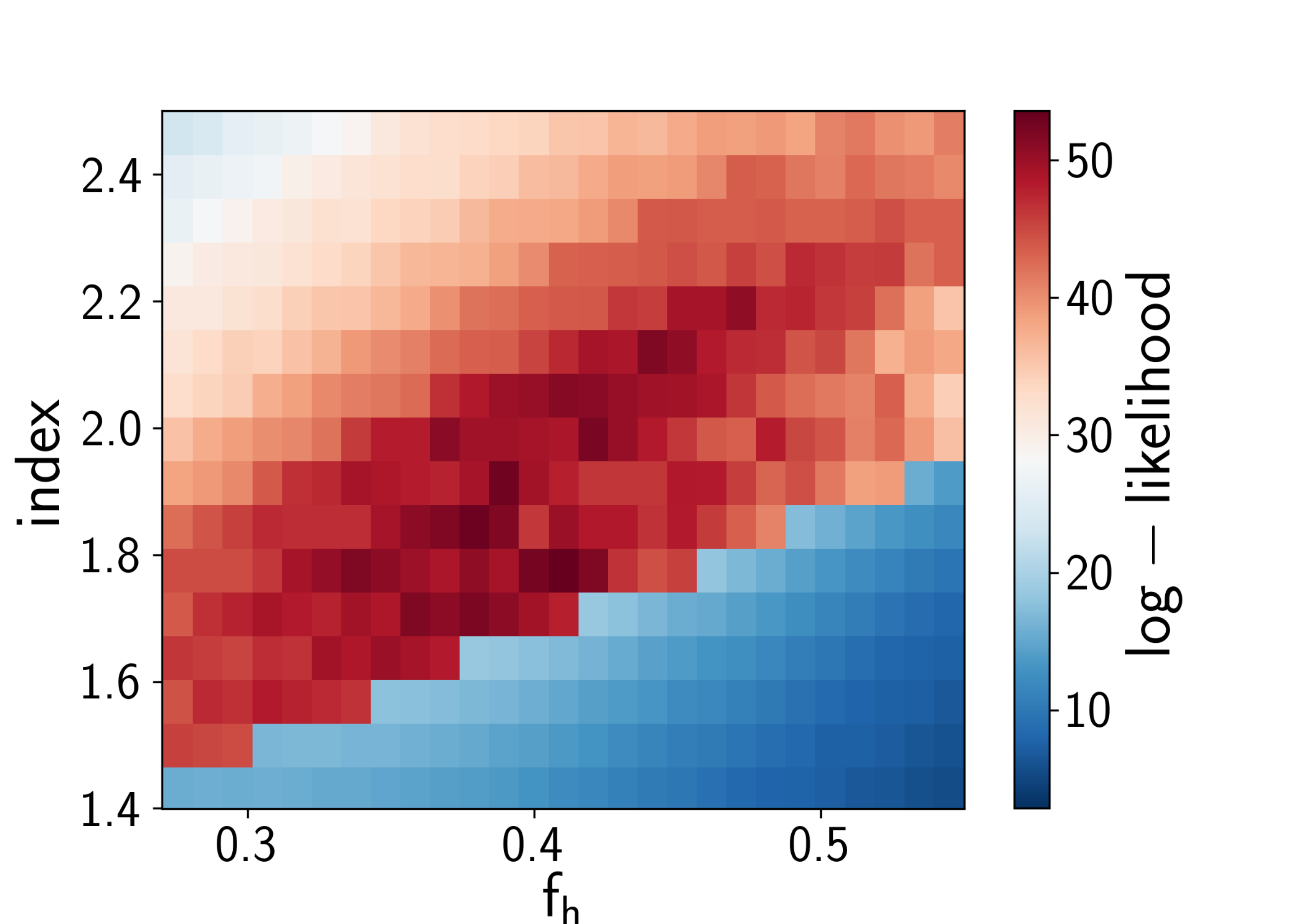}
        \label{real_data_likelihood_PL_115_15}
     \end{subfigure}%
     \begin{subfigure}
         \centering
         \includegraphics[width=.6\linewidth]{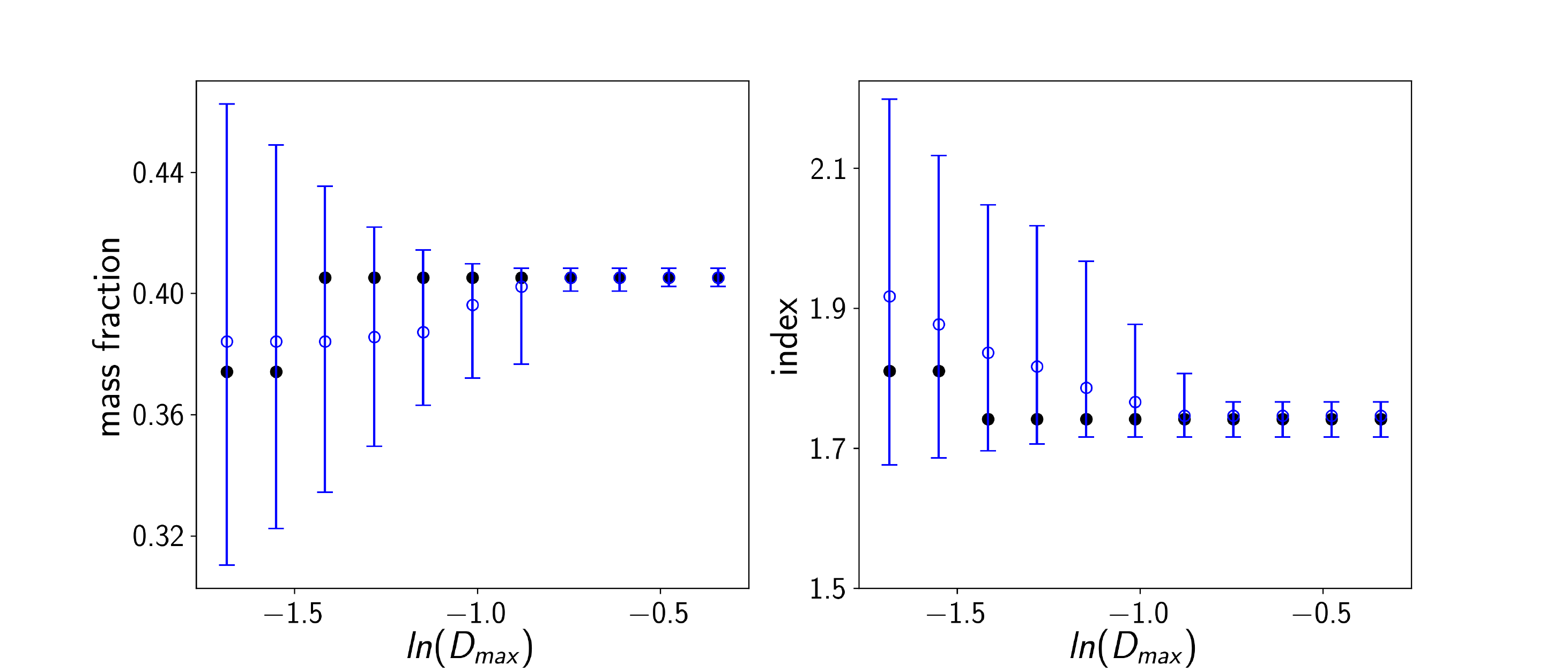}
         \label{error_bars_real_data_115_15}
     \end{subfigure}
\caption{ Top panel: likelihood test and error bar plot using sample stars from 9-11 kpc. Bottom panel: likelihood test and error bar plot using sample stars from 11.5-15 kpc. The maximum likelihood gives constraints on the parameters $f_h$ = 0.376, $\alpha$ = 1.974 for the first case and $f_h$ = 0.405, $\alpha$= 1.741 for the second case. In error bar plots on the right, the black points are the maxima of the likelihood. The median values are shown as blue hollow points with uncertainties determined from the posterior distribution of parameters (68\% confidence interval). We use ln $D_\textrm{max}$ = -1.14 for the likelihood plots in both samples.}
\label{real_data_likelihood}
\end{figure*}

\begin{figure*}
    \centering
    \begin{subfigure}
        \centering
        \includegraphics[width=1.0\linewidth]{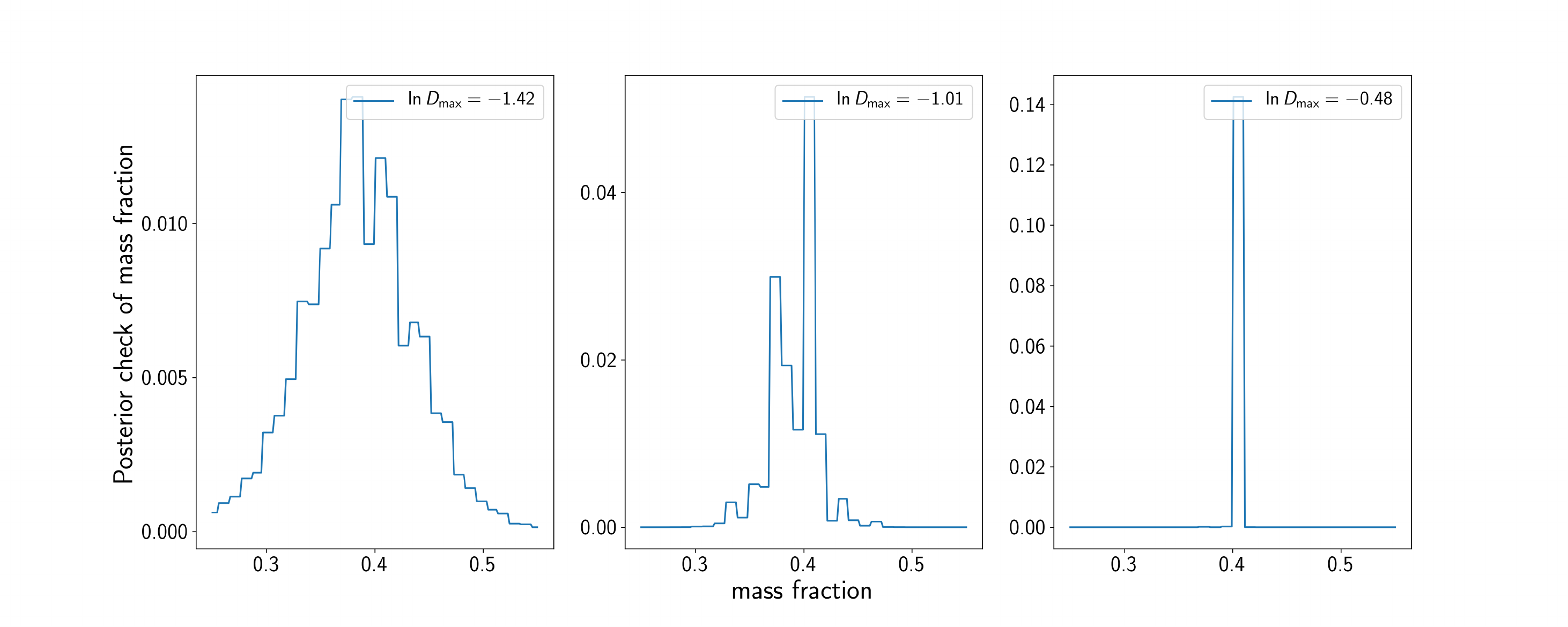}
        \label{posterior_real_data_fh_115_15}
    \end{subfigure}%
    \begin{subfigure}
        \centering
        \includegraphics[width=1.0\linewidth]{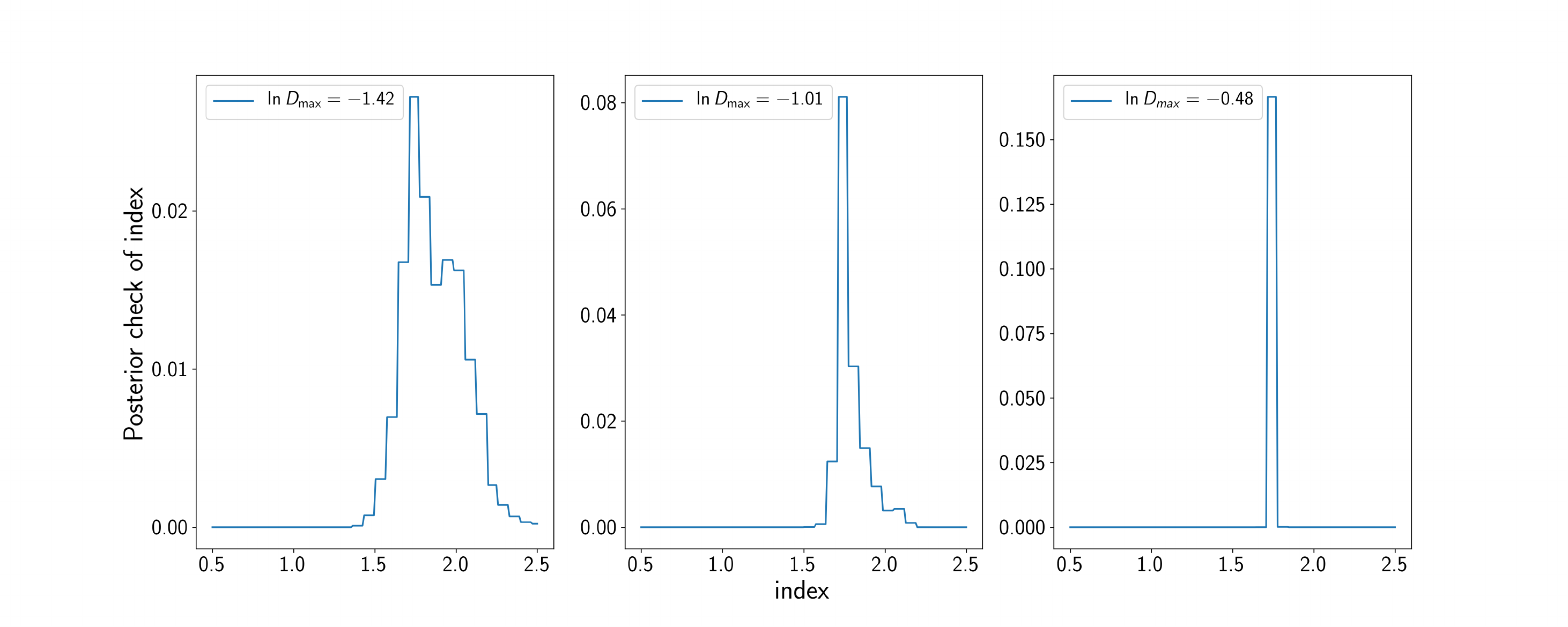}
        \label{posterior_real_data_index_115_15}
    \end{subfigure}
\caption{The posterior distribution of $f_h$ (upper panel) and $\alpha$ (lower panel) at three different values of $D_{\textrm{max}}$ calculated using stars with radial coverage from 11.5-15 kpc. Unlike simulation, the appearance of multiple peaks is obvious in the probability distribution and the peaks could vary with $D_{\textrm{max}}$ as well. This indicates that the paraboloid fitting cannot be used for uncertainties determination}
\label{posterior_real_data_115_15}
\end{figure*}

\begin{figure*}
    \centering
    \begin{subfigure}
        \centering
        \includegraphics[width=.3\linewidth]{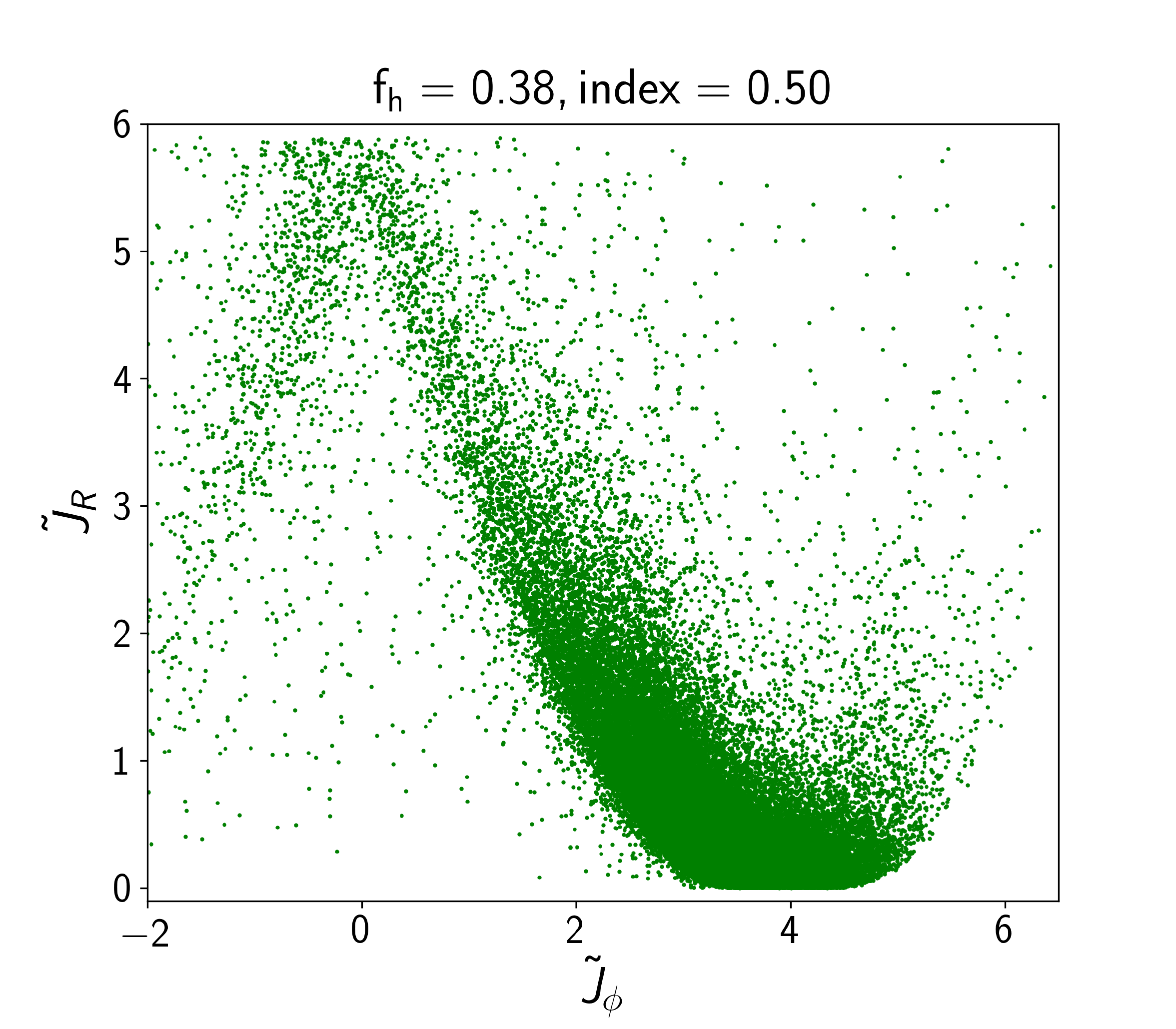}
        \label{action_var_f_h_fixed_p1_real_data_9_11}
    \end{subfigure}%
    \begin{subfigure}
        \centering
        \includegraphics[width=.3\linewidth]{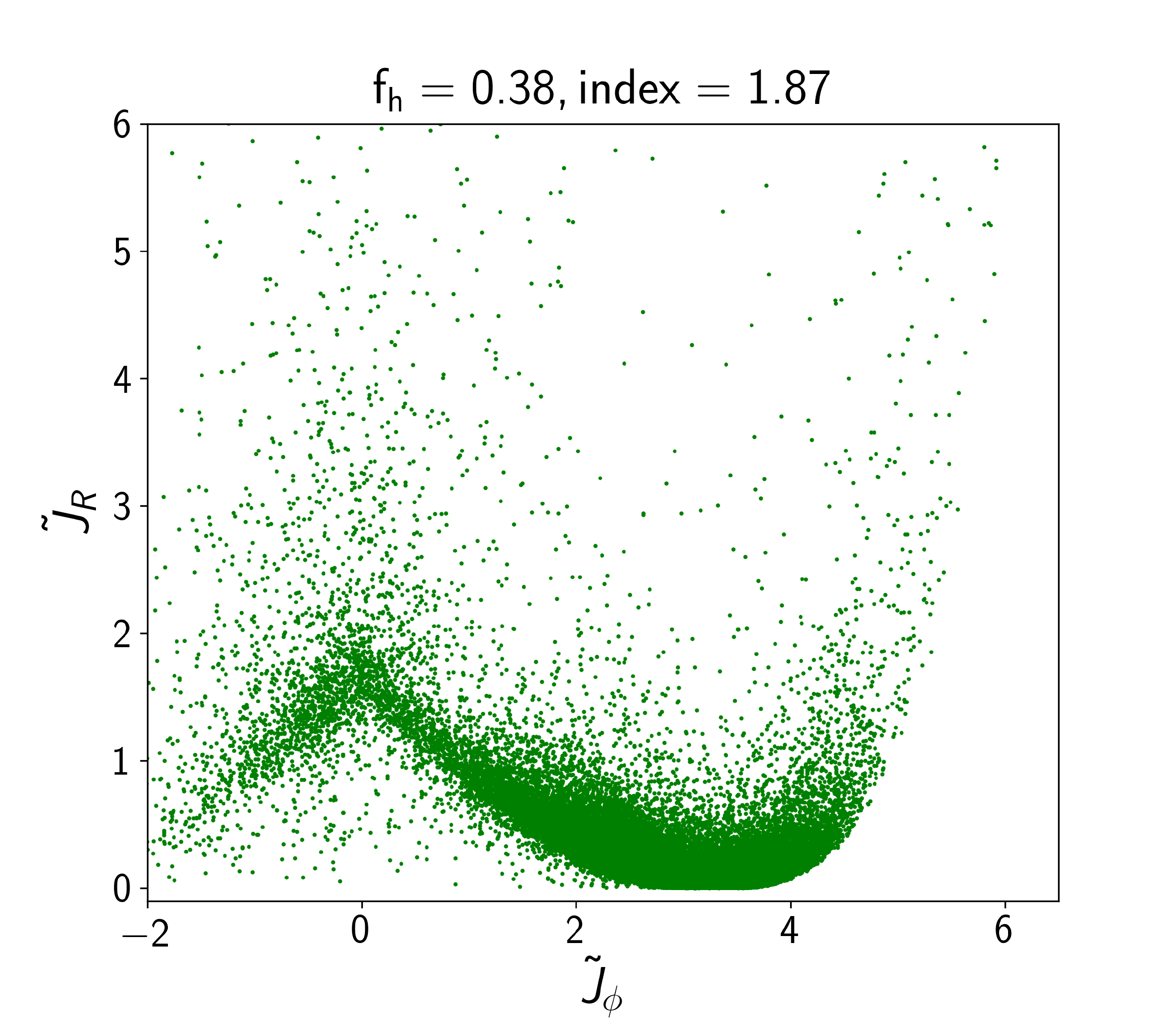}
        \label{action_var_f_h_fixed_p2_real_data_9_11}
    \end{subfigure}
    \begin{subfigure}
        \centering
        \includegraphics[width=.3\linewidth]{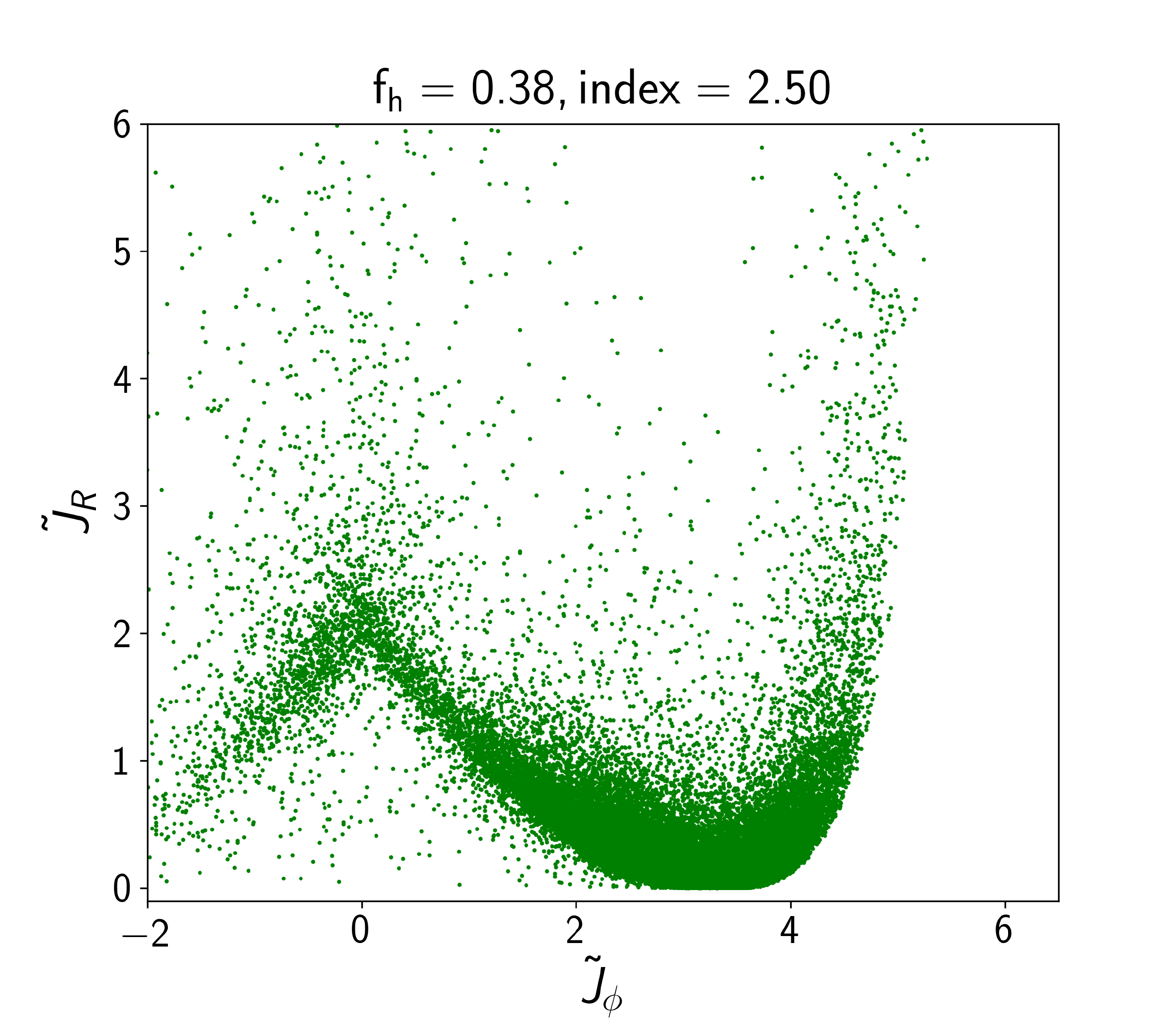}
        \label{action_var_f_h_fixed_p3_real_data_9_11}
    \end{subfigure}
    \begin{subfigure}
        \centering
        \includegraphics[width=.3\linewidth]{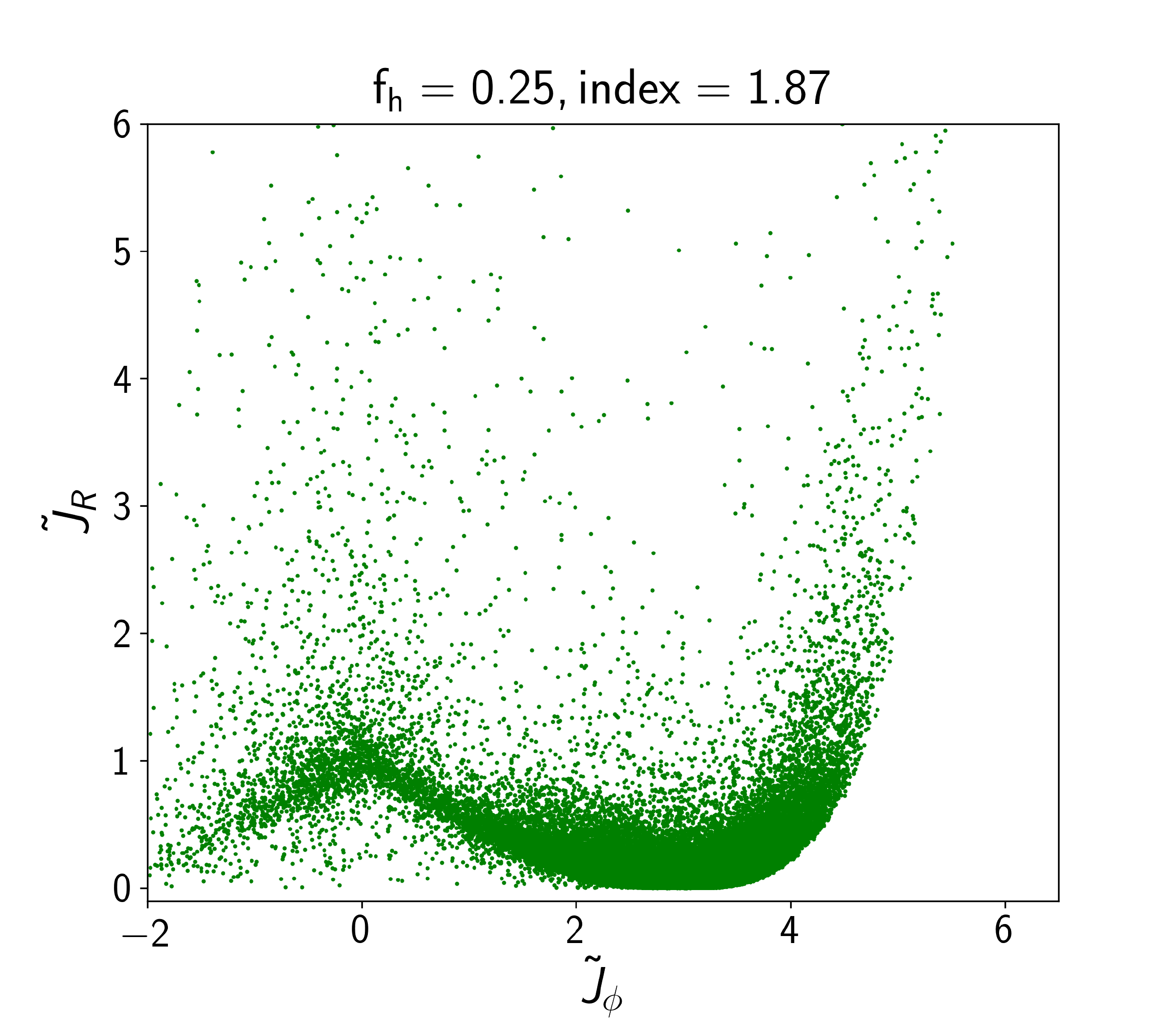}
        \label{action_var_index_fixed_p1_real_data_9_11}
    \end{subfigure}%
    \begin{subfigure}
        \centering
        \includegraphics[width=.3\linewidth]{Figures/action_var_central_real_data_9_11.pdf}
        \label{action_var_index_fixed_p2_real_data_9_11}
    \end{subfigure}
    \begin{subfigure}
        \centering
        \includegraphics[width=.3\linewidth]{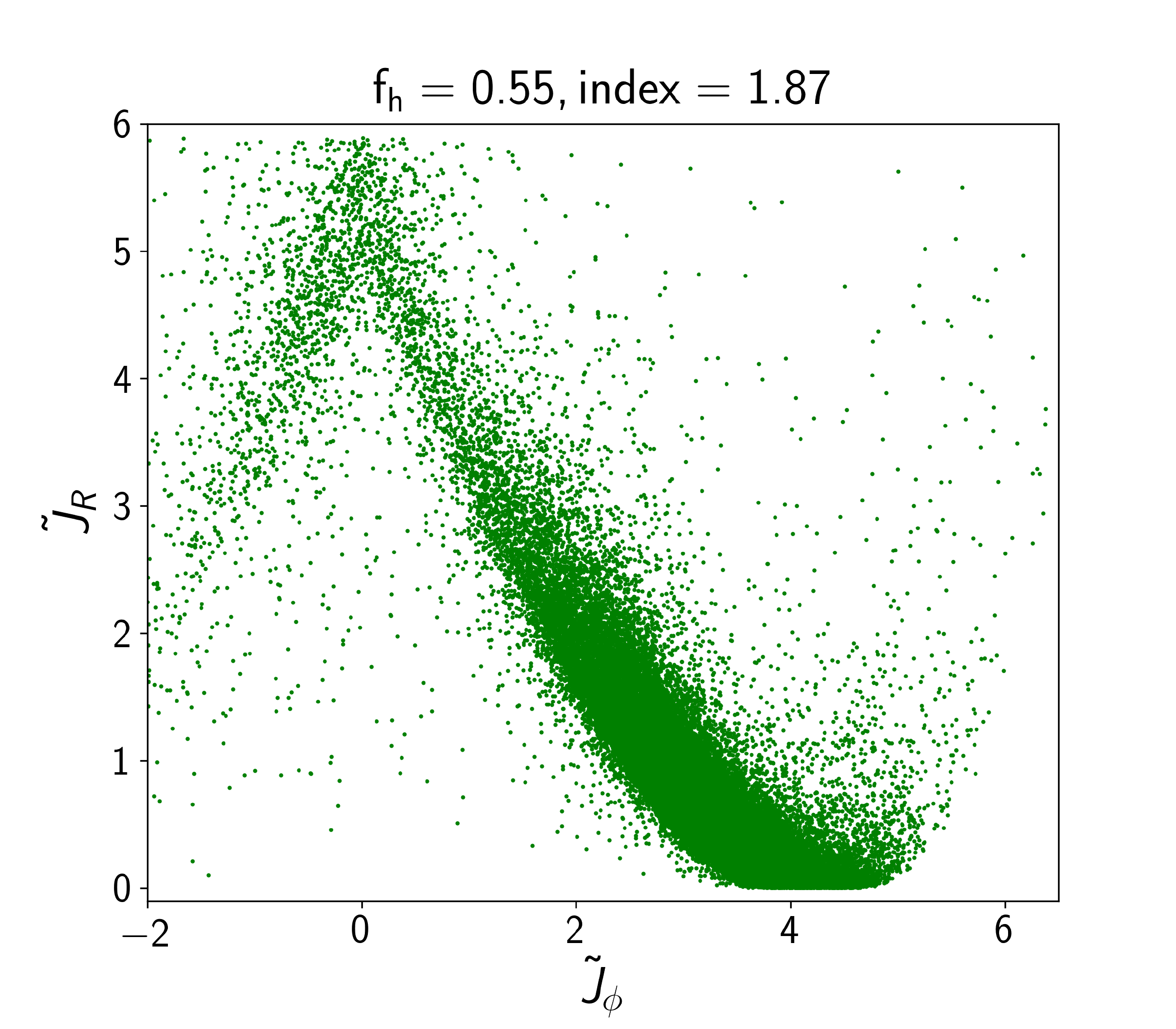}
        \label{action_var_index_fixed_p3_real_data_9_11}
    \end{subfigure}
    \begin{subfigure}
        \centering
        \includegraphics[width=.3\linewidth]{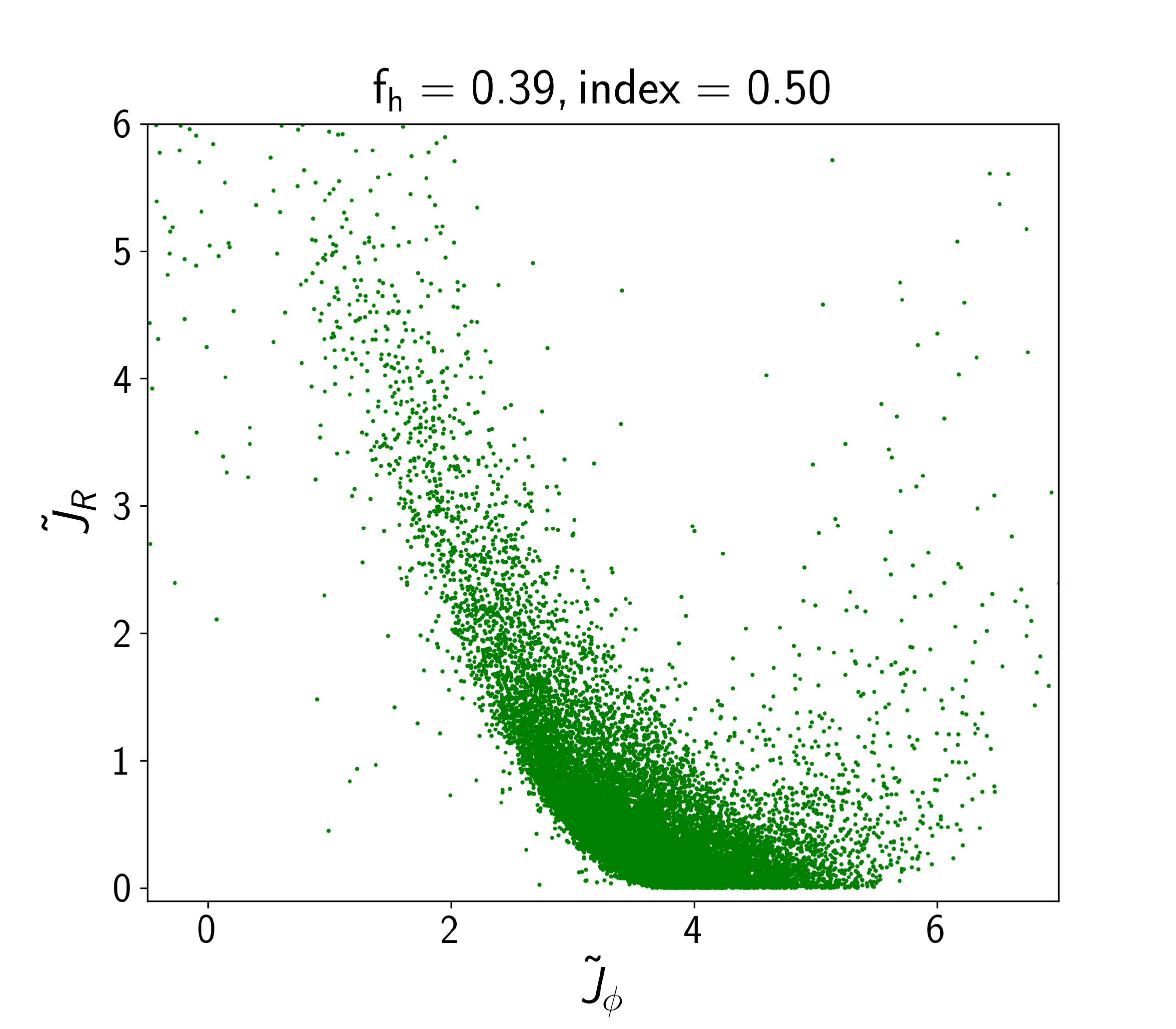}
        \label{action_var_f_h_fixed_p1_real_data_115_15}
    \end{subfigure}%
    \begin{subfigure}
        \centering
        \includegraphics[width=.3\linewidth]{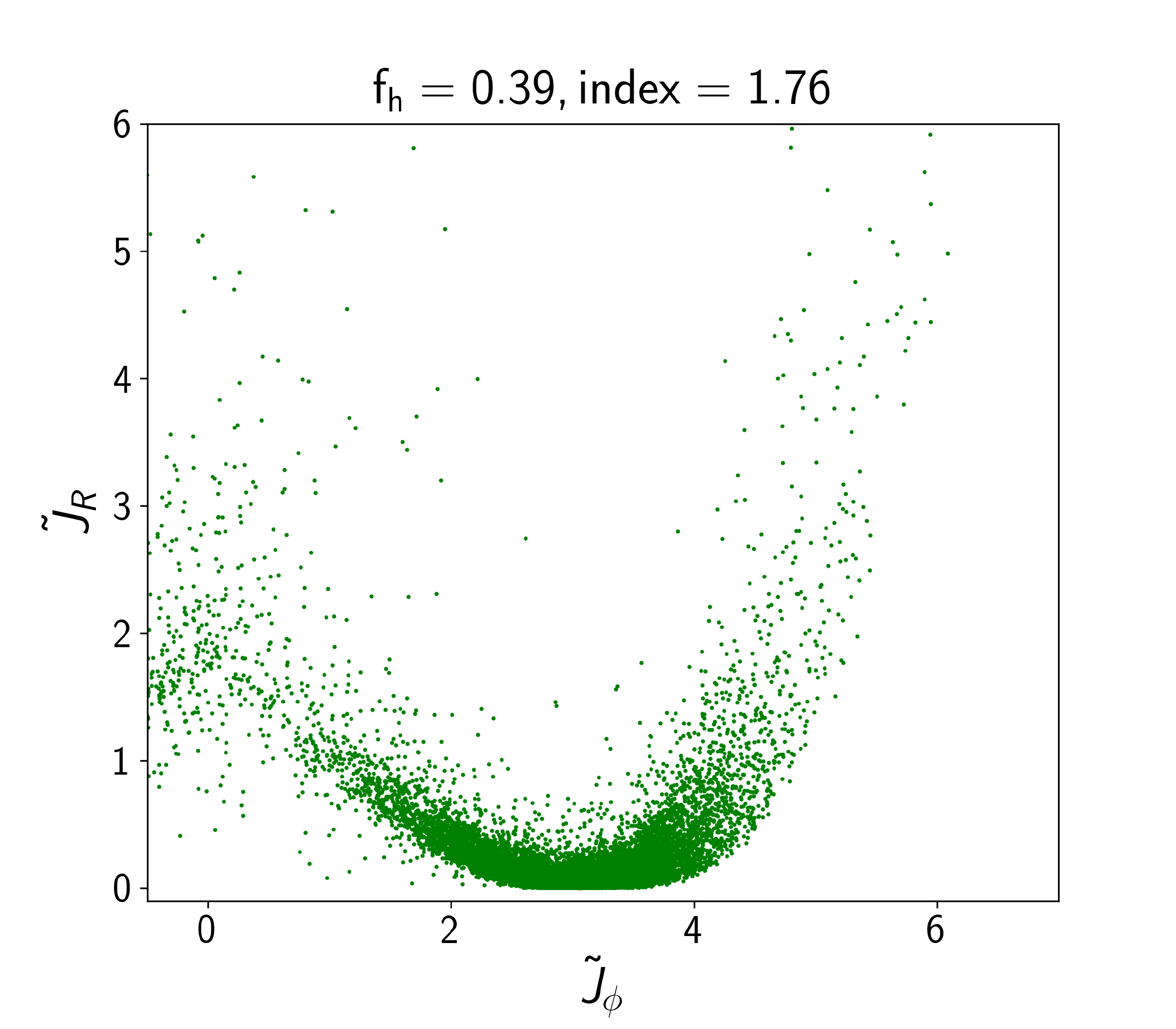}
        \label{action_var_f_h_fixed_p2_real_data_115_15}
    \end{subfigure}
    \begin{subfigure}
        \centering
        \includegraphics[width=.3\linewidth]{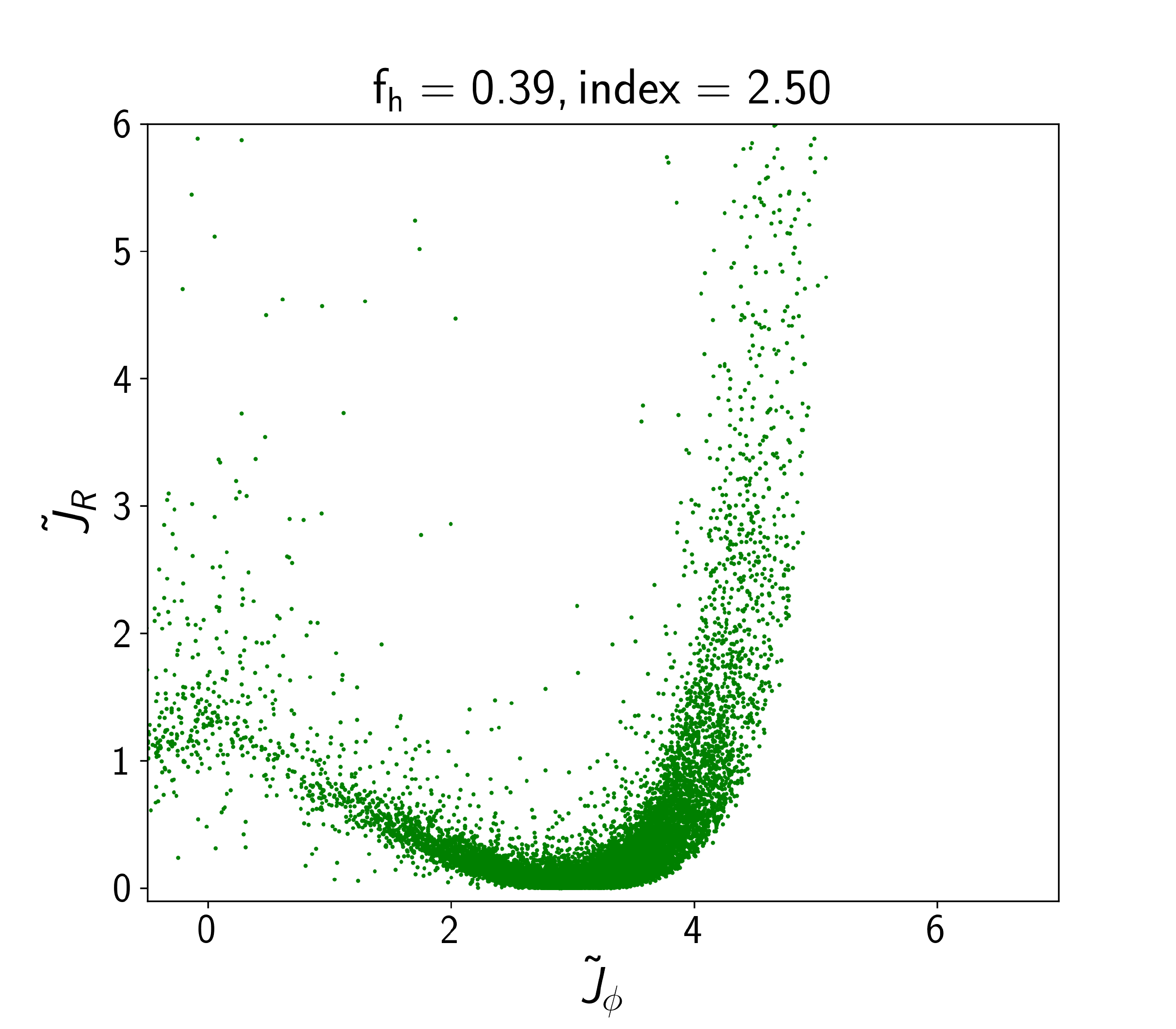}
        \label{action_var_f_h_fixed_p3_real_data_115_15}
    \end{subfigure}
    \begin{subfigure}
        \centering
        \includegraphics[width=.3\linewidth]{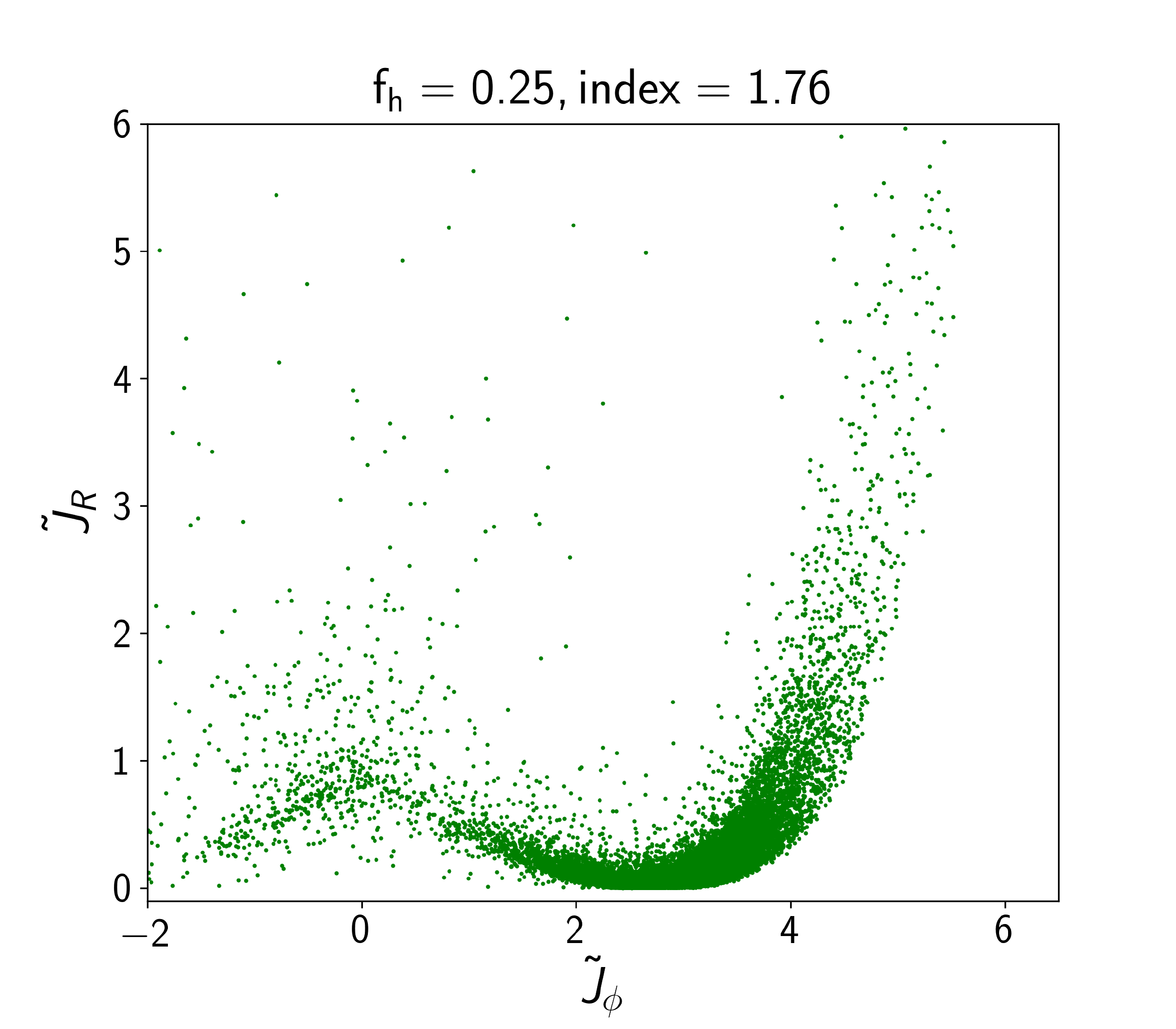}
        \label{action_var_index_fixed_p1_real_data_115_15}
    \end{subfigure}%
    \begin{subfigure}
        \centering
        \includegraphics[width=.3\linewidth]{Figures/action_var_central_real_data_115_15.pdf}
        \label{action_var_index_fixed_p2_real_data_115_15}
    \end{subfigure}
    \begin{subfigure}
        \centering
        \includegraphics[width=.3\linewidth]{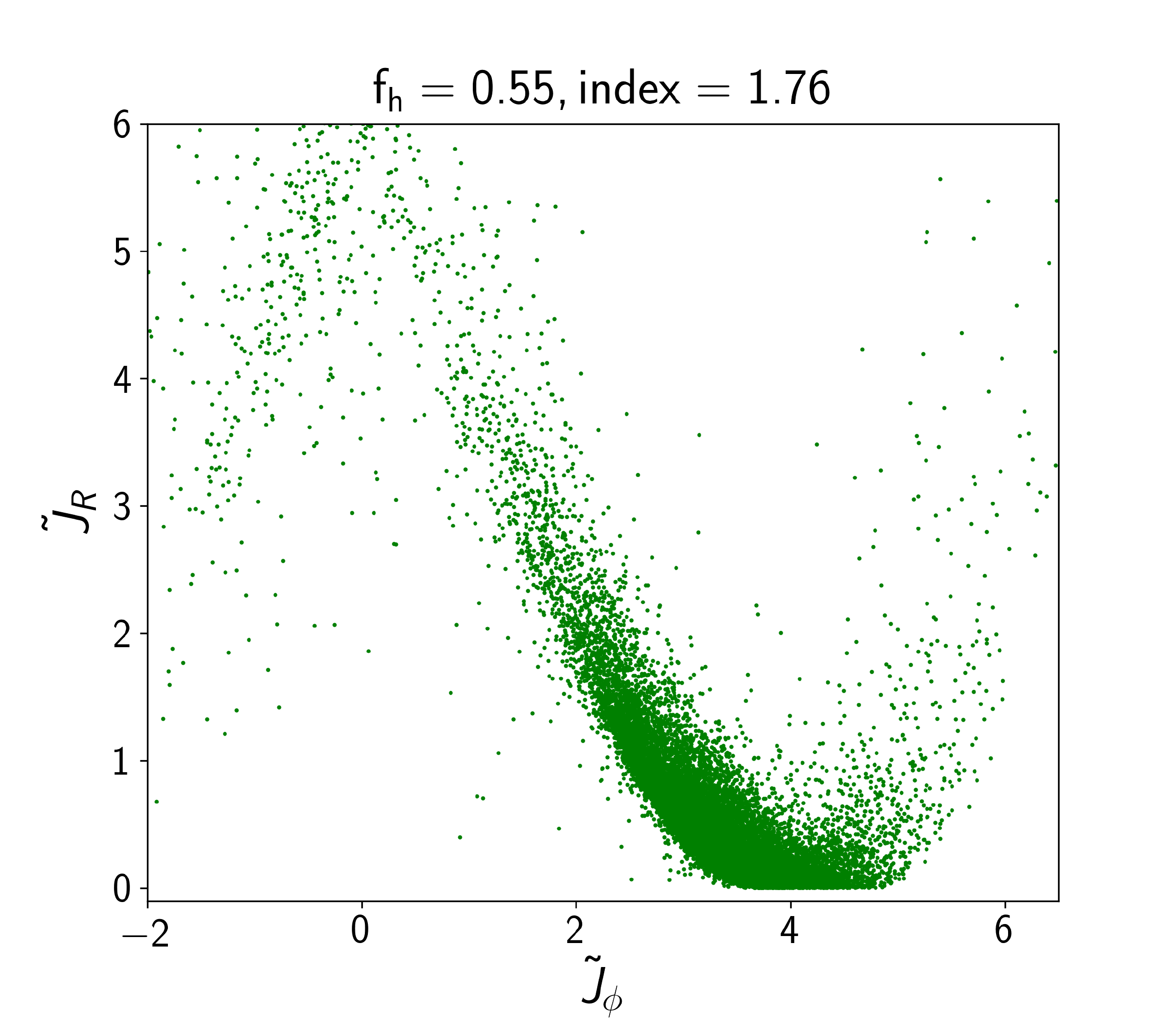}
        \label{action_var_index_fixed_p3_real_data_115_15}
    \end{subfigure}
\caption{Stellar distribution in the $\Tilde{J}_R$ and $\Tilde{J}_{\phi}$ 2D projected plane varying with different choices of potential, where $\Tilde{J}_R$ and $\Tilde{J}_\phi$ are defined as $J_{R}/\sigma_{J_R}$ and $J_{\phi}/\sigma_{J_{\phi}}$, i.e. radial and angular action variables normalized by their standard deviations over all stars in the sample. First two rows show the stellar distribution for the first case of real data
with fixed $f_h$ ($\alpha$) in the first (second) row. Last two rows show the result for the second case of real data. Interestingly, while approaching the potential that maximizes the likelihood, stars are tend to be more disk-like and display the properties of circular motion. Full movies are available online: \url{https://github.com/Supranta/GAIA_Potential/tree/master/Animation_movies}}
\label{real_data_action_var}
\end{figure*}

\begin{figure*}
    \centering
    \begin{subfigure}
        \centering
        \includegraphics[width=.45\linewidth]{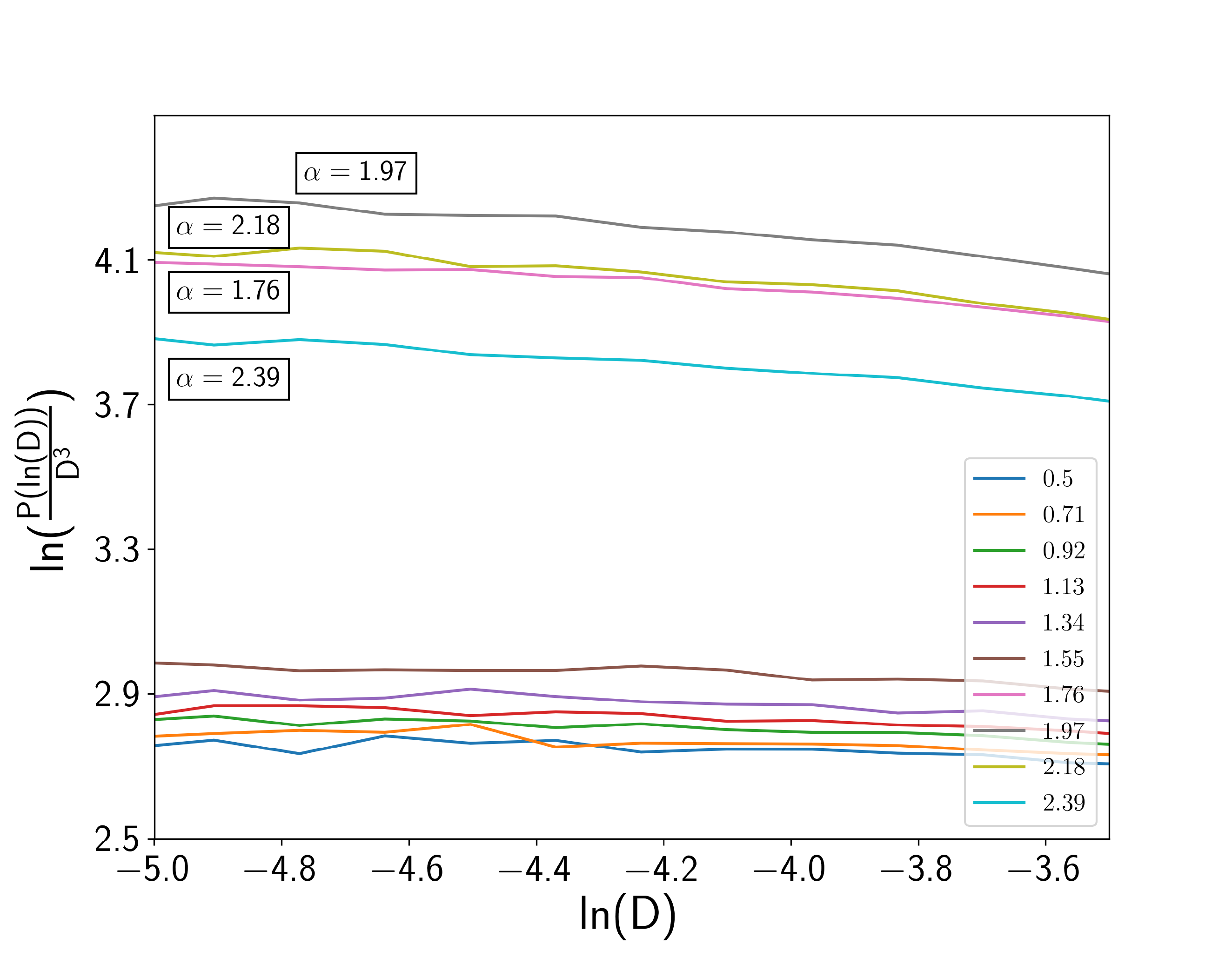}
        \label{probability_function_real_data_fixed_f_h_030_9_11}
    \end{subfigure}%
    \begin{subfigure}
        \centering
        \includegraphics[width=.45\linewidth]{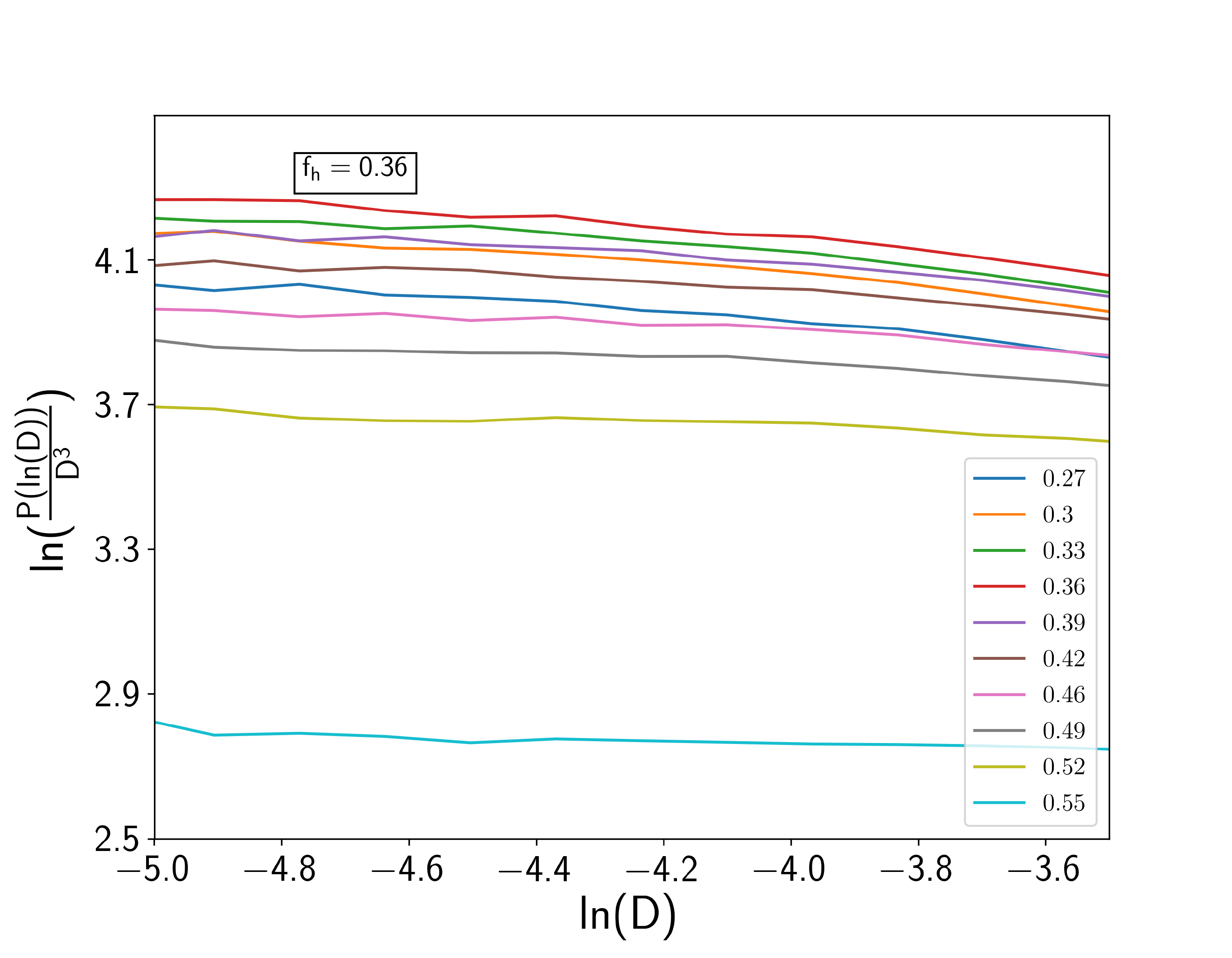}
        \label{probability_function_real_data_fixed_index_175_9_11}
    \end{subfigure}
    \centering
    \begin{subfigure}
        \centering
        \includegraphics[width=.45\linewidth]{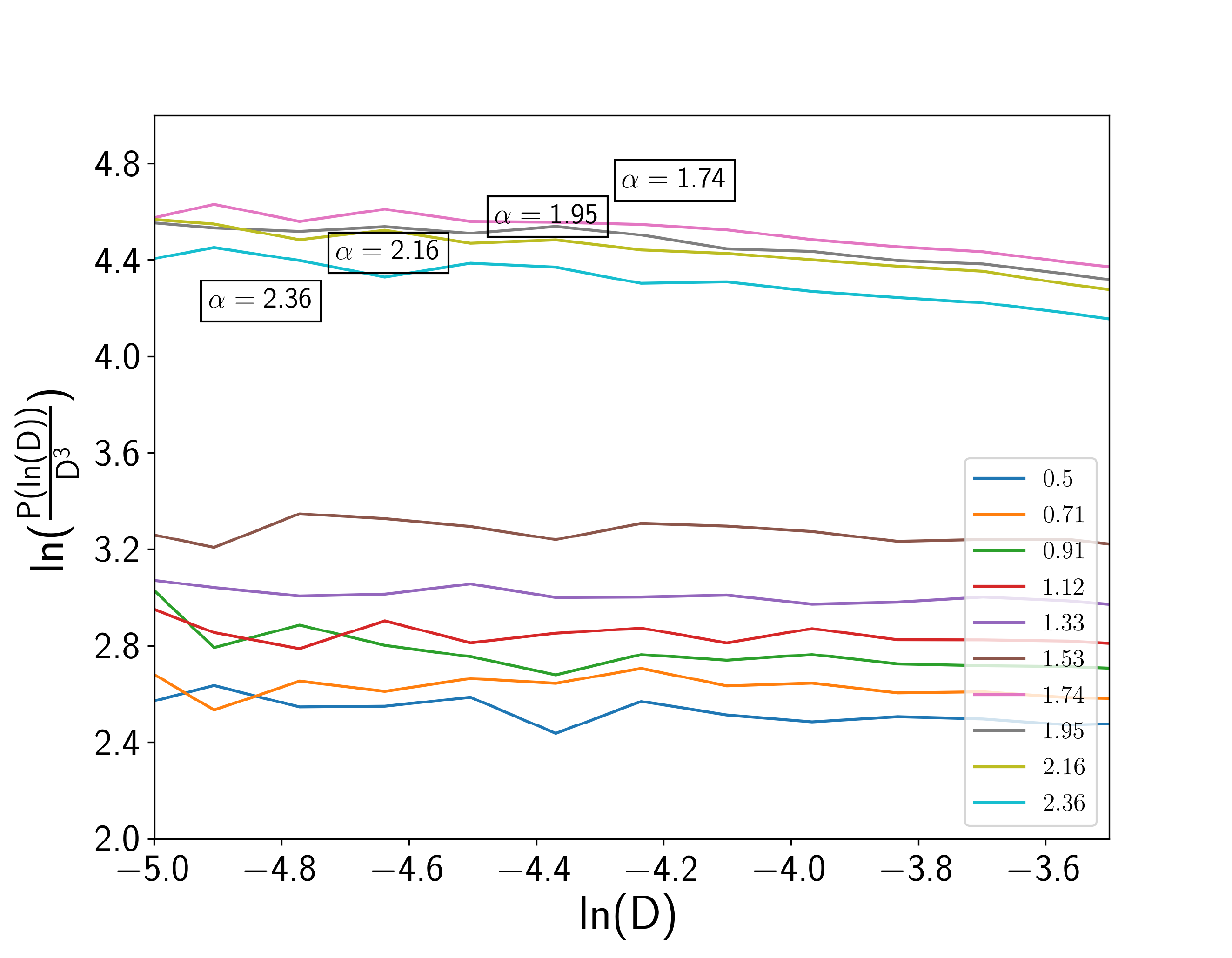}
        \label{probability_function_real_data_fixed_f_h_031_115_15}
    \end{subfigure}%
    \begin{subfigure}
        \centering
        \includegraphics[width=.45\linewidth]{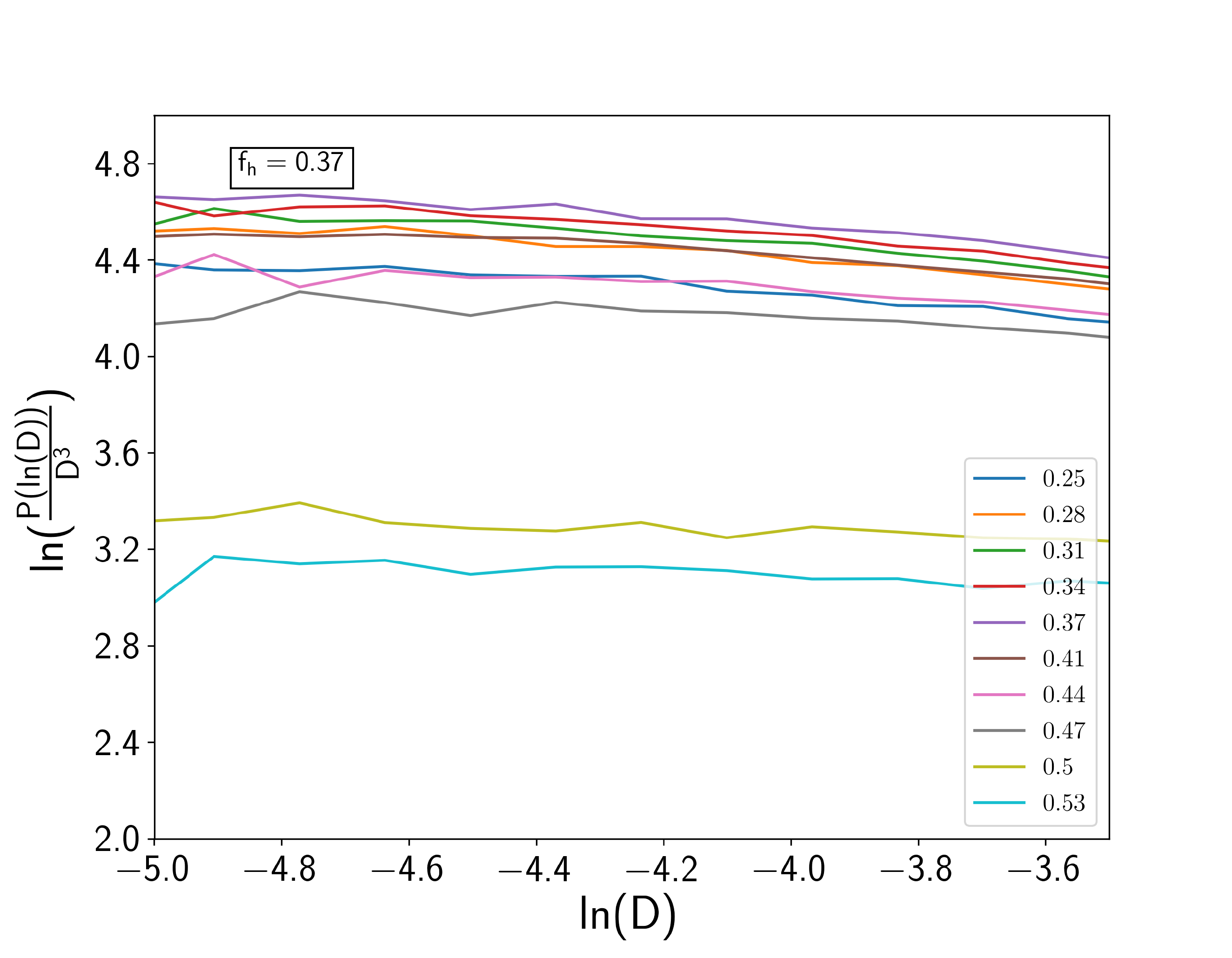}
        \label{probability_function_real_data_fixed_index_170_115_15}
    \end{subfigure}
\caption{Correlation function $\frac{P(\ln D)}{D^3}$ as a function of the distance in the action space in natural logarithm scale. The purpose of this figure is to check how
the two-point correlation function varies with different choices of potential and whether it is extremized around the set of parameter that maximize the likelihood function. Top panel: the behaviour of two-point correlation function for the first case of real data with fixed $f_h$ ($\alpha$) on the left (right). Bottom panel: the behaviour of two-point correlation function for the other case with fixed $f_h$ ($\alpha$) on the left (right).Different colors indicate the values of ln($\frac{P(\ln D)}{D^3}$) at different choices of potential.}
\label{probability_function_real_data}
\end{figure*}

The likelihood plots showing the constraints on the mass fraction and the index for both radial samples are shown in the left panels of Figure \ref{real_data_likelihood}. Although calculated within the same $\ln D_{\textrm{max}} \sim -1$, the log-likelihood values for ``real-data-9-11'' is larger than those of ``real-data-115-15'' as there are more stellar pairs included in the more nearby sample. Furthermore, the error bar plots in Figure \ref{real_data_likelihood} show how the likelihood peaks (black solid points) and median constraints on parameters (blue hollow circles with error bars) vary with different choices of $D_{\textrm{max}}$. We see that for $\ln D_{\textrm{max}} \lesssim -1$, the $f_h$ and $\alpha$ constraints are stable and robust to the choice of free parameter $D_{\textrm{max}}$. 
 
For the choice of $\ln D_{\rm max}$, we notice that for the more distant sample ``real-data-115-15'', there is a jump in the error bar plots for both parameters when ln $D_{\textrm{max}}$ is smaller than -1.5. Taking this into account, we treat -1.14 as our final choice of ln $D_{\textrm{max}}$. Results estimated at this point have the smallest uncertainties and the constraints are consistent (within error bars) for all ln $D_{\textrm{max}} \lesssim$ -1.14. We shall refer to this value as ln $D_{\textrm{max, optimum}}$. To be consistent, we use the same value of $D_{\rm max}$ for both Gaia samples. More discussions about choosing an appropriate $\ln D_{\textrm{max}}$ are presented in Section \ref{sec::discussion} and Appendix \ref{sec::simulation_with_bg}. 

Having seen statistical constraints on both parameters from the likelihood plots, we would like to evaluate the true uncertainties of the measurements. However, the determination of uncertainties is more subtle compared with the simulations. Unlike simulations, where we found the posterior distribution of parameters had a sharp gaussian peak, we notice that the likelihood 2D plots have multiple peaks for real Gaia data. This can be seen more clearly in the 1D posterior distributions in Figure \ref{posterior_real_data_115_15}, where (depending on the choice of $D_{\rm max}$) there can be multiple peaks. As a result, it is no longer appropriate to simply assume the likelihood distribution is approximated by a gaussian. In particular, the jump in $f_h$ around $\ln D_{\rm max} \simeq $ -1.5 in  ``real-data-115-15'' sample is due to the change in relative heights of the two main peaks in posteriors shown in the top panel of Figure \ref{posterior_real_data_115_15}.   
Therefore, we calculated the median of the parameters using the full posterior distribution within our prior range (Equation \ref{eqn::prior_range}), as it is a more robust statistical estimator than average whenever multiple peaks or outliers are presented in the distribution. The $68\%$ confidence interval (68\% CI) around the median, which can be also computed from the posterior distribution, is treated as the error on the parameter. 

As can be seen in the error bar plots in Figure \ref{real_data_likelihood}, the maxima of likelihood (black points) are all consistent with the median values within 68\% CI (blue hollow points with blue error bars), and the error bars become larger for smaller $D_{\textrm{max}}$, where fewer stellar pairs are included. When choosing ln($D_{\textrm{max}}$) as -1.14, the constraints we get under both situations are summarized in Table \ref{tab::results_with_error_cut}.

\begin{table}
  \begin{tabular}{lllll}
    \hline
    \multirow{2}{*}{\vtop{\hbox{\strut Error Type}\hbox{\strut ($|z|$ > 1 ~ {\rm kpc})}}} &
      \multicolumn{2}{c}{[9.0 ~\textrm{kpc} <R<11.0~ \textrm{kpc}]} &
      \multicolumn{2}{c}{[11.5 ~\textrm{kpc} <R<15.0~ \textrm{kpc}]} \\
    & \vtop{\hbox{\strut $f_h$}\hbox{\strut = 0.375}}  &  \vtop{\hbox{\strut $\alpha$}\hbox{\strut = 1.967}}& \vtop{\hbox{\strut $f_h$}\hbox{\strut  = 0.387}} & \vtop{\hbox{\strut $\alpha$}\hbox{\strut = 1.786}} \\
    \hline
    Stochastic & $^{+0.029}_{-0.038}$ & $^{+0.090}_{-0.070}$ & $^{+0.024}_{-0.027}$ & $^{+0.070}_{-0.181}$ \\
    \hline
    Systematic & $\pm$ 0.004 & $\pm$ 0.079 & $\pm$ 0.004 & $\pm$ 0.071 \\
    \hline
    Total & $\pm$ 0.034 & $\pm$ 0.112 & $\pm$ 0.026 & $\pm$ 0.144 \\
    \hline
  \end{tabular}
  \caption{Constraints on normalization and logarithmic slope of dark matter profile $(f_h,\alpha)$ for both radial samples with selection cuts. Here, we summarize all sources of measurement errors: the stochastic errors estimated from the posterior distribution of parameters, the systematic errors from simulation, and the total error given by the root of stochastic errors squared plus systematic errors squared.}
 \label{tab::results_with_error_cut}
 \end{table}

Furthermore, recall that from simulated measurements in Section  \ref{ssec::simulation}, we do expect an additional $4\%$ percent systematic discrepancy for index measurements and an $1\%$ off in mass fraction measurement. We do include these estimates in Table \ref{tab::results_with_error_cut} as systematic errors, which can be combined with our stochastic errors to obtain the total expected uncertainties.  

Let us now perform the same consistency checks we did in Section \ref{ssec::simulation} for simulated data, and see how stars from real data are distributed in the action space. Figures \ref{real_data_action_var} and \ref{probability_function_real_data} shows how (2D projections of) the stellar distribution in the action space, as well as its two-point correlation function change as we vary $f_h$ or $\alpha$ in the Milky Way halo potential. As expected, the correlation function values for both radial ranges are maximized while the potential approaches the parameters that maximize the likelihood. However, unlike in simulations, instead of a compact cluster, stars in the action space present a more ``lath-shaped'' distribution, where most stars concentrate around $J_R \sim$ 0 for the best-fit potential. In this situation, stars are extended along $J_{\phi}$ axis with no distinct $J_{R}, J_{z}$ contributions, which indicates the property of circular (or disk) motion for most stars. This is not surprising as can be seen from the tangential velocity distribution in Figure \ref{real_data_distribution}; the disk stars are still the dominant component in our real data samples even though we performed a $|z|$ > 1 kpc cut\footnote{Note that the inclusion of disk stars does not contradict with our assumption made in the likelihood. Disk is a mixture of correlated structures and uniform background, so the disk component can also contribute to the small-scale clustering signal. Also, our distance metric defined above has already included the effect of disk component (Equation \ref{eqn::distance_calculation_real_data}).}. Just as in the simulations, stars within the potential that maximizes the small-scale clustering statistics (Figure \ref{probability_function_real_data}) present the most compact distribution in the action space. The fact that stellar distribution reduces to circular motion for the best-fit potential is in practice consistent with the traditional assumption of circular motion for disk stars, in order to estimate the mass of Milky Way galaxy. However, our method does not explicitly make this assumption, and thus can account for deviations from circular motion, effectively combining (thin+thick) disk+halo stars.

\section{Discussion}\label{sec::discussion}

\begin{figure*}
    \centering
    \begin{subfigure}
        \centering
        \includegraphics[width=1.0\linewidth, height = 8cm]{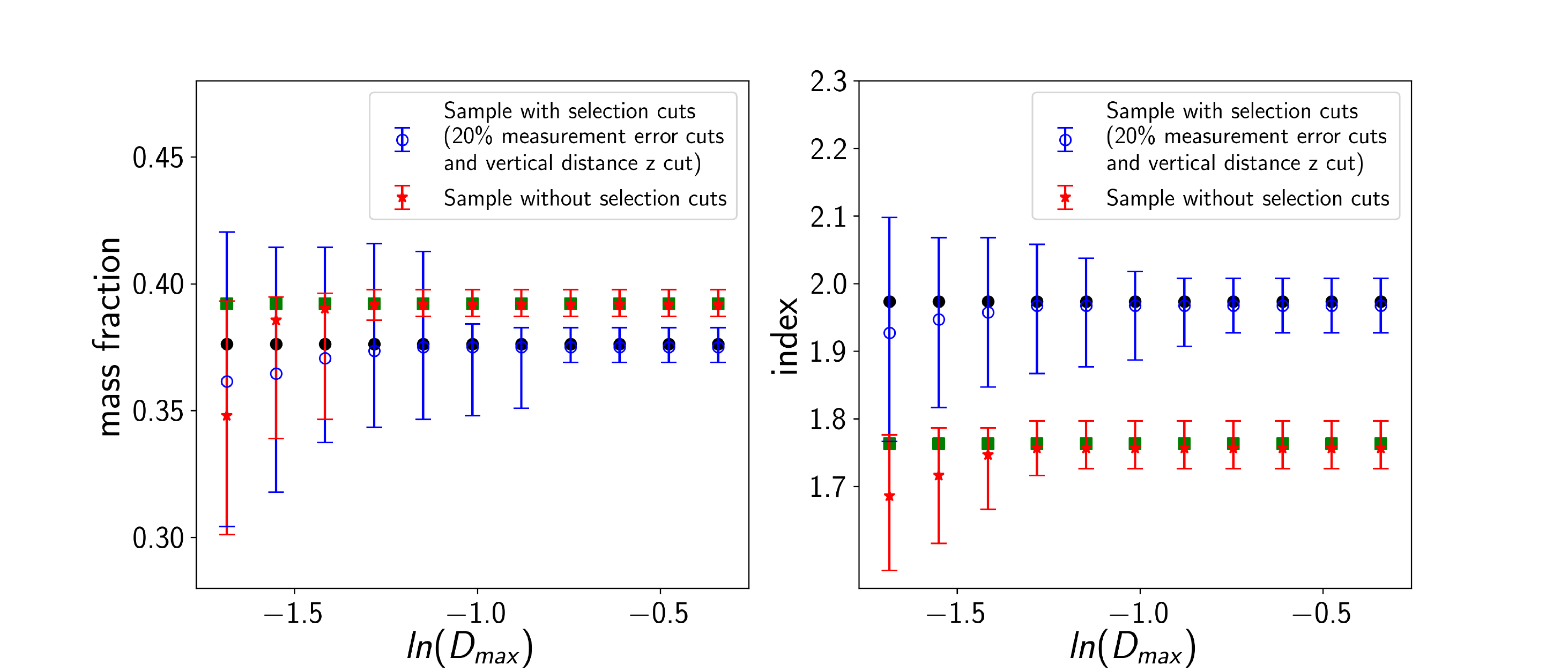}
        \label{error_bars_all_real_data_9_11}
    \end{subfigure}%
    \begin{subfigure}
        \centering
        \includegraphics[width=1.0\linewidth, height = 8cm]{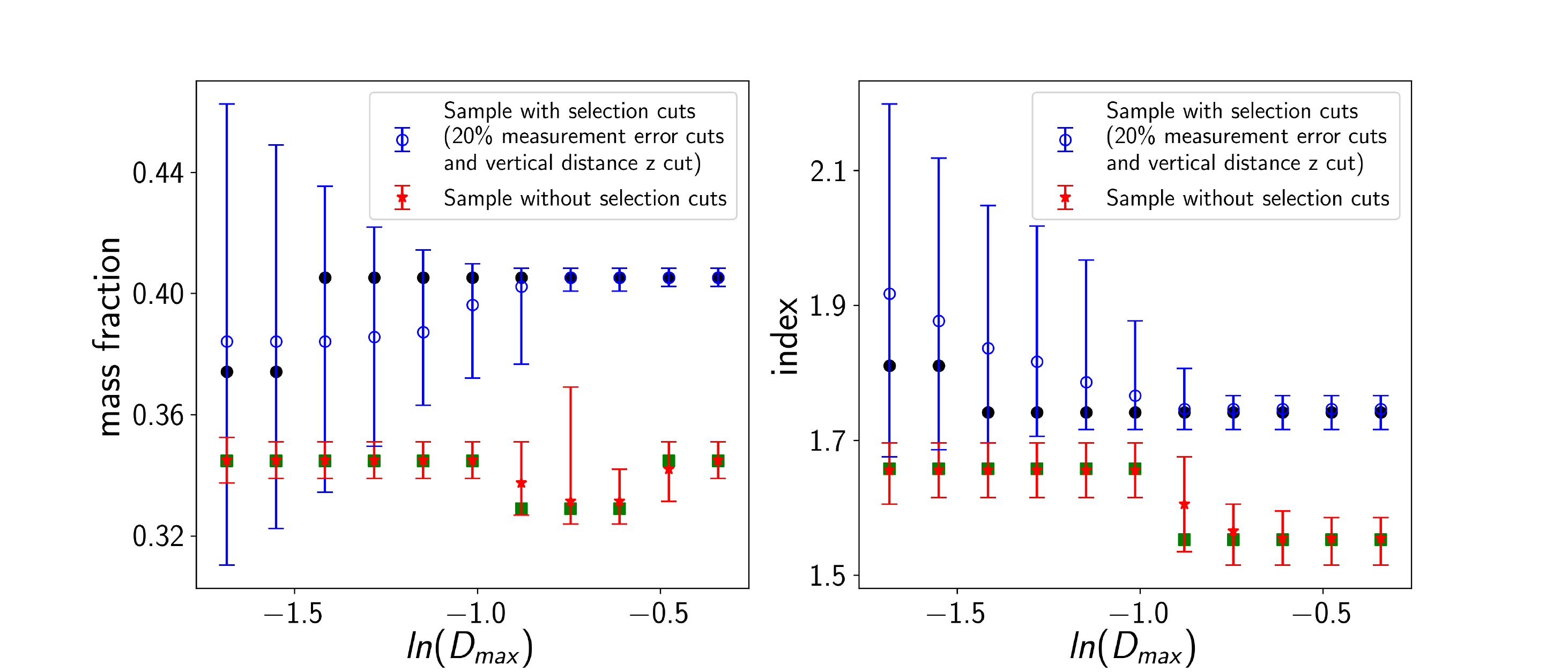}
        \label{error_bars_all_real_data_115_15}
    \end{subfigure}
\caption{Error bar plot using stars from 9 to 11 kpc (top) and 11.5 to 15 kpc (bottom) without selection cuts. Green square points shows the constraints to the parameters by directly finding the maximum from the likelihood plot, and at $\ln D_{\rm max}$ = -1.14, this gives $f_h$ = 0.392 and $\alpha$ = 1.763 for the nearby sample and $f_h$ = 0.345 and $\alpha$ = 1.658 for the other. The red star point and red line represent the median of parameter interpreted from the posterior distribution at each $D_{\textrm{max}}$. For comparison, the median values and 68\% CI error bars of two parameters determined from the sample with selection cuts are also over-plotted on the same figure, which are shown as 
blue hollow points with blue error bars (black solid circles are the maxima of the likelihood).}
\label{error_bars_all_real_data}
\end{figure*}

In the previous analysis, we used some measurement error cuts and a vertical distance cut to real data. However, selection cuts to the raw data could cause unexpected biases in the measured parameters. To investigate the degree to which our results are sensitive to an arbitrary choice of $z$-cut,  we randomly choose 90,000 stars from 9-11 kpc sample\footnote{There are 607,257 stars in total, but we only choose a subset of the catalogue due to the limitation of computational time.} and take all data from 11.5-15 kpc ($\sim$ 90,000 stars in total ), without imposing any of the previous error or distance cuts \footnote{However, we did apply some minimal cuts to the raw data in order to get rid of the unreliable observations, including $|z|$ < 10 kpc and the absolute values of all three components of velocity in cylindrical coordinates are smaller than 500 km/s}. The error bar plots are shown in Figure \ref{error_bars_all_real_data}. For comparison, we also overplot the results obtained before using the sample with selection cuts. Generally, at same value of $D_{\textrm{max}}$, the uncertainties on parameters are significantly reduced when using the data samples without selection cuts. To be consistent, for both radial ranges, we still take ln $D_{\textrm{max}}$ $\simeq$ -1.14 and check the corresponding constraints on $f_h$ and $\alpha$. The results are tabulated in Table \ref{tab::results_without_error_cut}.

\begin{table}
  \begin{tabular}{lllll}
    \hline
    \multirow{2}{*}{\vtop{\hbox{\strut Error Type}\hbox{\strut (no $|z|$ cut)}}} &
      \multicolumn{2}{c}{[9.0 ~\textrm{kpc} <R<11.0~ \textrm{kpc}]} &
      \multicolumn{2}{c}{[11.5 ~\textrm{kpc} <R<15.0~ \textrm{kpc}]} \\
    & \vtop{\hbox{\strut $f_h$}\hbox{\strut = 0.392}}  &  \vtop{\hbox{\strut $\alpha$}\hbox{\strut = 1.756}}& \vtop{\hbox{\strut $f_h$}\hbox{\strut = 0.345}} & \vtop{\hbox{\strut $\alpha$}\hbox{\strut = 1.656}} \\
    \hline
    Stochastic & $^{+0.004}_{-0.006}$ & $^{+0.030}_{-0.040}$ & $\pm$ 0.006 & $\pm$ 0.040 \\
    \hline
    Systematic & $\pm$ 0.004 & $\pm$ 0.070 & $\pm$ 0.003 & $\pm$ 0.066 \\
    \hline
    Total & $\pm$ 0.006 & $\pm$ 0.078 & $\pm$ 0.007 & $\pm$ 0.077 \\
    \hline
  \end{tabular}
  \caption{Constraints on normalization and logarithmic slope of dark matter profile $(f_h,\alpha)$ for both radial samples without selection cuts. Here, we summarize all source of measurement errors: the stochastic errors estimated from the posterior distribution of parameters, the systematic errors from simulation, and the total error given by the root of stochastic errors squared plus systematic errors squared.}
 \label{tab::results_without_error_cut}
 \end{table}

Compared the constraints at the same $D_{\textrm{max}}$ obtained previously but with error selection cuts with the results from full data set, we find a consistency in $f_h$ constraint for both radial ranges. However, for $\alpha$, we notice a 11$\%$ systematic discrepancy within 9-11 kpc, and an 7$\%$ discrepancy for the 11.5-15 kpc. For both radial ranges, the index estimates for the uncut sample are lower than those of the cut sample. 

This systematic shift is primarily due to the selection cut to the vertical distance, $z$. Although some measurement error cuts are also imposed on the raw data, $z$ distance cut seems to be the most severe: 80$\%$ of raw data survives the measurement error cuts, while only around 6$\%$ remain after the $|z|$>1 kpc cut is imposed. One possible reason for this systematic difference could be the inaccuracy of the simple analytic model for the disk potential used in Equation \ref{eqn::Disk_pot}. It remains to be seen whether a more realistic model (e.g. using other datasets), or including the disk parameters in the likelihood marginalization, could lead to more consistent (and realistic) estimates. In order to account for additional potential systematic errors due selection cuts, we use the probability function defined in Appendix \ref{sec::posterior_results_determination}, which yields our final constraints on the mass fraction of dark matter and the index in the localized density profile in Table \ref{tab::results_summary_table}.

\begin{table}
  \begin{tabular}{lll}
    \hline
    \multirow{1}{*}{} &
      \multicolumn{1}{c}{\vtop{\hbox{\strut [9.0 ~\textrm{kpc} <R< }\hbox{\strut 11.0~ \textrm{kpc}]}}} &
      \multicolumn{1}{c}{\vtop{\hbox{\strut [11.5 ~\textrm{kpc} <R< }\hbox{\strut 15.0~ \textrm{kpc}]}}} \\
      \hline
      \hline
    \vtop{\hbox{\strut $f_h$}\hbox{\strut}} & $0.375^{+0.029}_{-0.038}$ & $0.387^{+0.024}_{-0.027}$ \\
    \vtop{\hbox{\strut $\alpha$}\hbox{\strut ($|z|$>1~\textrm{kpc},}\hbox{\strut stochastic error only)}} & $1.967^{+0.090}_{-0.070}$ & $1.786^{+0.070}_{-0.181}$ \\
    \hline
    \vtop{\hbox{\strut $f_h$}\hbox{\strut}} & $0.392^{+0.004}_{-0.006}$ & 0.345$\pm$ 0.006 \\
    \vtop{\hbox{\strut $\alpha$}\hbox{\strut (no $|z|$ cut,}\hbox{\strut stochastic error only)}} & $1.756^{+0.030}_{-0.040}$ & 1.656$\pm$ 0.040 \\
    \hline
    \vtop{\hbox{\strut $f_h$}\hbox{\strut}} & 0.391$\pm$ 0.009 & 0.351$\pm$ 0.012 \\
    \vtop{\hbox{\strut $\alpha$}\hbox{\strut (combined, }\hbox{\strut stochastic+systematic error)}} & 1.835$\pm$ 0.092 & 1.687$\pm$ 0.079 \\
    \hline
  \end{tabular}
  \caption{Constraints on dark matter halo parameters for both samples with/without selection cuts. The top four rows compare results for different cuts (including only, mostly independent, stochastic errors), while the final two rows is an attempt to combine these results, including the systematic errors introduced due to selection, as detailed in Appendix \ref{sec::posterior_results_determination}.}
 \label{tab::results_summary_table}
 \end{table}

\begin{figure*}
    \centering
    \begin{subfigure}
        \centering
        \includegraphics[width=0.45\linewidth]{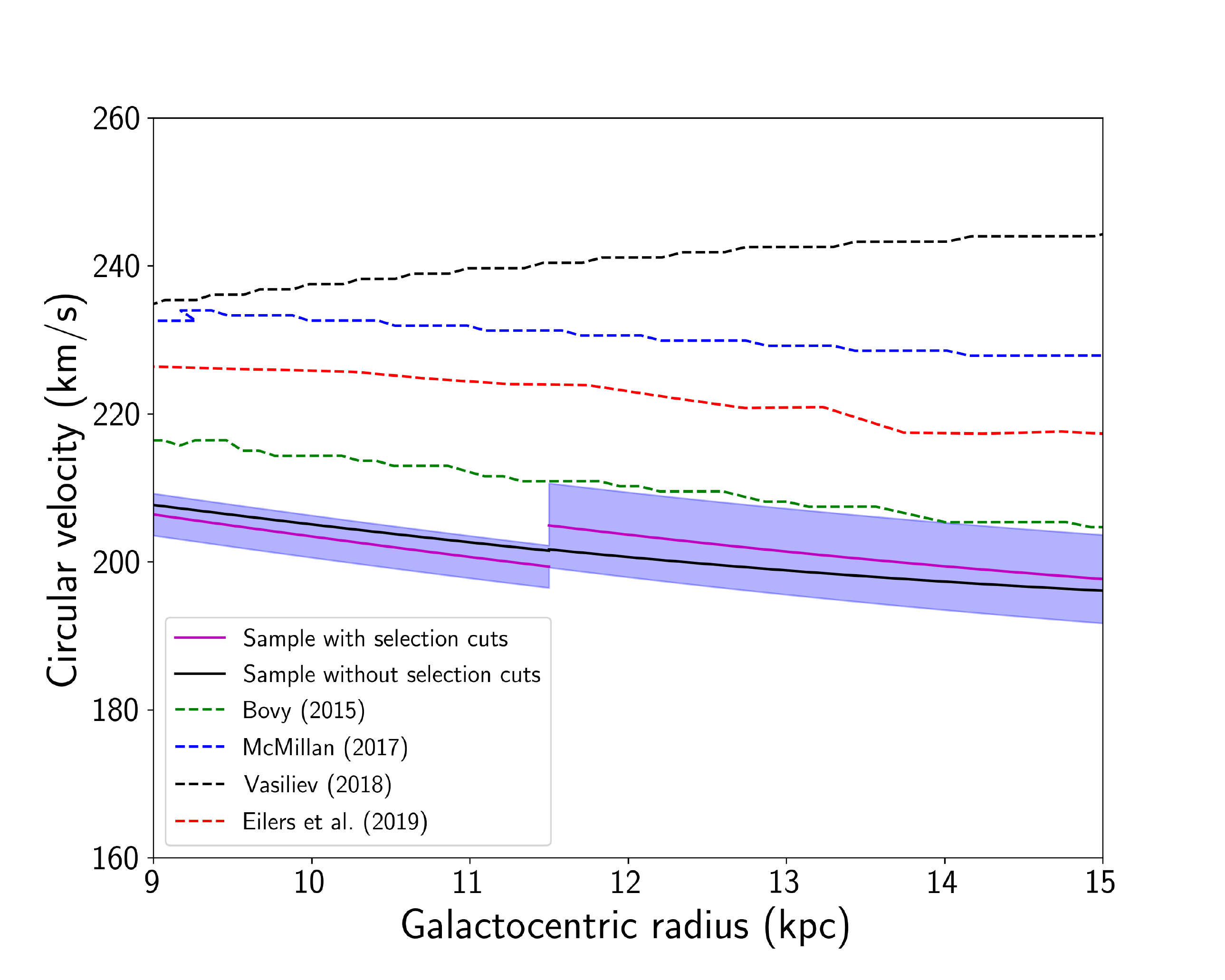}
        \label{velocity_check_total_Bovy_solar_info}
    \end{subfigure}%
    \begin{subfigure}
        \centering
        \includegraphics[width=0.45\linewidth]{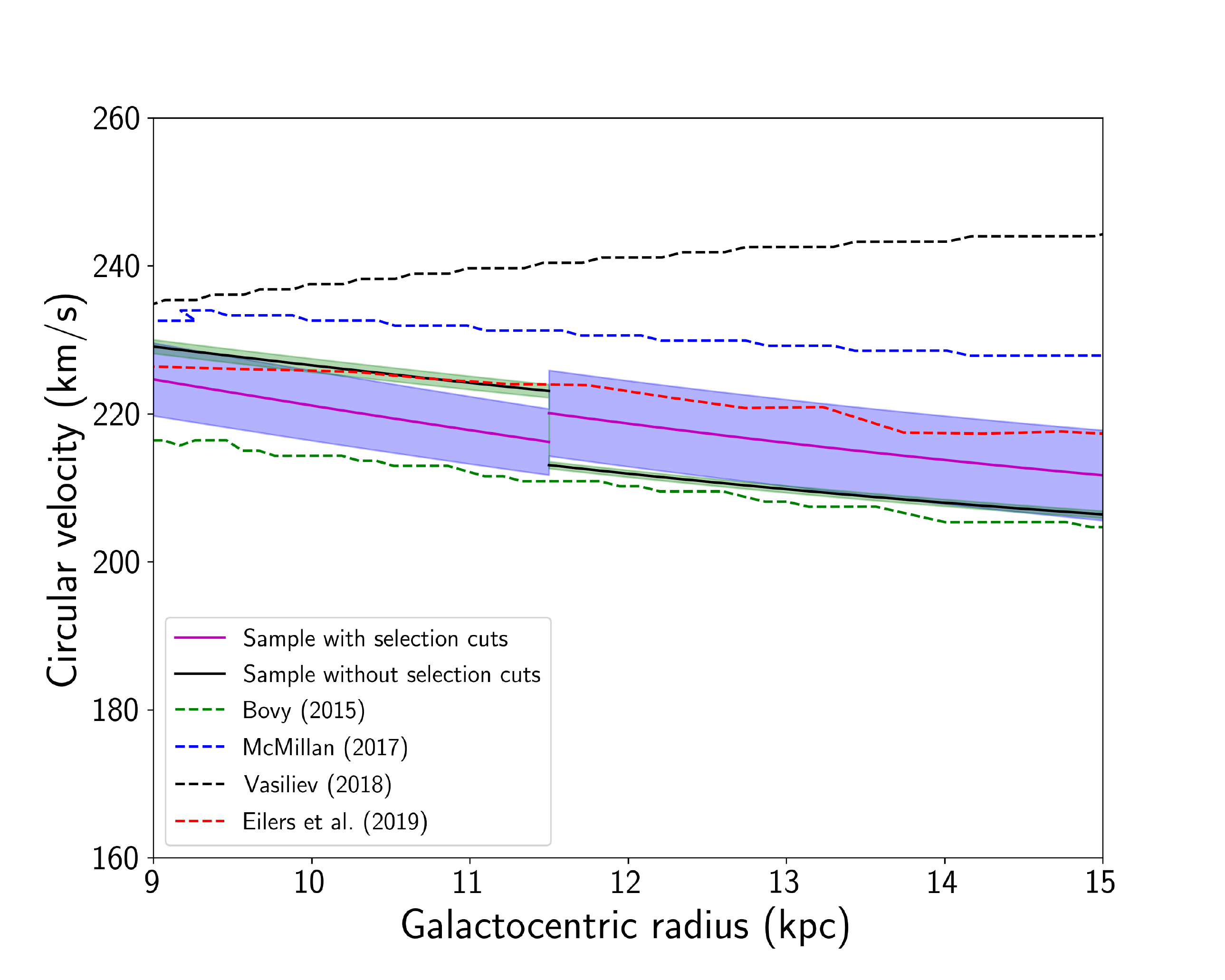}
        \label{velocity_check_total}
    \end{subfigure}
\caption{The total rotation curves calculated from different potential models. Results for this work are indicated as purple solid line (for the sample with selection cuts) and black solid line (or the sample without selection cuts). The shaded area indicates the 1-$\sigma$ uncertainty calculated based on the weighted likelihood (see text for more details). For comparison, results estimated from other works are over-plotted on the same figure. \textit{Left:} curve calculated using the solar information in \citet{Galpy_Paper} while converting the coordinates. The curve is comparable to that of  \citet{Galpy_Paper} (green dashed line) but are systematically lower than three other studies by 9-17\%. \textit{Right:} curve calculated using the solar information used by \citet{Eilers_paper} (GRAVITY measurements). Results are slightly above \citet{Galpy_Paper}'s curve and comparable to \citet{Eilers_paper}'s curve (red dashed line). But even for this curve, they are still systematically lower than the other two studies by 5-10$\%$  [\citet{McMillian_v_rot}: blue dashed line and \citet{globular_cluster_Vasiliev}: black dashed line]}
\label{velocity_check}
\end{figure*}

After obtaining the constraints on both parameters in the dark matter halo density profile, we can translate them to less model-dependent constraints by computing the rotation curve (circular Keplerian velocity) of the Milky Way, as a function of distance from the centre. This result can then be compared to other studies that use different parametrizations and methods. To obtain a more robust estimation to the circular rotation curve, we evaluate the average and standard deviation of $v_{\rm circ.}(R) \equiv \sqrt{\frac{\partial\Phi_{\rm tot}(R,z=0)}{\partial \ln R}}$ given the likelihoods found from our different Gaia samples (Equation \ref{eqn::likelihood}) over our prior range of $f_h$ and $\alpha$ (Equation \ref{eqn::prior_range}). 

As the expression of disk potential is fixed (where we also fix $z$ = 0 in Equation \ref{eqn::Disk_pot}), its contribution  to the total rotation curve (as well as that of the bulge) can be simply added to the halo part in quadrature. Right panel of Figure \ref{velocity_check} displays the circular velocity curve obtained from this work (purple and black solid line) and its corresponding uncertainty shaded area. For comparison, the results obtained from \cite{Galpy_Paper}, \cite{McMillian_v_rot}, \cite{globular_cluster_Vasiliev}, and \citet{Eilers_paper} are also shown in the same figure. \cite{McMillian_v_rot} used kinematic data from maser observations with (expected)  near-circular motion to fit a Milky Way model with an NFW spherical halo, a stellar and gas disk plus a central bulge. Using a nearly identical model, \citet{globular_cluster_Vasiliev} assumed Jeans equilibrium of Milky Way globular clusters in Gaia data to constrain the gravitational potential. \citet{Eilers_paper} also used Jeans equilibrium for Gaia luminous red-giant stars to determine the circular velocity of the Milky way over radial range 5 kpc < R < 25 kpc. Although we approximate the localized halo density profile as a simple power law, our result is consistent with the estimation from \citet{Eilers_paper} and relatively close to (but around 4\% higher than) the best-fit NFW dark matter potential found in \citet{Galpy_Paper} (\texttt{MWPotential2014}). However, the circular velocity (radial force) is about 5-10\% (10-17\%) smaller than the other two studies. Compared with these studies, our method might be more robust as it does not rely on assumptions of circular motion or Jeans equilibrium, and can be equally applied to halo or disk stars.

An important consideration for comparison to other measurements of circular velocity is our choices of the solar coordinates in the coordinate transformation. If we make the choice of solar coordinates consistent with the analysis of \citet{Galpy_Paper} ($V_{\phi,\odot}=220$ km/s and $R_\odot = 8$ kpc), as shown in the left panel of Figure \ref{velocity_check}, the recovered circular velocity curve is comparable to \citet{Galpy_Paper}'s estimation but $\sim$ 9-17\% lower than other three curves. However, the measurements of ${V_\odot}$ and $R_\odot$ have been progressively improving. For example, if we take the GRAVITY results \citep{GRAVITY_paper} used in \citet{Eilers_paper} to convert the coordinates, it does bring our curve close to their measurements. As shown in the right panel of the figure, measurements using GRAVITY solar coordinates are higher than \citet{Galpy_Paper}'s and comparable to those of \citet{Eilers_paper} from Gaia DR2, but are still significantly lower than maser and globular cluster measurements by $\sim$ 5-10\%. 

Let us now comment on our choice of distance (or metric) in the action space (Equation \ref{eqn::distance_calculation_real_data}). The reason why we normalize action variables by their standard deviation to compute distance is partly due to the assumption we made in the likelihood derivation in Appendix \ref{sec::likelihood_modified}. Our derivation starts from a uniformly distributed background plus gaussian fluctuations which model clustering in action space. Therefore, the structures we consider should be on smaller scale than the background distribution in the action space. Since the extent of the background could be different in different directions in the action space, the normalization has the effective role of making the distribution homogeneous and isotropic, at least for $D \ll 1$, i.e. close pairs.

Note that here we ignore the covariance between different action variables. As a sanity check, we modified our distance definition to $D^2 = \sum_{i,j = 1}^3 \Delta J_{i}\Delta J_{j} \sigma_{ij}^{-2}$ (where $\sigma_{ij}^{-2}$ is the inverse covariance matrix of $J_i$'s over the entire sample) accounting for the correlation between action variables. We checked this using both our real data and simulations, and found no significant change in our results (e,g., for 11.5$<$R$<$15 kpc and same $\ln D_{\rm max}$, the relative changes in $f_h$ and $\alpha$ are less than 1\%). 

For the choice of the free parameter $\ln D_{\textrm{max}}$, the main criterion is that we do not expect the constraints on the parameters to significantly vary with $\ln D_{\textrm{max}}$. Therefore, when $\ln D_{\rm max} \lesssim \ln D_{\rm max, optimum}$, the constraints on both parameters should not be a strong function of $\ln D_{\textrm{max}}$ and also be self-consistent within error bars. Meanwhile, the measurements are better to be the least uncertain at the $\ln D_{\rm max, optimum}$. One of the reasons why we plot the median values (with error bars) and two-point correlation function, $\ln\left[\frac{P(\ln D)}{D^3}\right]$, as a function of $\ln D_{\rm max}$ (or generally, $\ln D$) is to see which $\ln D_{\rm max}$ value can give us stable {\it and} reliable constraints. Therefore, we do not expect the optimum choice of $\ln D_{\rm max}$ to be necessarily the same for different systems. This criterion is further explored in Appendix \ref{sec::simulation_with_bg}, where we include a background in our simulations, leading to a different $\ln D_{\rm max, optimum}$.

Here, for real data analysis, we choose ln $D_{\textrm{max, optimum}} = $ -1.14 as mentioned in previous section, and we also choose the same range for the stream-only simulation for consistency (as under this specific case, neither parameters drastically change with ln $D_{\textrm{max}}$). To give more intuition about what the chosen $D_{\rm max}$ physically means, we calculate the values of two-point correlation function using the $\emph{Gaia-Enceladus}$ globular clusters data \citep{globular_cluster_data}, which is shown as orange dashed line in Figure \ref{probability function}. By doing this comparison, we can see that maximum separations of the pairs we considered  (within the considered $D_{\rm max}$) are slightly smaller, but comparable to the size of $\emph{Gaia-Enceladus}$ globular clusters distribution (also known as Gaia Sausage) in the action space, which indicate that the largest size of the structure contributing to the likelihood estimate is close to the characteristic size of the $\emph{Gaia-Enceladus}$ structure. We suggest that the choice of $\ln D_{\rm max}$ needs to be inferred from the behavior of two-point correlation function (as a point of transition from clustered streams to background) and $f_{h,\rm fit}$($\alpha_{\rm fit}$) as a function of $\ln D$ or $\ln D_{\rm max}$, which does help to determine the point where our method and the estimations of parameters are still reliable.

As the assumption made in our likelihood derivation (Appendix \ref{sec::likelihood_modified}) is a uniform background plus a random gaussian field, one might be skeptical about the validity of systematic error estimations with a stream-only simulation as presented in Section \ref{ssec::simulation}. To improve this, we conducted another simulation with the inclusion of a background. Background stars are directly taken from Gaia DR2 and their action variables are calculated in a simulated host potential with [$f_h$ = 0.35, $\alpha$ = 1.70] using the `St\"ackel approximation'. Based on this, we subsequently randomize the stellar distribution in action space by adding a random gaussian scatter to each $J_{i}$ respectively, then stars with randomized action distribution are transformed back to ($\Vec{X}$,$\Vec{V}$) in cylindrical coordinate using the TorusMapper code \citep{torus_mapper_citation}. We combine three original streams with the simulated background and evolve them with different choices of ($f_h$, $\alpha$). We would like to test whether the parameters recovered by the likelihood function is consistent with the initial input of the simulated host potential (which is [$f_h$ = 0.35, $\alpha$ = 1.70]). A detailed analysis is presented in Appendix \ref{sec::simulation_with_bg}. As can be seen from the error bar plots (Figure \ref{error_bar_probability_plots_sim_with_bg}), wherever the constraints do not significantly depend on $\ln D_{\rm max}$, the inclusion of background actually improves the measurements (at $\ln D_{\rm max, optimum}$, the constraint we get is $f_h = 0.352 \pm 0.003 $, $\alpha = 1.678 \pm 0.058$ from quadratic fit, while for a stream-only simulation, we have $f_h = 0.352 \pm 0.001 $, $\alpha = 1.634 \pm 0.014$). Therefore, we conclude that the systematic error estimated from a stream-only simulation should be conservative and can be propagated to further analysis.

As we noted in Sec. \ref{sec::intro}, there are other proposals to use the action-angle (or similar) variables to constrain the potential. \citet{stream_orbit_misalign} use the correlations in the angle-frequency\footnote{For an integrable system, frequencies can be simply thought as another coordinate system in the action space.} space for stars of a single stream to constrain gravitational potential. For a true potential, the angle and frequency differences of stars in a long narrow stream should lie along a straight line. An incorrect potential could cause a misalignment between the stream orbit and the underlying progenitor orbit. By minimizing this misalignment, which is potential-dependent, they manage to recover the expected constraints to a spherical logarithm potential using a simulated tidal stream. While this method uses more information (i.e. angle variables) than ours, and thus can be potentially more precise, it requires identifying only stream stars and relies on the assumption of a cold stream, which does limit its precision and accuracy. \citet{Magorrian_paper} relates the clumpiness of the stellar action-space distribution to the potential and define a likelihood, where the stellar action-space distribution is drawn from a Dirichlet process. This study justifies the viability of constraining the potential using stellar action-space distribution. However, instead of assuming a specific functional form of stellar distribution in the action space, our method is independent of f($\bf J$) as we marginalized over all possible distribution of f($\bf J$) in the derivation of likelihood function (or more precisely, what we assumed here is the probability functional P\{f($\bf J$)\} is gaussian with an arbitrary 2-point function that only depends on $D$.). Methods introduced in \cite{energy_clustering}, \cite{action_space_clustering}, \cite{Sanderson_action_clustering}, and \cite{Applying_Gaia_louville_theorem} are the closest compared with our methodology, which {\it minimize} relative entropy (or KLD) of a system in the space of action variables (or more generally, integrals of motion). However, these studies do not directly connect their statistical representations to the two-point correlation function in the action space as proposed in this study. Indeed, our derivation in Appendix \ref{sec::likelihood_modified} suggests that relative entropy of the distance distribution $P(\ln D)$ [rather than $f({\bf J})$] is more directly related to the likelihood. On a more practical note, given that the density of stars (or pairs of stars) is discrete, the answer does depend on the coarse-graining procedure. However, since there are many more stellar pairs than stars ($N(N-1)/2$ vs $N$), our likelihood computation is much more robust to coarse-graining. Furthermore, to our knowledge, none of these methods have yet been applied to real data.

Finally, to be fair, we should also highlight some of the caveats in our study. Several assumptions are made in the derivation of our likelihood test in Appendix \ref{sec::likelihood_modified}, most importantly that of a uniform background with statistically uniform gaussian fluctuations in the action space. How much do gravitationally bound structures or non-uniformity of the background can bias our finding? While the latter effect is partially captured by the dependence on $D_{\rm max}$, a more systematic test using numerical simulations of galaxy formation may be more satisfactory. Another point of concern is the dependence of the best-fit parameter on the selection cuts. While, this could signal the inadequacy of our current potential model (either for stellar disk or dark matter halo, which may need more free parameters), it could also signal deeper problems such as errors in computing the action variables for Milky Way potential (using methods in Sec. \ref{sec::theory}), or their non-adiabatic evolution. 

\section{Conclusion}\label{sec::conclusion}

In this work, we develop a novel method to constrain gravitational potentials from small-scale clustering in the space of action variables, and use it to provide precise constraints on the Milky Way dark matter halo potential within 9-15 kpc from Gaia DR2. 
We first derive the likelihood function for different host potentials, with the assumption that the stellar distribution in the action space is a uniformly distributed background with correlated gaussian fluctuations on small scales, showing that it can be written as an integral over the two-point correlation function evaluated in the action space. The main advantage of our method is that, contrary to past studies, it does not require identification of any kind of compact structures or streams beforehand, assume the circular orbits, or any equilibrium state of the distribution. We first check the viability of our method in simulations of streams with different host potentials, showing that it recovers the normalization (slope) of the host potential with less than 1\% (4\%) systematic error (while stochastic errors shrink with the number of stars in the sample). We then apply our method to analyze two samples from Gaia DR2 over radial ranges of 9-11 kpc and 11.5-15 kpc, and studied the effect of selection cuts on the final results. Including all the known systematic errors, we find the parameters $(f_h,\alpha)= (0.391\pm 0.009,  1.835\pm 0.092) $  and $(0.351\pm 0.012,1.687\pm 0.079)$, for 9-11 kpc and 11.5-15 kpc respectively, for the median and 68\% CI uncertainty from the posterior distribution. For both simulations and real data, we can visually confirm that the potential that maximizes the likelihood function does indeed correspond to the largest two-point correlation function and most compact distribution in the action space, which again, demonstrates the reliability of our method.

We would like to clarify that the fraction of DM, $f_h$, characterizes the fraction of halo component contribution to the radial force {\it extrapolated to} the position of the Sun. Given the uncertainty, our final index constraints obtained at two radial bins are consistent within $\sim 1.2 \sigma$. Based on the NFW prediction, the absolute value of index gets larger at outer radii. Therefore, a power law extrapolation back to the solar radius would lead to a larger local halo density. Therefore, the fact that we find a larger value of $f_h$ in outer radii is consistent with the expectations from NFW (or any profile that gets steeper at larger radii).

To our knowledge, this is the first study that constrains the halo potential of the Milky Way using the action space clustering with real data. While more work is needed to fully understand the systematic error of this method (as discussed in Section \ref{sec::discussion}), its sheer statistical power is formidable as it scales with the number of all the stars in the sample, and with proper calibrations can provide exquisite constraints on dark matter potential. Further improvements (or checks) may come from identification of streams beforehand or other criteria to separate disk and halo components \citep{Gaia_halo_compoent_2017, box_of_chocolate_2016, Halo_substructure_Gaia_SDSS_2018, halo_velocity_dist_2018}. 
Additionally, in this study, we only varied the parameters in the local dark matter halo density profile but kept the stellar disk potential fixed. More robust constraints, left for future work, requires varying the parameters in the disk potential as well, and possibly include other probes of stellar density. 

\section*{Acknowledgements}
We would like to thank Ana Bonaca, Jo Bovy, Charlie Conroy, and David Hogg for useful discussion.

This work was supported by the University of Waterloo, Natural Sciences and Engineering Research Council of Canada (NSERC), and the Perimeter Institute for Theoretical Physics. Research at the Perimeter Institute is supported by the Government of Canada through Industry Canada, and by the Province of Ontario through the Ministry of Research and Innovation.

This work has made use of data from the European Space Agency (ESA) mission {\it Gaia} (\url{https://www.cosmos.esa.int/gaia}), processed by the {\it Gaia} Data Processing and Analysis Consortium (DPAC,
\url{https://www.cosmos.esa.int/web/gaia/dpac/consortium}). Funding for the DPAC has been provided by national institutions, in particular the institutions participating in the {\it Gaia} Multilateral Agreement.
\bibliographystyle{mnras}
\bibliography{draft}

\appendix
\onecolumn
\section{Derivation of the Likelihood Function}\label{sec::likelihood_modified}

In this section, we discuss the derivation of the likelihood which we use to constrain the parameters of the Milky Way potential. Using Bayes' Theorem, we have,
\begin{equation}\label{eqn::posterior_likelihood_appendix}
    \mathcal{P}(\Theta|{\cal D}) = \frac{\mathcal{P}({\cal D}|\Theta)\mathcal{P}(\Theta)}{\mathcal{P}({\cal D})}
\end{equation}
where $\Theta$ are the parameters we want to constrain, while ${\cal D}$ is the data we have available. In particular, in the context of constraining the Milky Way potential which is our goal here, $\Theta$ represents the parameters of the potential, $(f_h,\alpha)$.

Let us divide the action space into $M$ small bins such that the average number of stars {\it per bin} (over all of action space) is $\bar{n} \ll 1  $. However, because of hierarchical structure formation, the expected number counts of stars in different bins will not be independent of each other. In order to capture this, we assume that the star count in bin $a$ is a Poisson sampling of a mean, $\bar{n}+\bar{n}\chi_a$, where $\chi_a$'s are correlated random gaussian variables. 
Therefore, the probability of measuring star counts $\{\nu_a\}$ is given by:
\begin{equation}\label{eqn::marginalised_prob}
    \mathcal{P}(\{\nu_a\},\{\chi_a\}) = \frac{\exp\left(-\frac{1}{2}\sum_{a,b}\chi_a\xi^{-1}_{ab}\chi_b\right)}{\sqrt{\text{det}(\xi)}(2\pi)^{M/2}} \prod_a \frac{(\bar{n}+\bar{n}\chi_a)^{\nu_a}\exp(-\bar{n}-\bar{n}\chi_a)}{\nu_a!}.
\end{equation}

From this definition, it follows that
\begin{equation}\label{eqn::F_def}
    \langle\chi_a\chi_b\rangle = \xi_{ab},   
\end{equation}
is the covariance matrix of the random gaussian variables $\{\chi_a\}$. Since we do not directly observe $\chi_a$'s, we should marginalize over them. Therefore, in the limit of $M \to \infty$ (i.e. when $\nu_a =0$ or $1$) the posterior (Equation \ref{eqn::posterior_likelihood_appendix}) is given by:
\begin{equation}
    \mathcal{P}(f_h,\alpha|\{a_k\}) \propto \mathcal{P}_{\rm prior}(f_h,\alpha) \int  \frac{\exp\left(-\frac{1}{2}\sum_{a,b}\chi_a\xi^{-1}_{ab}\chi_b\right)}{\sqrt{\text{det}(\xi)}(2\pi)^{M/2}} \left\{\prod_k \left[1+\chi_{a_k}\right] \right\}\prod_a d\chi_a,
\end{equation}
where $a_k$ is the action-space bin in which the {\it k}-th star lies, and we assume $\sum_a \chi_a =0$ over the entire action space. Using Wick's theorem, the above Gaussian integral can be expressed in terms of a sum of the product of 2-point functions over all possible pairings of stars. This yields the likelihood (defined as the ratio of posterior to prior):
\begin{equation}
    \mathcal{L}({\rm data}|f_h,\alpha) \equiv \frac{\mathcal{P}(f_h,\alpha|\{a_k\})}{\mathcal{P}_{\rm prior}(f_h,\alpha)} \propto \left\langle\prod_k \left[1+\chi_{a_k}\right] \right\rangle \propto \sum_{\rm pairings} \prod_{\rm pairs}\left[1+\xi_{\rm pair}\right] \propto \left\langle\exp\left[ \sum_{\rm pairs}\ln\left(1+\xi_{\rm pair}\right)\right]\right\rangle_{\rm pairings}.
\end{equation}
Now, for a large number of pairs, we can expect that the exponent in this expression have smaller and smaller relative fluctuations around its mean for different possible pairings. This is often known as the {\it mean-field approximation} in statistical mechanics (where the sum represents the partition function and the exponent is proportional to the energy), and allows us to move the average inside the exponent:   
\begin{equation}
\ln\mathcal{L}({\rm data}|f_h,\alpha) \approx  \left\langle \sum_{\rm pairs}\ln\left(1+\xi_{\rm pair}\right) \right\rangle_{\rm pairings}
\end{equation}
This equation defines our log-likelihood formula adopted in Equation \ref{eqn::likelihood} in the main text (and subsequent statistical analyses), where we further assume that the 2-point function $\xi$ only depends on the normalized distance $D$ (Equation \ref{eqn::distance_calculation_real_data}) in the action space.  

\section{Sanity check for simulations}\label{sec::sanity_check_simulation}

\begin{figure*}
    \centering
    \begin{subfigure}
        \centering
        \includegraphics[width=.3\linewidth]{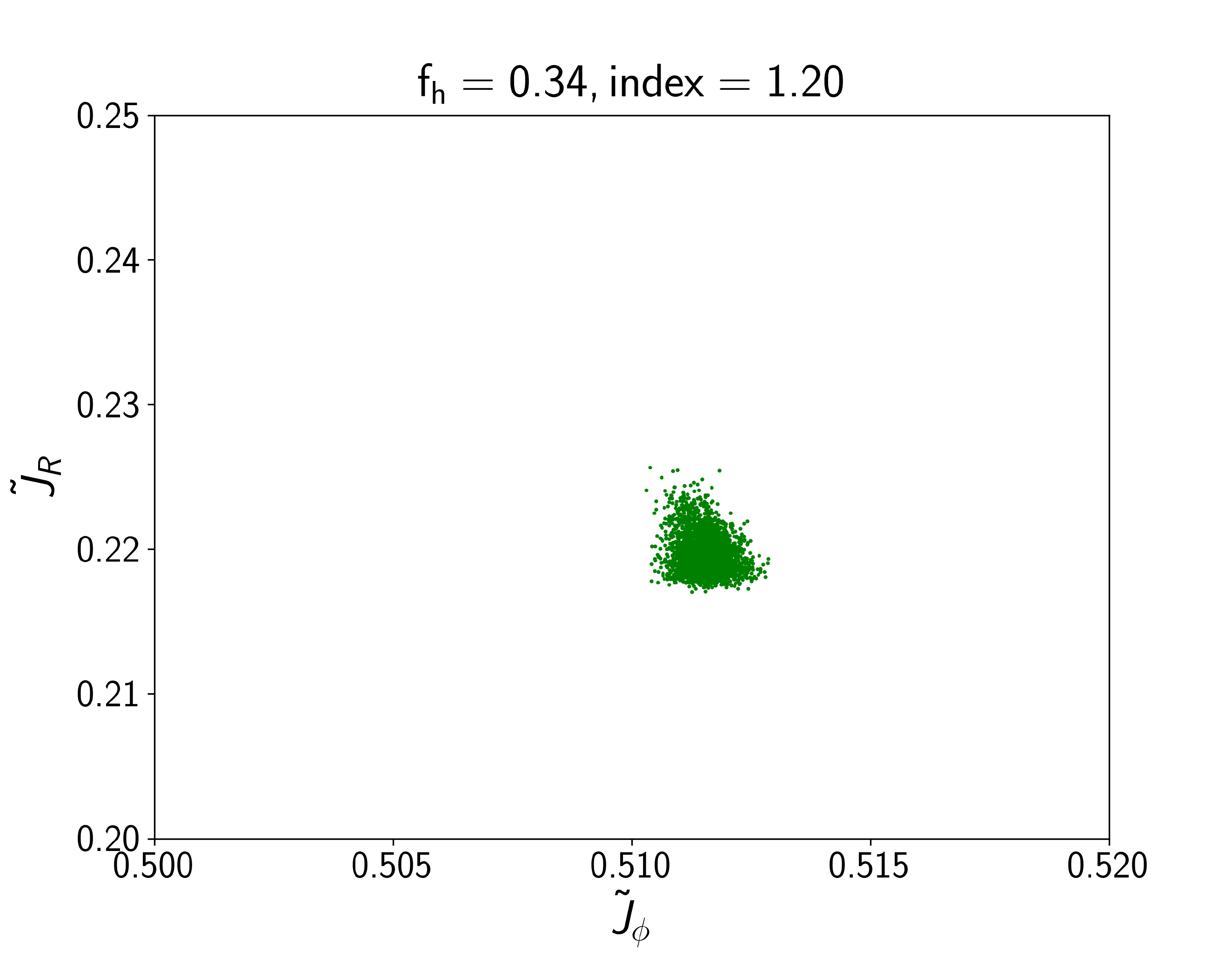}
        \label{action_var_f_h_fixed_p1_simulation_3517}
    \end{subfigure}%
    \begin{subfigure}
        \centering
        \includegraphics[width=.3\linewidth]{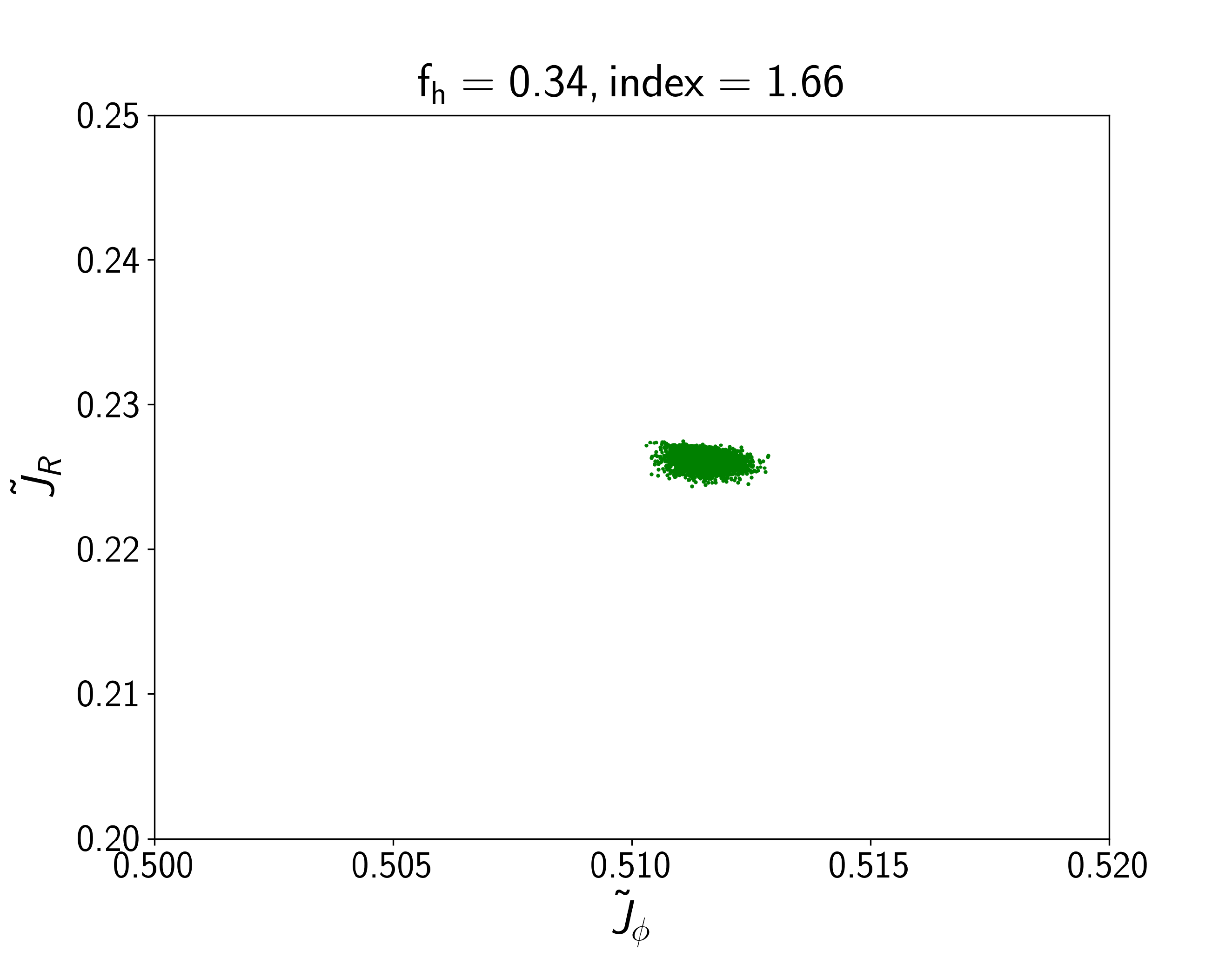}
        \label{action_var_f_h_fixed_p2_simulation_3517}
    \end{subfigure}
    \begin{subfigure}
        \centering
        \includegraphics[width=.3\linewidth]{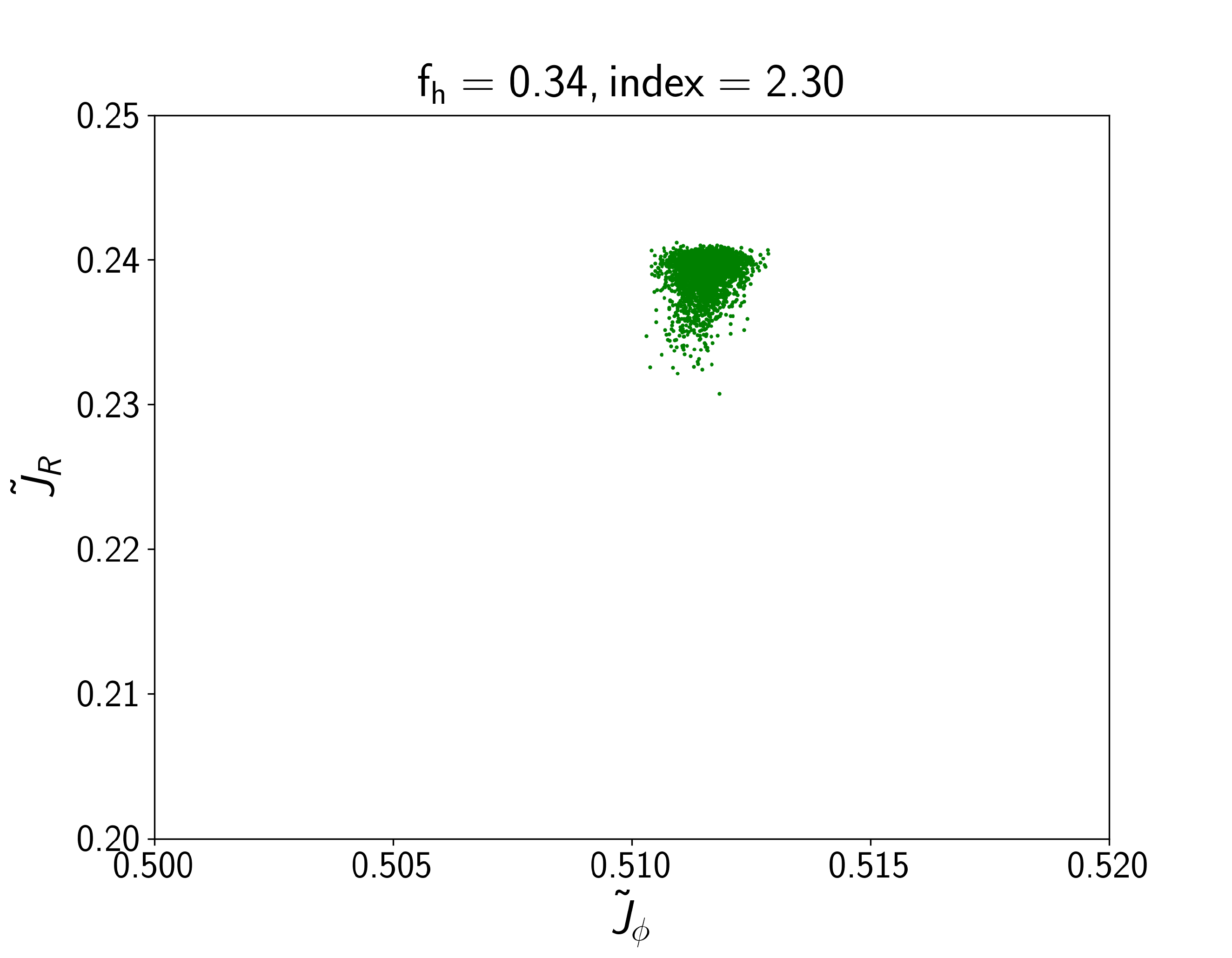}
        \label{action_var_f_h_fixed_p3_simulation_3517}
    \end{subfigure}
    \begin{subfigure}
        \centering
        \includegraphics[width=.3\linewidth]{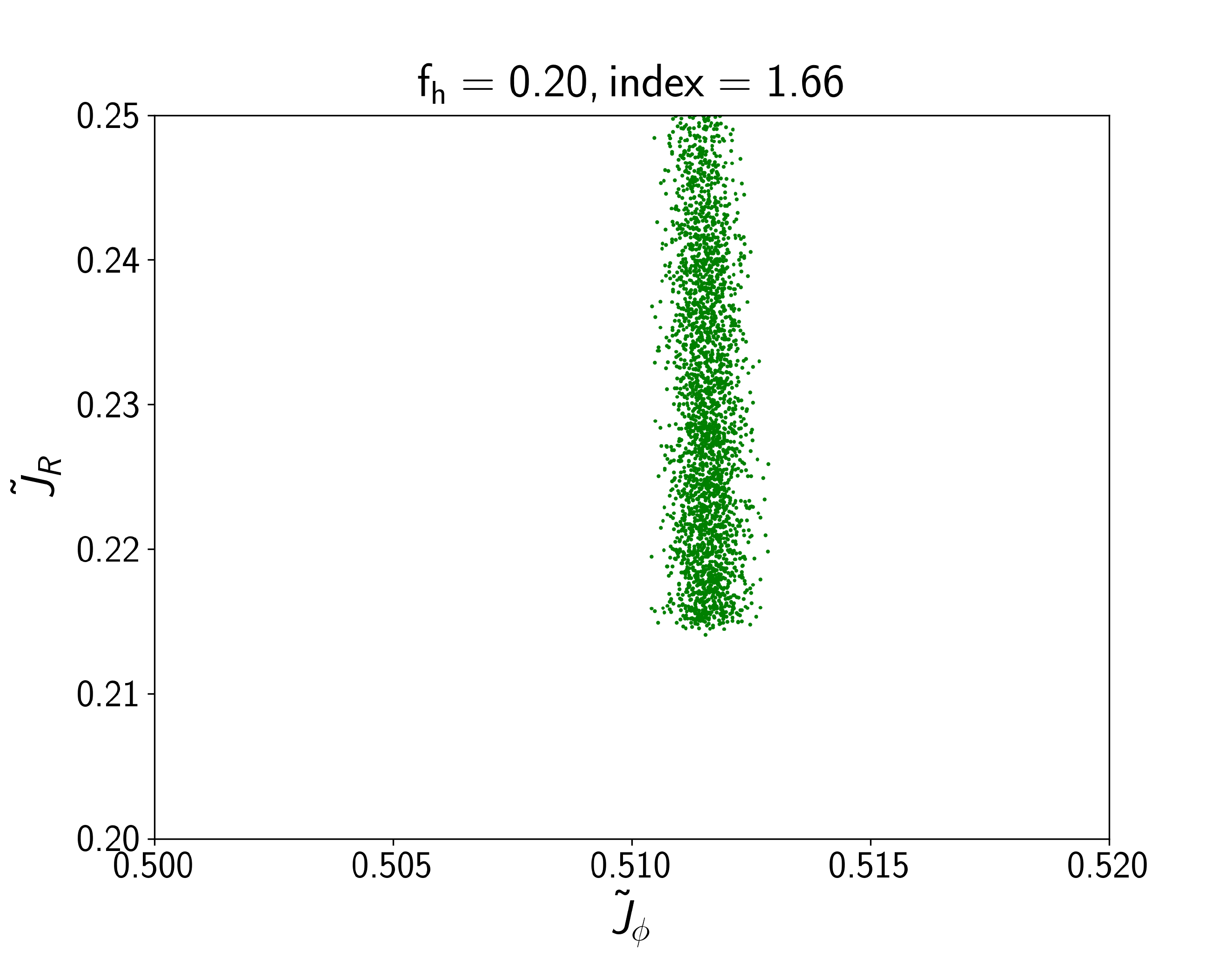}
        \label{action_var_index_fixed_p1_simulation_3517}
    \end{subfigure}%
    \begin{subfigure}
        \centering
        \includegraphics[width=.3\linewidth]{Figures/action_var_central_simulation_3517.pdf}
        \label{action_var_index_fixed_p2_simulation_3517}
    \end{subfigure}
    \begin{subfigure}
        \centering
        \includegraphics[width=.3\linewidth]{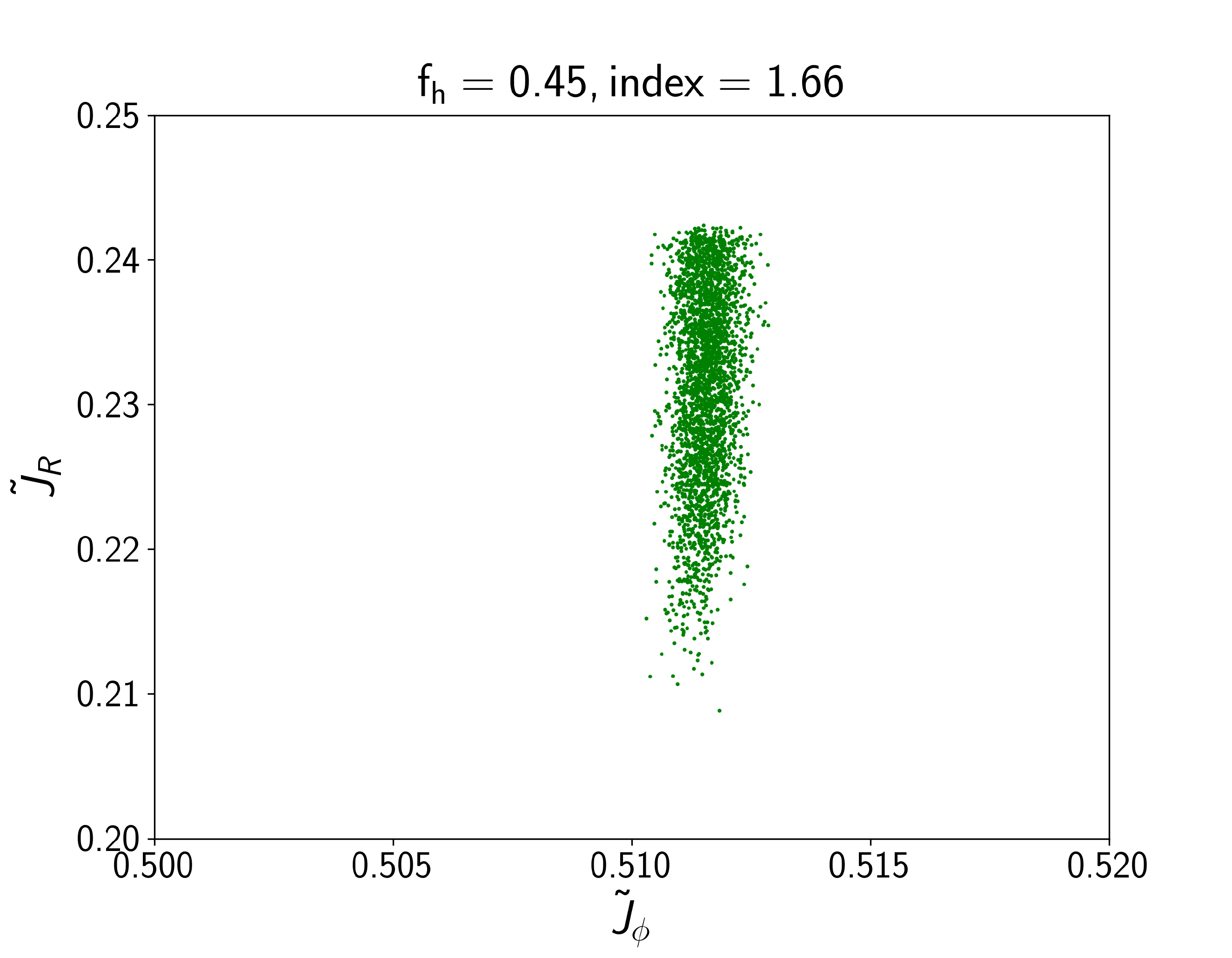}
        \label{action_var_index_fixed_p3_simulation_3517}
    \end{subfigure}
    \centering
    \begin{subfigure}
        \centering
        \includegraphics[width=.3\linewidth]{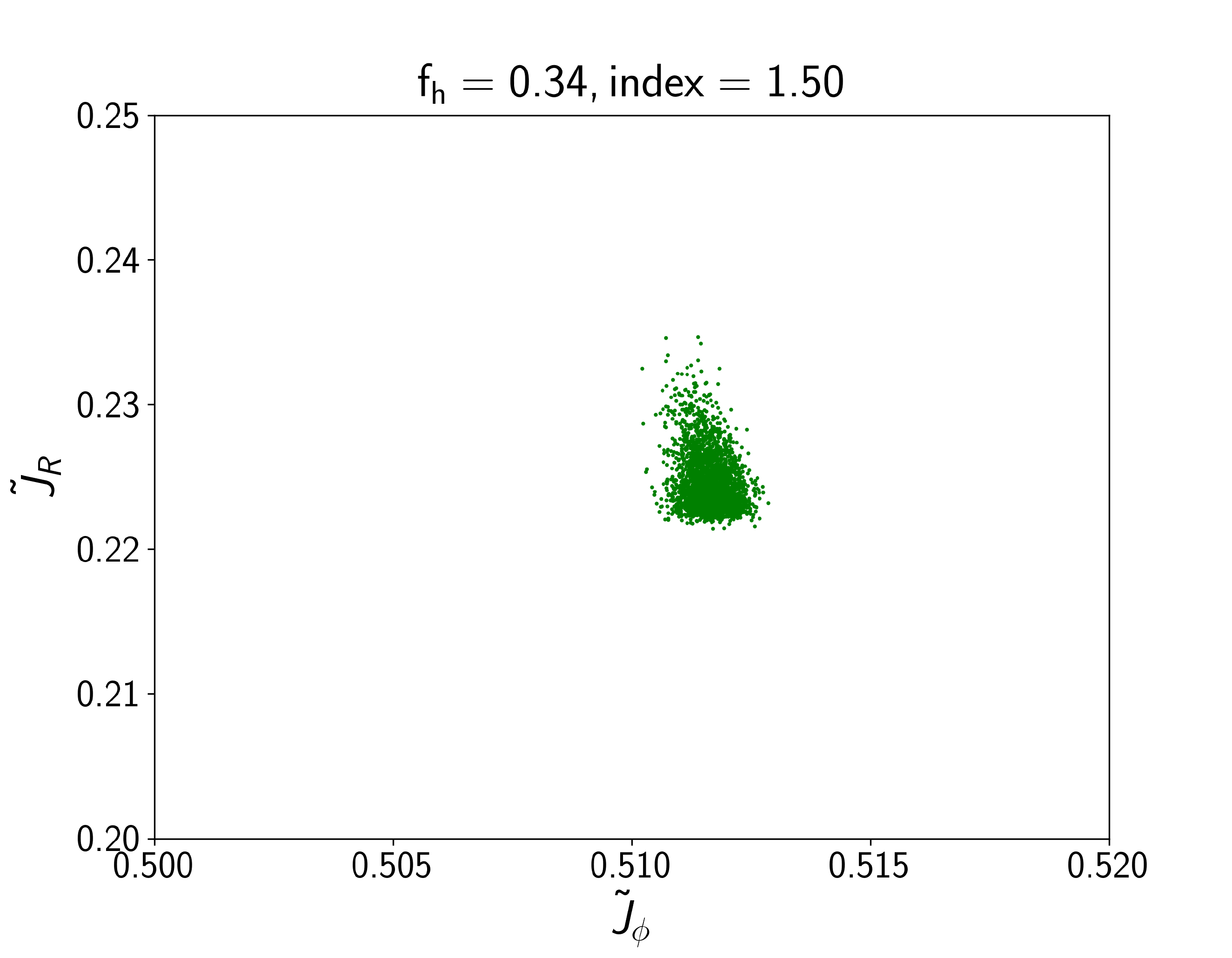}
        \label{action_var_f_h_fixed_p1_simulation_3520}
    \end{subfigure}%
    \begin{subfigure}
        \centering
        \includegraphics[width=.3\linewidth]{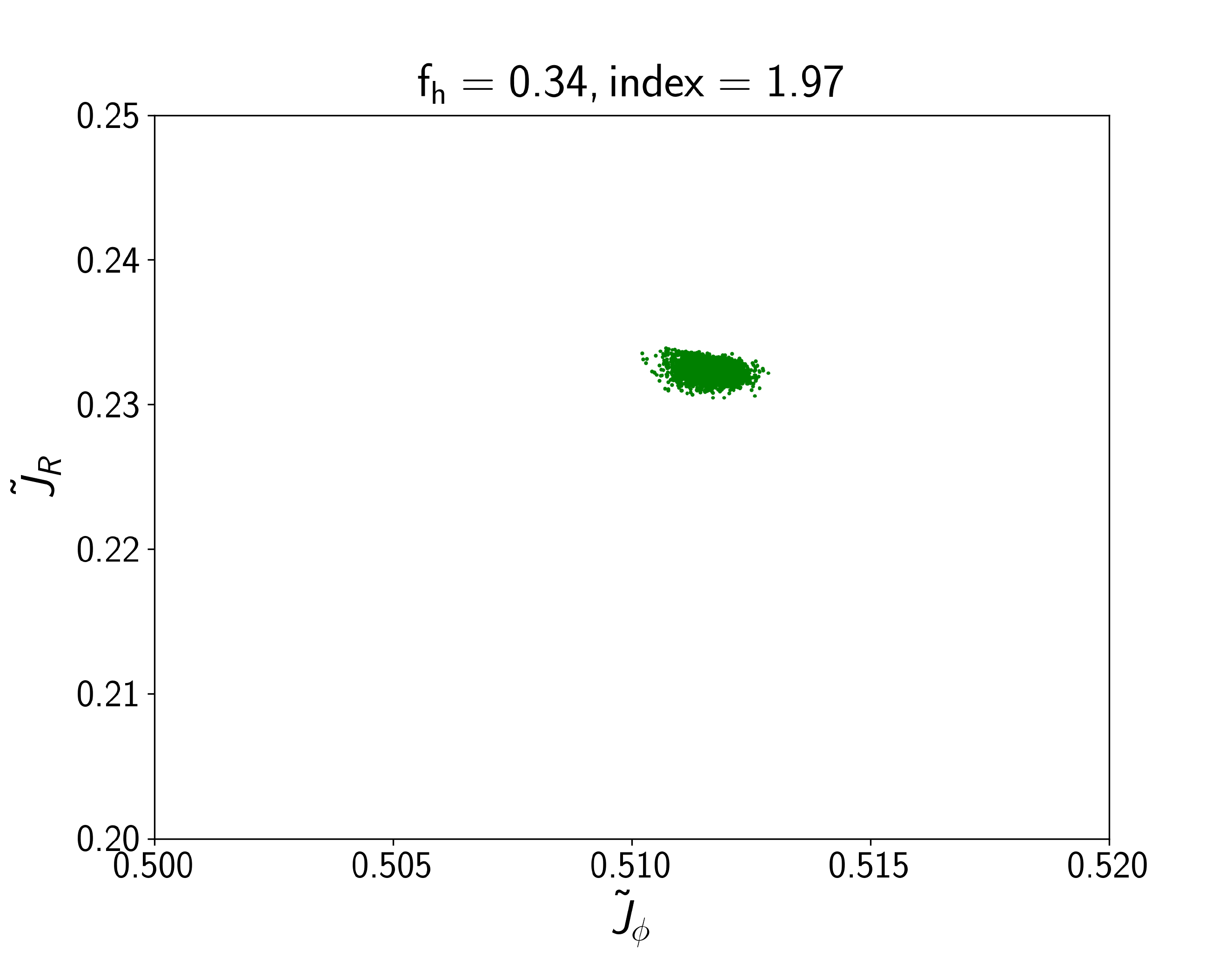}
        \label{action_var_f_h_fixed_p2_simulation_3520}
    \end{subfigure}
    \begin{subfigure}
        \centering
        \includegraphics[width=.3\linewidth]{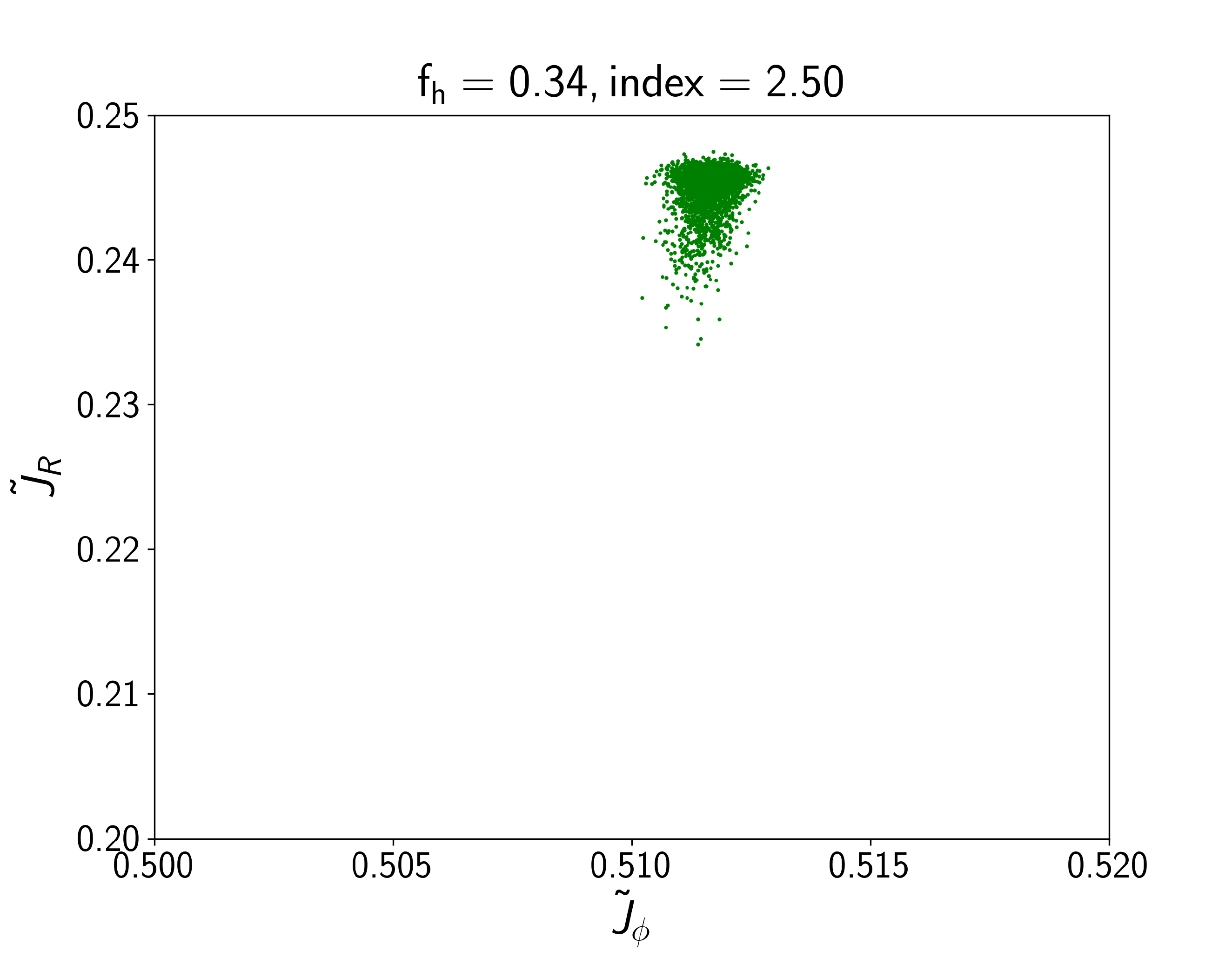}
        \label{action_var_f_h_fixed_p3_simulation_3520}
    \end{subfigure}
    \begin{subfigure}
        \centering
        \includegraphics[width=.3\linewidth]{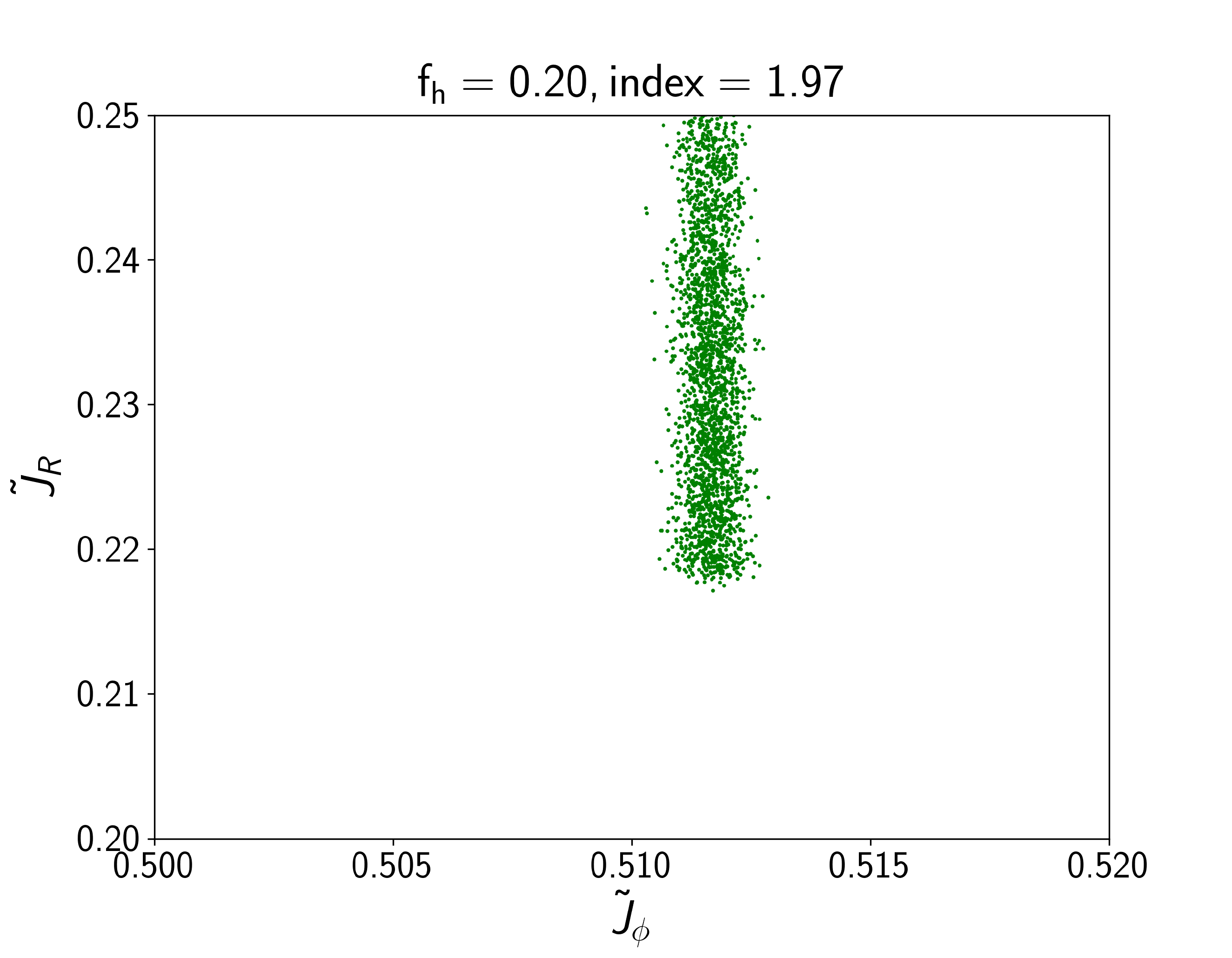}
        \label{action_var_index_fixed_p1_simulation_3520}
    \end{subfigure}%
    \begin{subfigure}
        \centering
        \includegraphics[width=.3\linewidth]{Figures/action_var_central_simulation_3520.pdf}
        \label{action_var_index_fixed_p2_simulation_3520}
    \end{subfigure}
    \begin{subfigure}
        \centering
        \includegraphics[width=.3\linewidth]{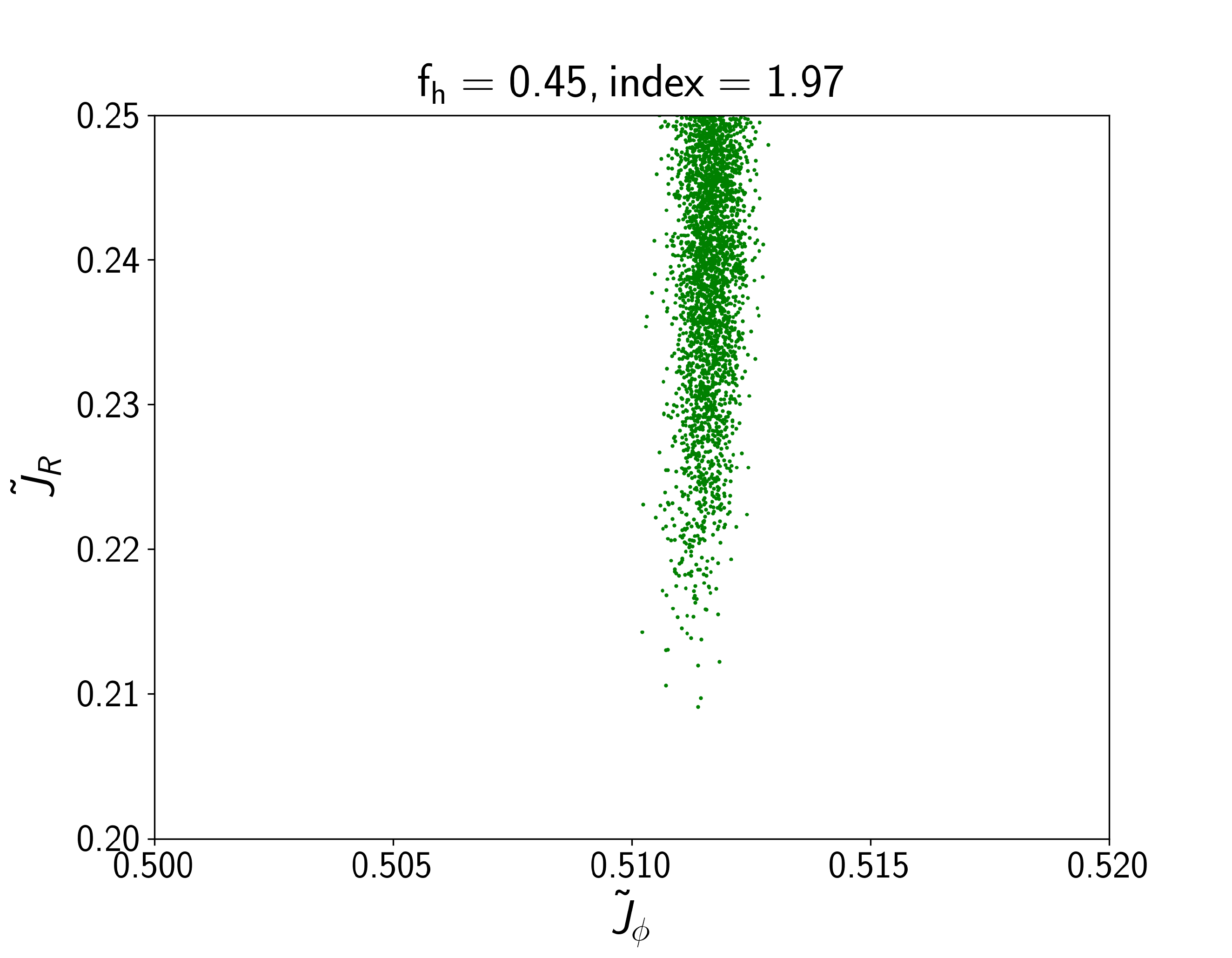}
        \label{action_var_index_fixed_p3_simulation_3520}
    \end{subfigure}
\caption{Stellar distribution of one stream in the $\Tilde{J}_R$ and $\Tilde{J}_\phi$ 2D projected plane, where $\Tilde{J}_R$ and $\Tilde{J}_\phi$ are defined as $J_{R}/\sigma_{J_R}$ and $J_{\phi}/\sigma_{J_{\phi}}$. This figure is aimed at presenting a general view of how the "compactness" (clustering behaviour) of stars in the action space varying with different choices of potential. First two rows show the action distribution for the simulation with [$f_h$ = 0.35, $\alpha$ = 1.70], varying $f_h$ and $\alpha$ used in the calculation of action variable. Last two rows show the same thing the simulation with [$f_h$ = 0.35, $\alpha$ = 2.00]. As expected, stars appear to be most clustered if the correct parameters are used to compute the actions. Full movies are available online: \url{https://github.com/Supranta/GAIA_Potential/tree/master/Animation_movies}}
\label{sim_action_var}
\end{figure*}

\begin{figure*}
    \centering
    \begin{subfigure}
        \centering
        \includegraphics[width=.45\linewidth]{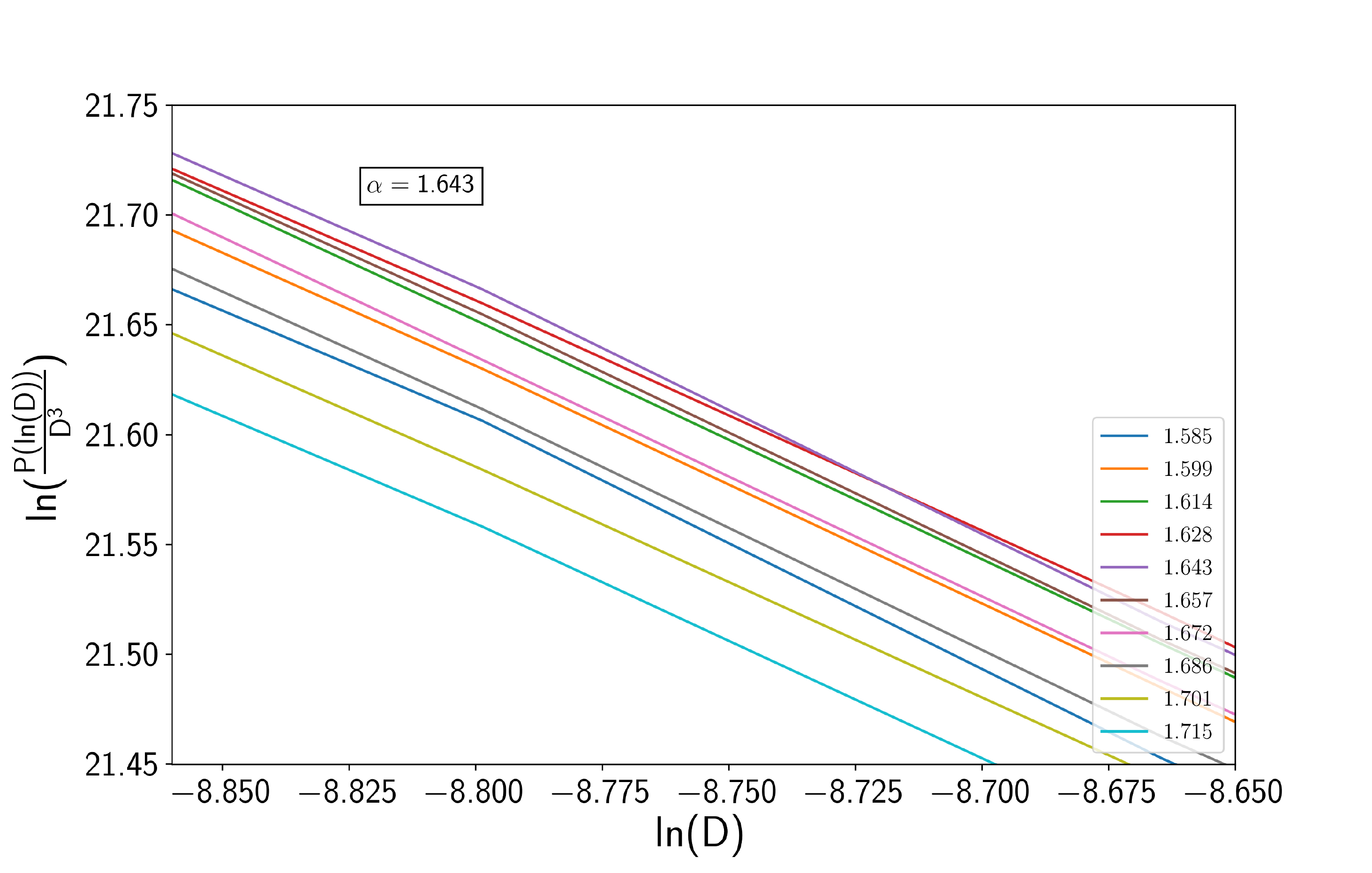}
        \label{probability_function_simulation_fixed_f_h_035_3517}
    \end{subfigure}%
    \begin{subfigure}
        \centering
        \includegraphics[width=.45\linewidth]{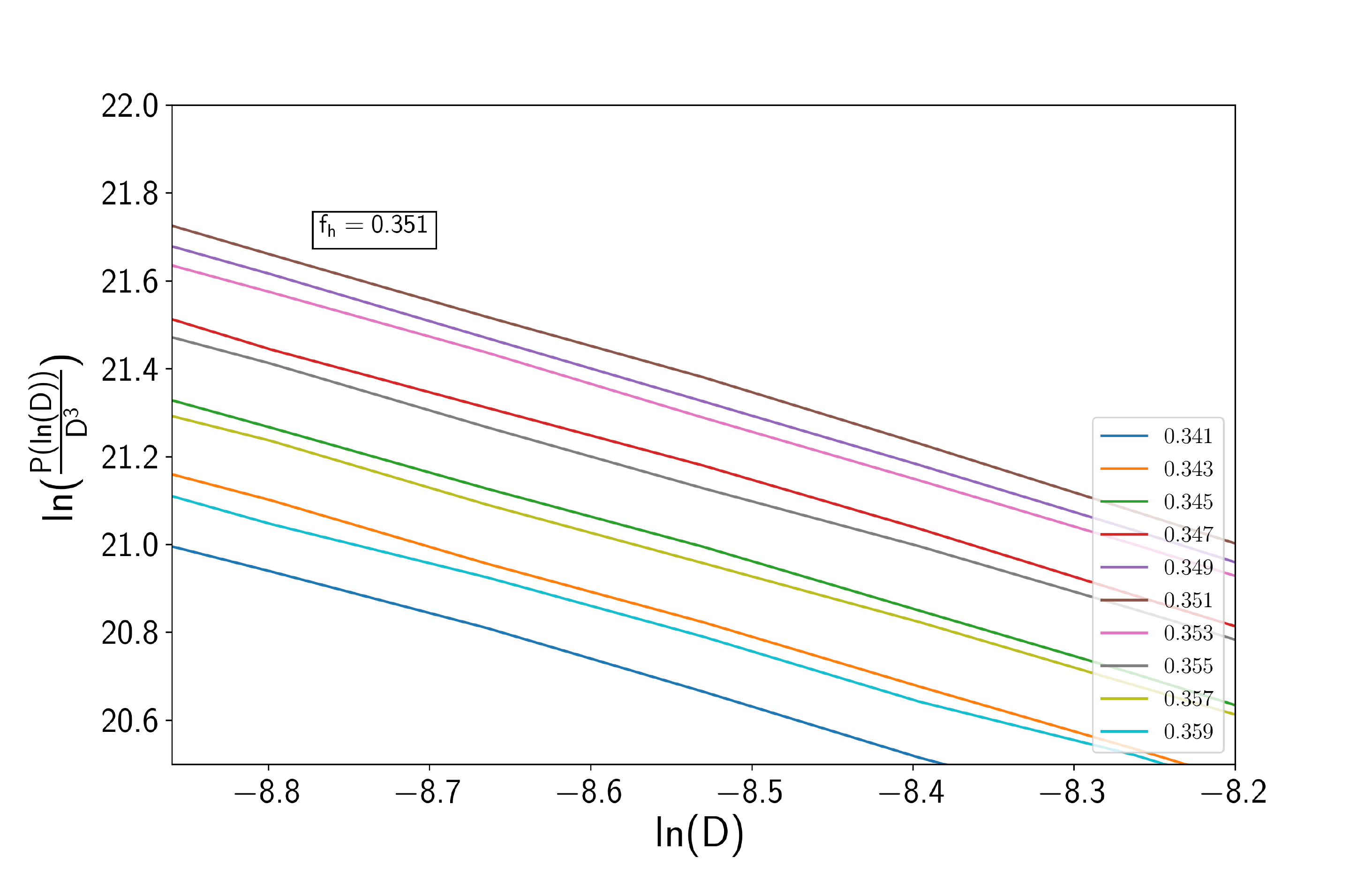}
        \label{probability_function_simulation_fixed_index_1633_3517}
    \end{subfigure}
    \begin{subfigure}
        \centering
        \includegraphics[width=.45\linewidth]{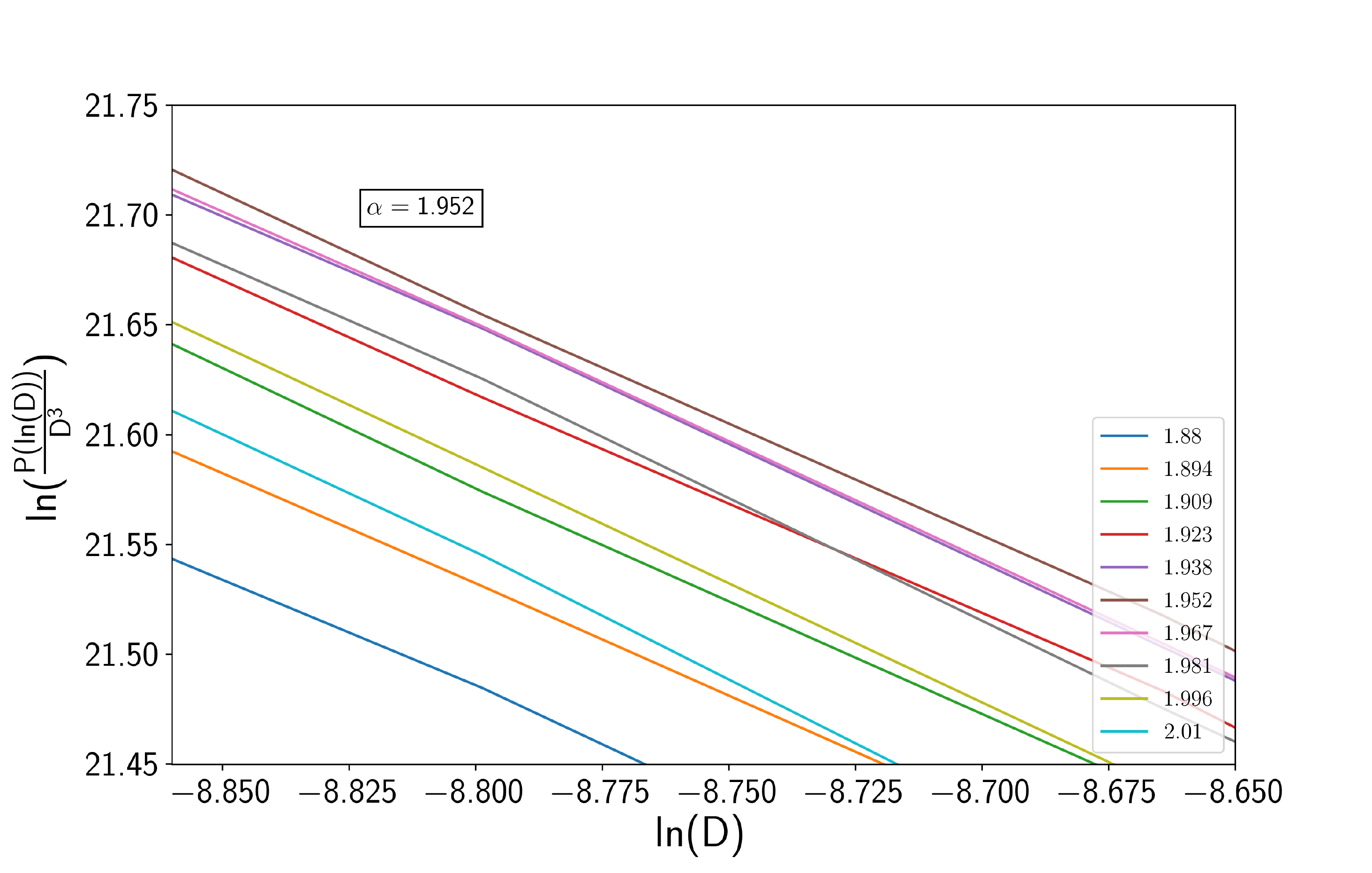}
        \label{probability_function_simulation_fixed_f_h_035_3520}
    \end{subfigure}%
    \begin{subfigure}
        \centering
        \includegraphics[width=.45\linewidth]{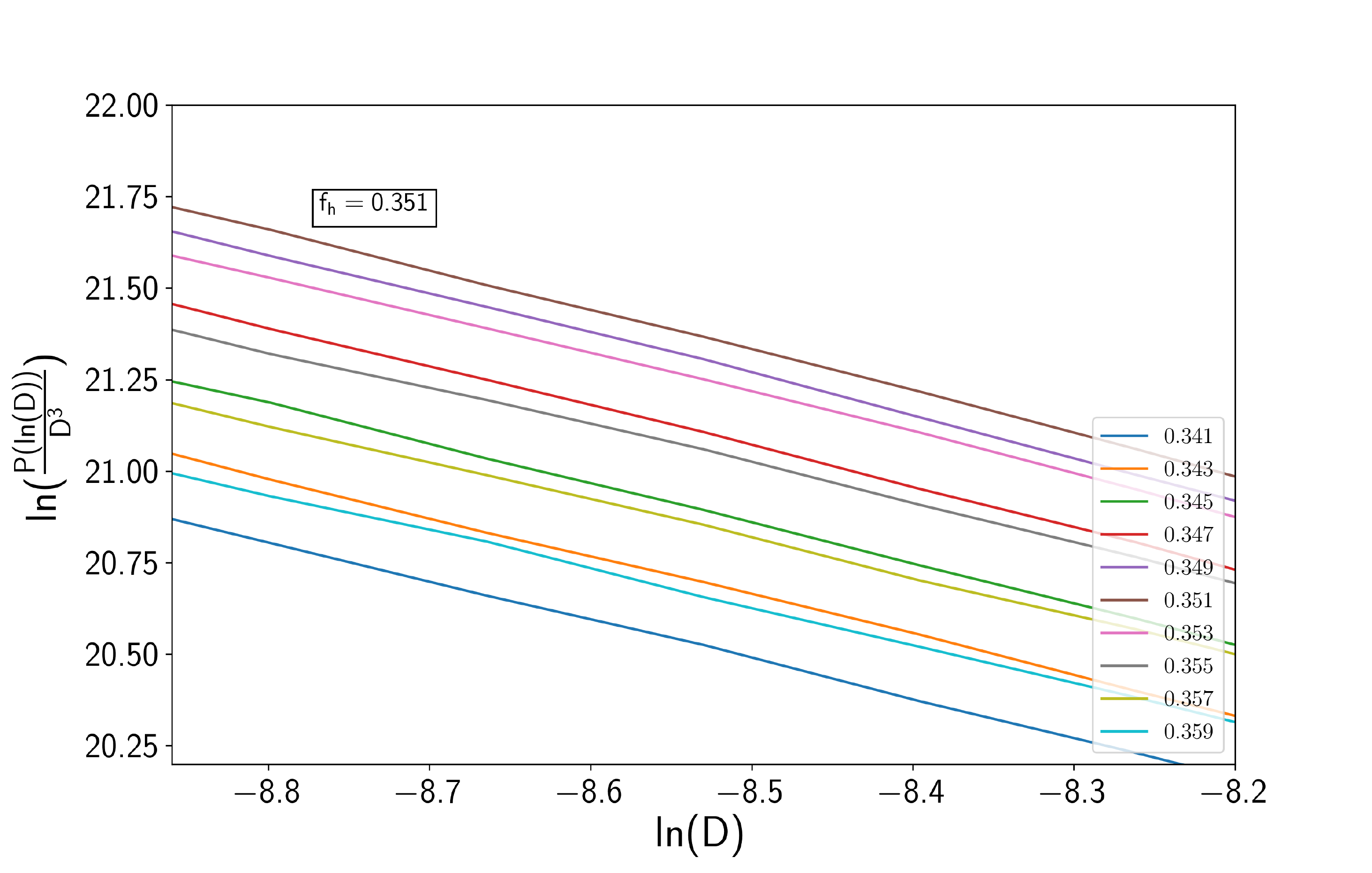}
        \label{probability_function_simulation_fixed_index_1957_3520}
    \end{subfigure}
\caption{Correlation function $\frac{P(\ln D)}{D^3}$ as a function of the distance in the action space in natural logarithm scale. The purpose of this figure is to check
how the two-point correlation function varies with different choices of potential, and whether the correlation function is maximized at the correct set of parameter. Top panel: the behaviour of two-point correlation function for case [$f_h$ = 0.35, $\alpha$ = 1.70] with fixed $f_h$ ($\alpha$) on the left (right). Different colors indicate the values of ln($\frac{P(\ln D)}{D^3}$) at different choices of potential.Bottom panel: the behaviour of two-point correlation function for case [$f_h$ = 0.35, $\alpha$ = 2.00] with fixed $f_h$ ($\alpha$) on the left (right).}
\label{probability_function_simulation}
\end{figure*}

To check whether the maximum likelihood test does correspond to the most clustering in the action space for either simulations, we plot the 2D projection of stellar distribution and two-point correlation function varying the choice of potential parameters $f_h$ and $\alpha$ used in the action computation. Here, we summarize the results. Figure \ref{sim_action_var} shows how a 2D projection of stellar distribution in the action space varies with different choices of potential for one of the simulations. In these figures, $f_h$ ($\alpha$) is fixed, while  $\alpha$ ($f_h$) is varying across its correct value. As expected, for both simulations, the most compact distributions occur when parameters approach the correct values for the simulation (middle panel in both figures). This is also verified in the behavior of the two-point correlation function in Figure \ref{probability_function_simulation}, where one of the parameters is fixed and the other one is varying. For both simulations, we see that the two-point correlation function is indeed maximized around the expected value, which proves the viability of our method.

\begin{figure*}
    \centering
    \begin{subfigure}
        \centering
        \includegraphics[width=1.0\linewidth]{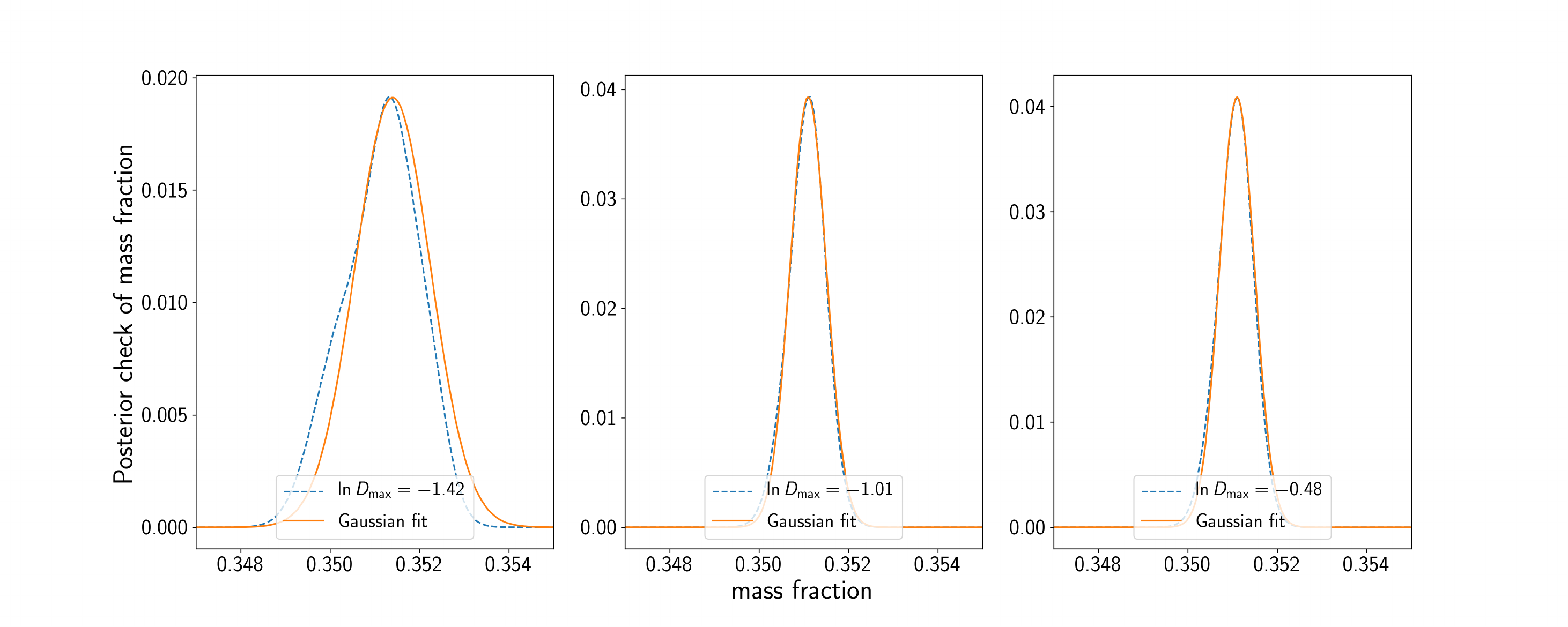}
        \label{posterior_sim_fh_3517}
    \end{subfigure}%
    \begin{subfigure}
        \centering
        \includegraphics[width=1.0\linewidth]{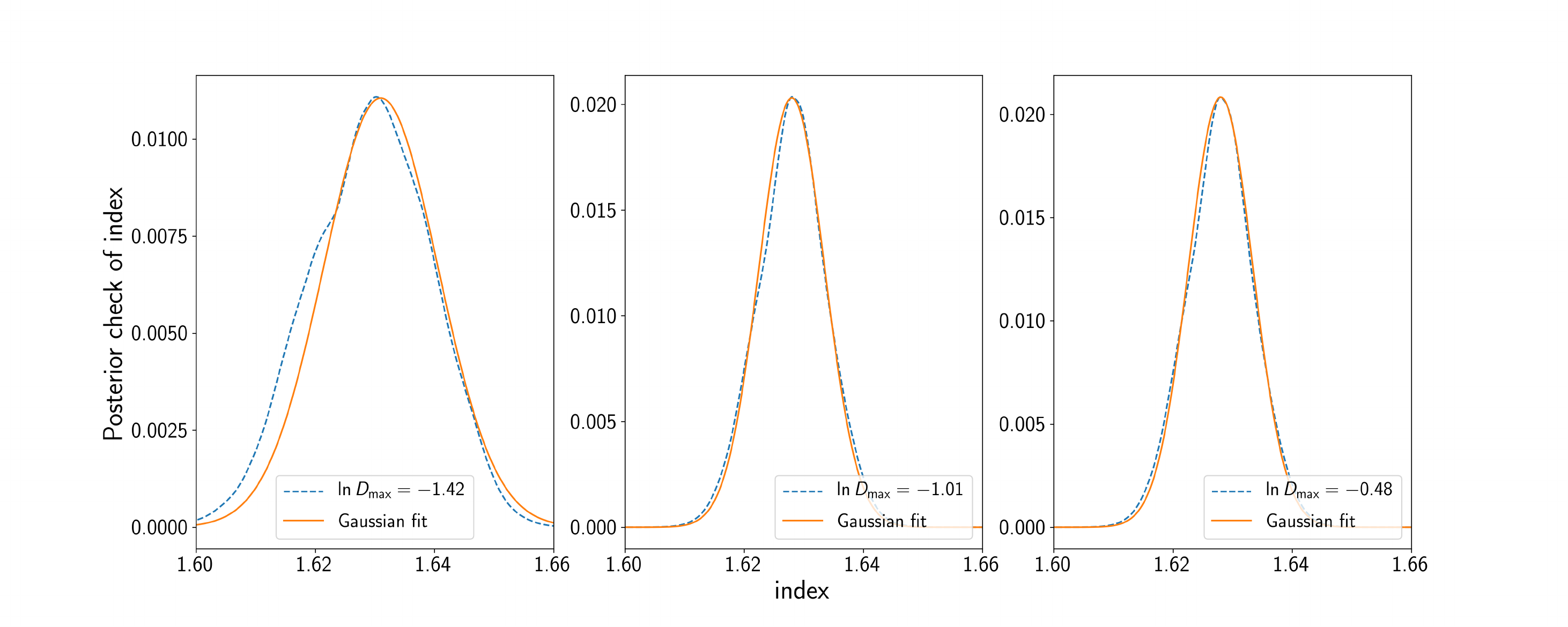}
        \label{posterior_sim_index_3517}
    \end{subfigure}
\caption{The posterior distribution of $f_h$ (upper panel) and $\alpha$ (lower panel) for case [$f_h$ = 0.35, $\alpha$ = 1.70] in simulation. The distributions are evaluated at three different values of $D_{\textrm{max}}$. A Gaussian fit (blue dashed line) has also been over-plotted on each panel for comparison. We see that the posterior distribution is well approximated by a Gaussian.}
\label{posterior_sim_3517}
\end{figure*}

\section{Simulation with the inclusion of a background}\label{sec::simulation_with_bg}

In Section \ref{ssec::simulation}, we conducted a simulation using a system that is entirely composed of stream stars. The discrepancies between the initial simulated host potentials and the parameters recovered by simulation are regarded as the systematic errors, and the systematic errors are further propagated to the real data analysis. However, as the assumption made in the likelihood derivation (Appendix \ref{sec::likelihood_modified}) is that the stellar distribution in action space is a uniform background plus gaussian fluctuations, in this section, we present the results obtained from another set of simulations with a realistic background.
 
To generate a uniform background in the action space, we take real observations from Gaia DR2 and then calculate their actions using a power law potential with [$f_h$ = 0.35, $\alpha$ = 1.70]. Then, we randomize the action distribution by adding a random gaussian scatter to each $J_{i}$. The scatter is generated from a gaussian distribution $N(0, \sigma^2_{\rm scatter})$ , where $\sigma^2_{\rm scatter}= \frac{1}{3}<J_i^2>D_{\rm max}^2$ with $\ln D_{\rm max}$ = -1 (equivalent to coarse-graining the action distribution using a gaussian filter of width $\ln D_{\rm max}$ = -1). Then, the randomized action variables are transformed back to the position and velocity in the cylindrical coordinates using the TorusMapper code developed by \citet{torus_mapper_citation}, which is implemented using \texttt{galpy} package. Following these steps, a set of background stars, which are generated by randomizing the action distribution calculated from Gaia real data, can be produced. Therefore, we can then combine this background with three stream stars that are evolved in the same host potential for further analysis. Using the likelihood function defined in Equation \ref{eqn::likelihood}, we will then test whether the recovered potential that corresponds to the most clustering distribution in action space is the same as the initial input.

\begin{figure*}
    \centering
    \begin{subfigure}
        \centering
        \includegraphics[width=1.0\linewidth]{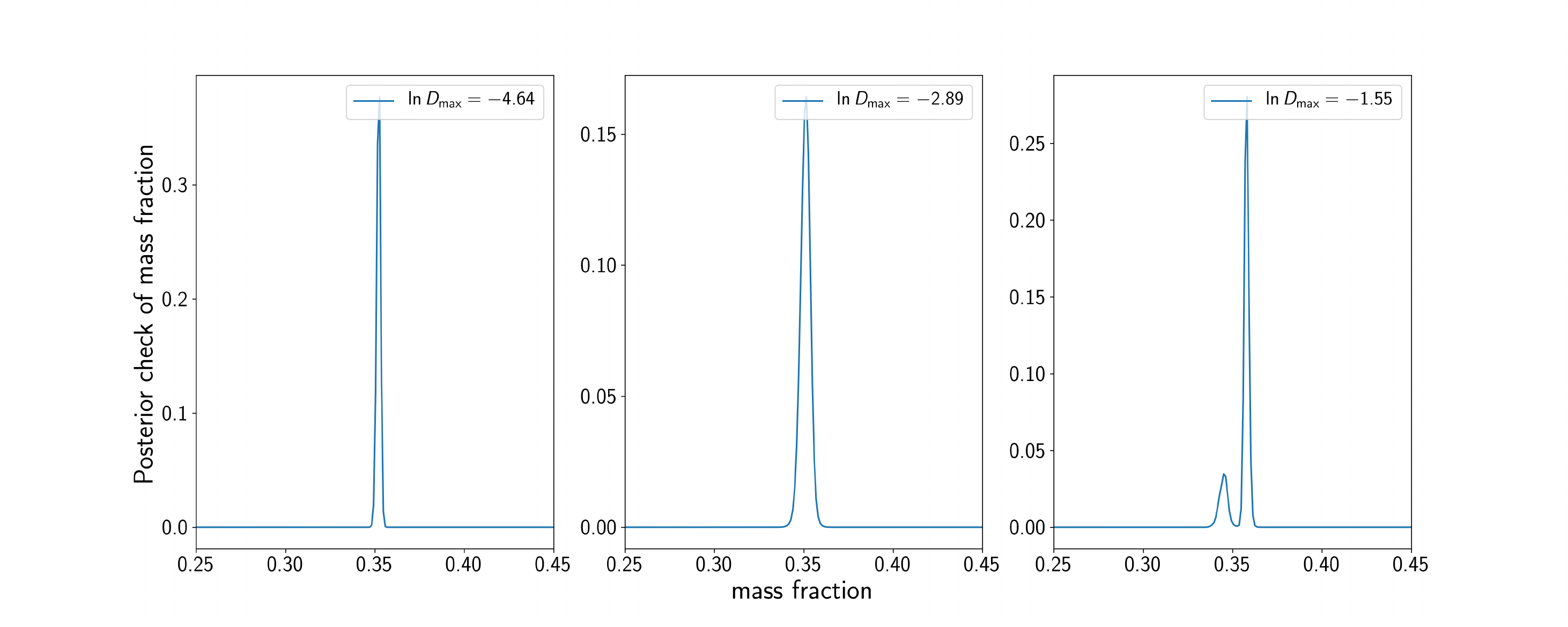}
        \label{posterior_sim_fh_with_bg}
    \end{subfigure}%
    \begin{subfigure}
        \centering
        \includegraphics[width=1.0\linewidth]{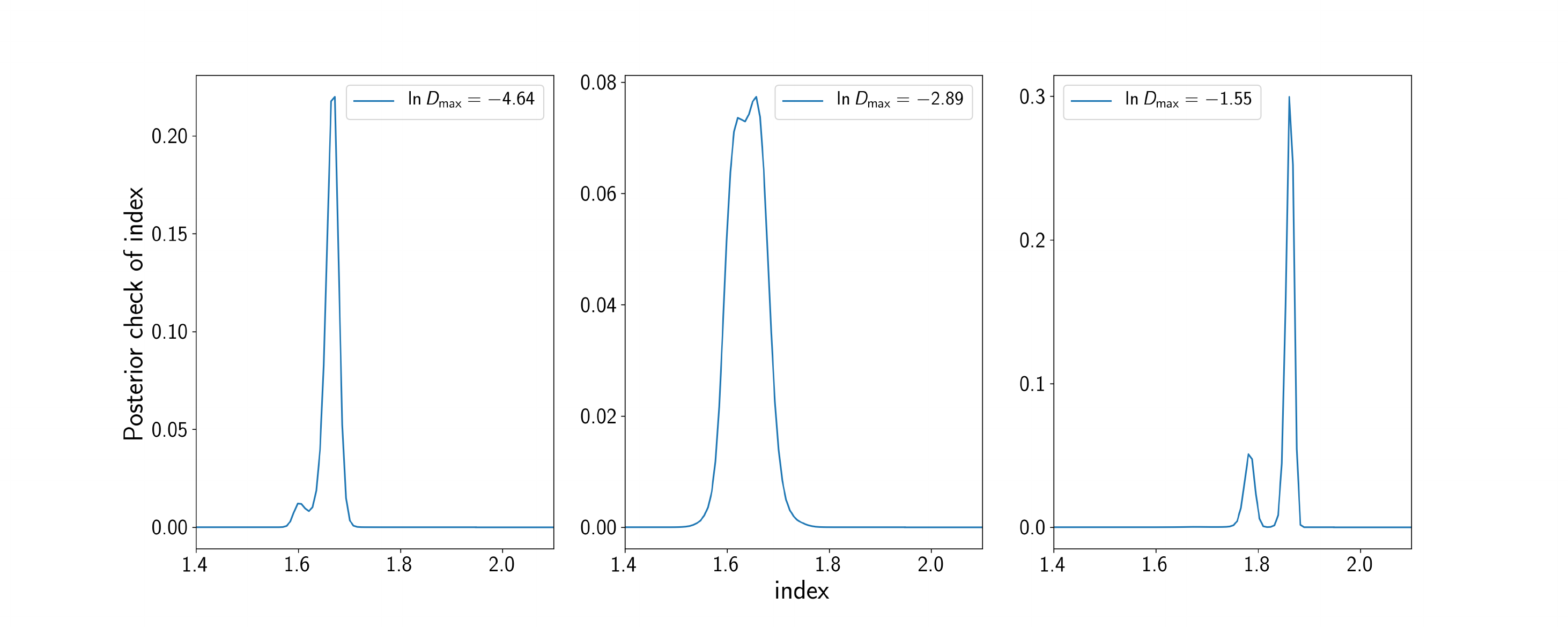}
        \label{posterior_sim_index_with_bg}
    \end{subfigure}
\caption{The posterior distribution of $f_h$ (upper panel) and $\alpha$ (lower panel) for case [$f_h$ = 0.35, $\alpha$ = 1.70] in simulation with the inclusion of background stars. The distributions are evaluated at three different values of $\ln D_{\textrm{max}}$.}
\label{posterior_sim_with_bg}
\end{figure*}

\begin{figure*}
\begin{subfigure}
        \centering
        \includegraphics[width=.50\linewidth]{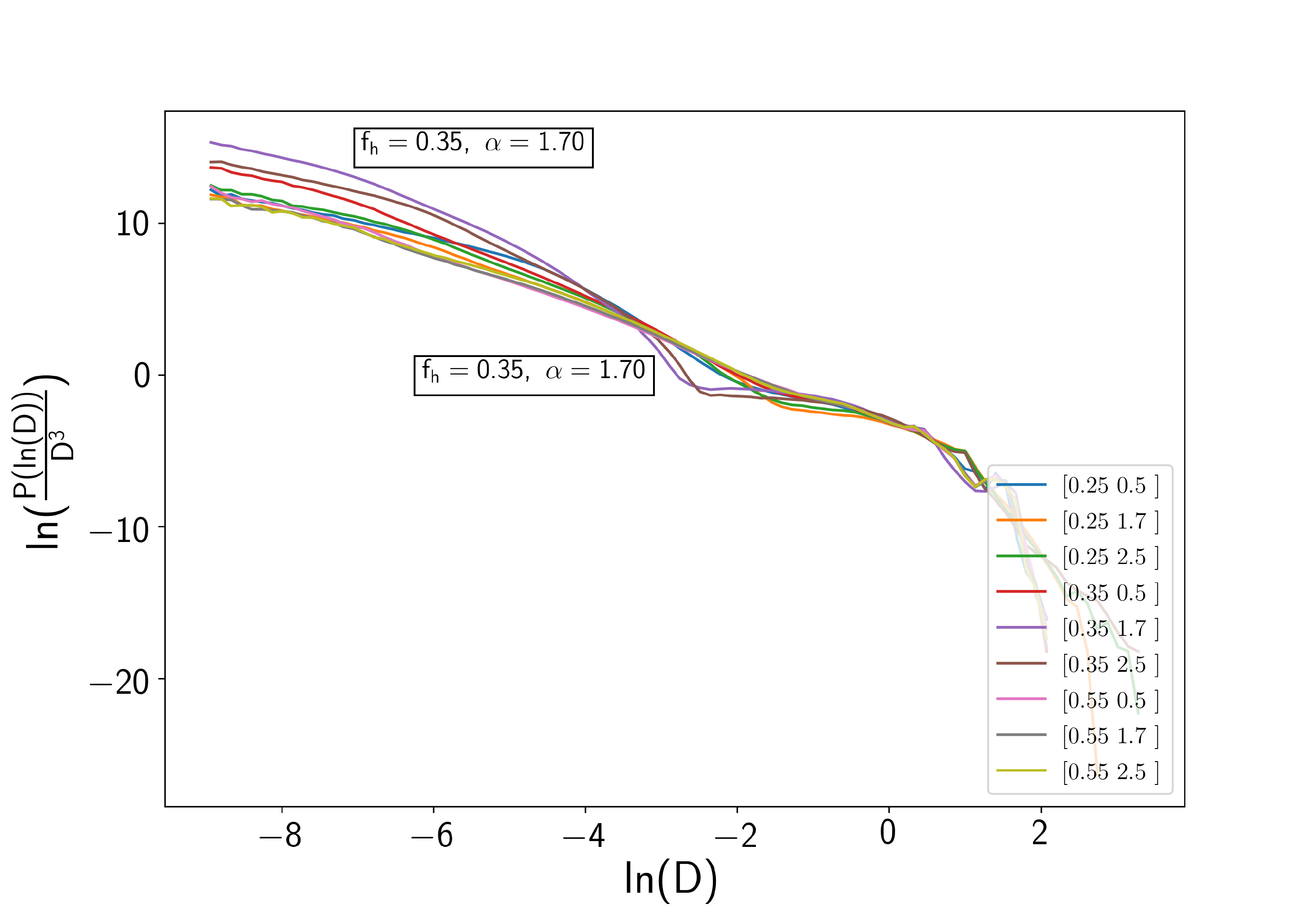}
    \end{subfigure}
    \begin{subfigure}
        \centering
        \includegraphics[width=.60\linewidth]{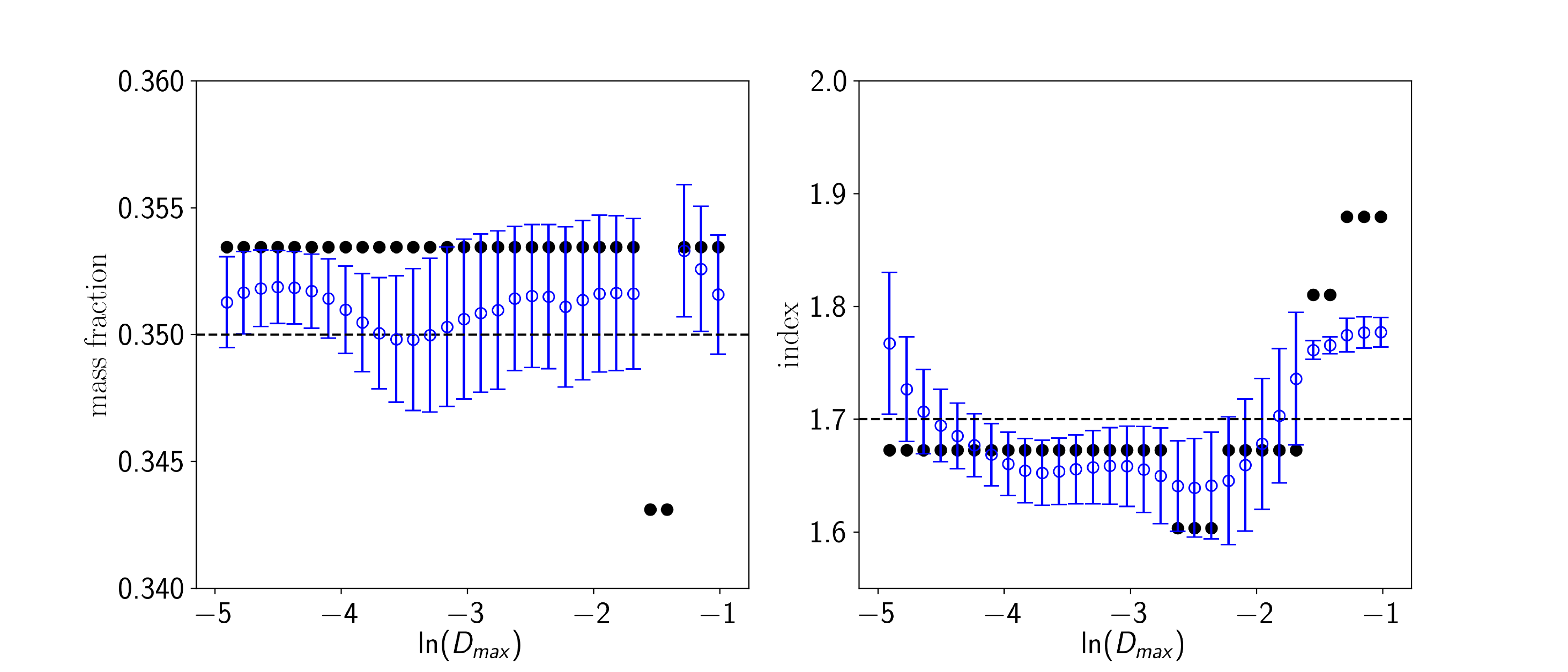}
    \end{subfigure}
    \begin{subfigure}
        \centering
        \includegraphics[width=.60\linewidth]{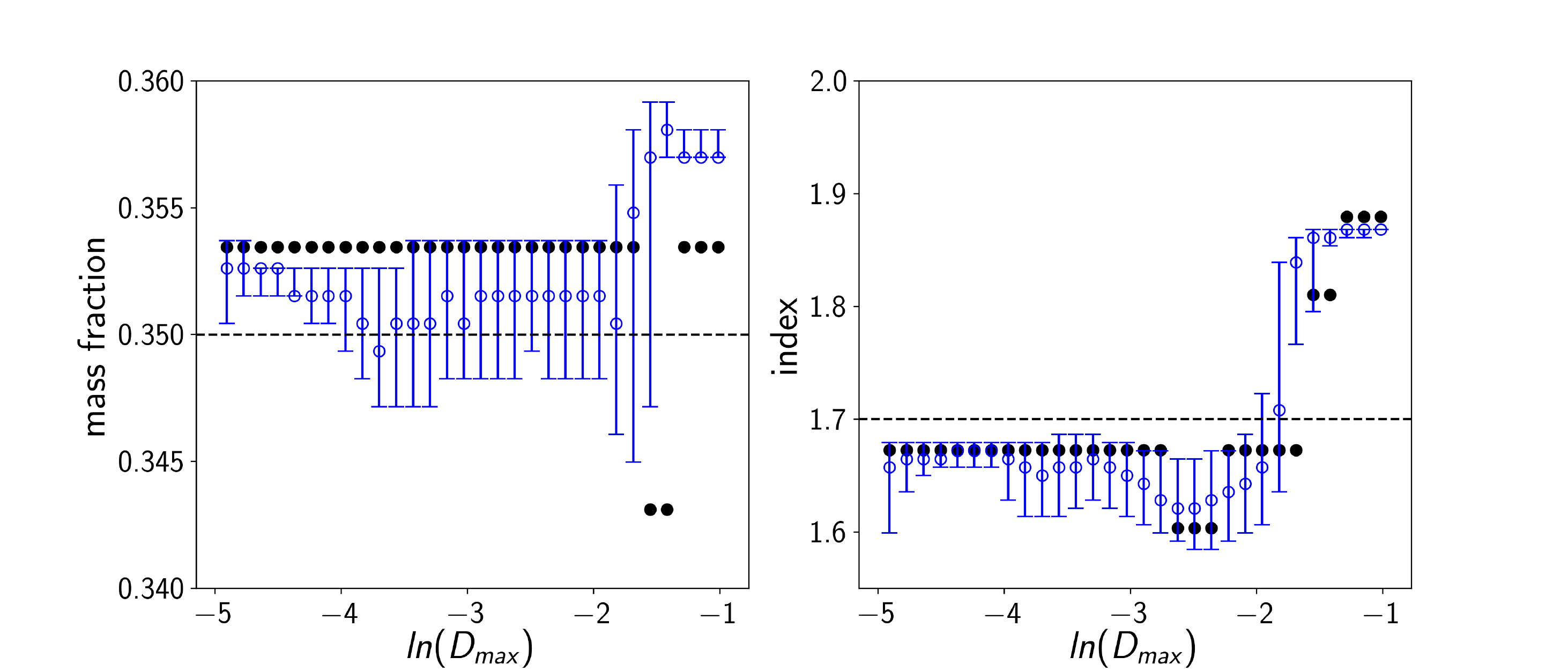}
    \end{subfigure}%
\caption{Top panel: Correlation function $\frac{P(\ln D)}{D^3}$ as a function of the distance in the action space in natural logarithm scale. As expected, the two-point correlation function is maximized while approaching the correct values. Different colors indicate the values of ln($\frac{P(\ln D)}{D^3}$) at different choices of potential. Middle panel: Error bar plot using data from the simulation with the inclusion of background stars. Maximum and errors (blue hollow points) are determined by using the same quadratic fit procedures outlined in Section \ref{ssec::simulation}. Black solid points shows the constraints to the parameters by directly finding the maximum from the likelihood plot. Bottom panel: Error bar plot using the same data from the simulation. Maximum and errors (blue hollow points) are determined by finding the median from the posterior distribution of parameters outlined in Section \ref{ssec::real_data}. As can be seen from these three plots, under this circumstance, neither of $f_h$ nor $\alpha$ estimation is stabilized until $\ln D_{\rm max}$ is smaller than $\sim$ -2, which is also reflected in the correlation function plot.}
\label{error_bar_probability_plots_sim_with_bg}
\end{figure*}

Here, we include a wider prior (Equation \ref{eqn::prior_range}) to capture whether there might be any double peak features that are presented in real data analysis. Figure \ref{posterior_sim_with_bg} shows the posterior distribution of two parameters. Although the posterior distribution of each parameter is dominated by a single peak around the correct value, at some values of $\ln D_{\rm max}$, the inclusion of the background stars seems to cause the distribution present a double peak features. In Figure \ref{error_bar_probability_plots_sim_with_bg}, we present the correlation function as a function of $\ln D$ and the error bar plots, where the error bars are determined by the quadratic fit (as introduced in Section \ref{ssec::simulation}) and from the posterior distribution (as introduced in Section \ref{ssec::real_data}), respectively. From the top panel, we can see that the two-point correlation is maximized when approaching the correct values. To be noted that there is a ``dip'' presented in the correlation function, and this is also reflected in the error bar plots: both of the $f_h$ and $\alpha$ estimation do not drastically change with $\ln D_{\rm max}$ when $\ln D_{\rm max} \lesssim$ -2. This justifies our criteria for choosing the free parameter $\ln D_{\rm max}$. We want to choose a value of $\ln D_{\rm max}$ which gives the least uncertain measurements, and at $\ln D_{\rm max} \lesssim \ln D_{\rm max, optimum}$, the estimations of both parameters should be stabilized (not a strong function of $\ln D_{\rm max}$) and the constraints need to be all consistent with each other (within error bars). For this specific set of simulation, it should be $\ln D_{\rm max, optimum}$ $\sim$ -2 (while in real data, we chose $\ln D_{\rm max, optimum}$ $\sim$ -1). That is the point where our method is still valid and the constraints can be safely obtained with confidence. Also, compare the systematic errors obtained from this simulation with those got from Section \ref{ssec::simulation}, it seems like the inclusion of background stars improve the constraints, ``bringing'' the constraints recovered from likelihood function closer to the correct values (for example, at $\ln {D_{\rm max}} \sim$ -2, the constraint we get here is $f_h = 0.352 \pm 0.003 $  $\alpha = 1.678 \pm 0.058$ from quadratic fit and $f_h = 0.352 \pm 0.003$ $\alpha = 1.657^{+0.066}_{-0.051} $ from median and posterior distribution, while for a stream-only simulation, we have $f_h = 0.352 \pm 0.001 $  $\alpha = 1.634 \pm 0.014$ at same $\ln {D_{\rm max}}$ from quadratic fit only). Therefore, we here conclude that the systematic errors estimated from a stream-only simulation should be conservative, and we thus propagate those systematic errors to the real data analysis.

\section{Combining measurements with unknown systemics}\label{sec::posterior_results_determination}

Here, we discuss how to combine measurements $x_i$ (of a single quantity $x$) that have independent known stochastic gaussian errors $\sigma_i$, as well as an unknown (but independent) systematic gaussian error $\sigma_{\rm sys}$. The joint likelihood is given by:
\begin{equation}\label{eqn::joint_prob_results_determination}
    {\cal L}(x,\sigma_{\rm sys}|\{x_i,\sigma_i\})= \prod_i \frac{\exp\left[-\frac{(x-x_i)^2}{2(\sigma_i^2+\sigma_{\rm sys}^2)}\right]}{\sqrt{2\pi(\sigma_i^2+\sigma_{\rm sys}^2)}}.
\end{equation}
Now, assuming a flat prior on $\sigma_{\rm sys}$, up to some maximum $\sigma_{\rm sys, max}$, we can find the posterior on the parameter $x$:
\begin{equation}\label{eqn::posterior_results_determination}
    P(x) \propto \int_0^{\sigma_{\rm sys, max}} d\sigma_{\rm sys}  \prod_i \frac{\exp\left[-\frac{(x-x_i)^2}{2(\sigma_i^2+\sigma_{\rm sys}^2)}\right]}{\sqrt{2\pi(\sigma_i^2+\sigma_{\rm sys}^2)}}.
\end{equation}

In practice, based on the difference in results we find with and without the cuts, we make conservative choices of $\sigma_{\rm sys, max} =$ 0.02 and 0.2 for $f_h$ and $\alpha$ determinations, respectively. 

\end{document}